# Ancient DNA from 120-Million-Year-Old Lycoptera Fossils Reveals Evolutionary Insights

Wan-Qian Zhao[1,11*], Zhan-Yong Guo[2], Zeng-Yuan Tian[1], Tong-Fu Su[3], Gang-Qiang Cao[1], Zi-Xin Qi[4], Tian-Cang Qin[5], Wei Zhou[6], Jin-Yu Yang[7], Ming-Jie Chen[8], Xin-Ge Zhang[9], Chun-Yan Zhou[10,15], Chuan-Jia Zhu[11], Meng-Fei Tang[1], Di Wu[1], Mei-Rong Song[3], Yu-Qi Guo[11*], Li-You Qiu[12*], Fei Liang[13], Mei-Jun Li[14*], Jun-Hui Geng[1], Li-Juan Zhao[11], Shu-Jie Zhang[11]

**Affiliations:**

[1]School of Agricultural Sciences, Zhengzhou University, Zhengzhou, China

[2]College of Agronomy, Henan Agricultural University, Zhengzhou, China

[3]College of Sciences, Henan Agricultural University, Zhengzhou, China

[4]College of Animal Science and Technology, Henan University of Animal Husbandry and Economy, Zhengzhou, China

[5]Zhengzhou ToYou Bioengineering Co., Ltd, Zhengzhou, China

[6]Jiangsu Institute of Product Quality Supervision and Inspection, Nanjing, China

[7]Haining BoShang Biotechnology Co., Ltd, Haining, China

[8]NewCore Biotechnology Co., Ltd, Shanghai, China

[9]College of Tobacco, Henan Agricultural University, Zhengzhou, China

[10]Jiesai Biotechnology Co., Ltd, Haining, China

[11]School of Life Sciences, Zhengzhou University, Zhengzhou, China

[12]College of Life Sciences, Henan Agricultural University, Zhengzhou, China

[13]College of Paleontology, Shenyang Normal University, Shenyang, China

[14]National Key Laboratory of Petroleum Resources and Engineering, College of Geosciences, China University of Petroleum (Beijing), Beijing, China

[15]Jiangsu Jiayu Biomedical Technology Co., Ltd, Jiangyin, China

*Corresponding authors and emails:

wqzhao@zzu.edu.cn; guoyuqi@zzu.edu.cn; qliyou@henau.edu.cn; meijunli@cup.edu.cn

**Abstract:** High-quality ancient DNA (aDNA) is essential for molecular paleontology. Due to DNA degradation and contamination by environmental DNA (eDNA), current research is limited to fossils less than 1 million years old. The study successfully extracted DNA from *Lycoptera davidi* fossils from the Early Cretaceous period, dating 120 million years ago. Using high-throughput sequencing, 1,258,901 DNA sequences were obtained. We established a rigorous protocol known as the "mega screen method." Using this method, we identified 243 original *in situ* DNA (oriDNA) sequences, likely from the Lycoptera genome. These sequences have an average length of over 100 base pairs and show no signs of "deamination". Additionally, 10 transposase coding sequences were discovered, shedding light on a unique self-renewal mechanism in the genome. This study provides valuable DNA data for understanding ancient fish evolution and advances paleontological research.

Keywords：ancient DNA，environmental DNA, original *in situ* DNA, deamination, transposase

## Introduction

The acquisition of high-quality aDNA data is crucial for molecular paleontological research. While scientists have successfully extracted DNA from fossils dating back about 1 million years, extracting amplifiable DNA from most fossils remains a huge challenge[1]. Much of the oriDNA has been lost, degraded, or crosslinked with macromolecules, rendering it unsuitable for use as amplification templates. Lithological fossils can adsorb external DNA molecules; some adsorbed molecules may be further embedded or replaced by eDNA[2,3,4]. Most fossils contain predominately bacterial eDNA, which can overshadow oriDNA sequencing signals[5,6]. Furthermore, studies of fossilized ancient primates and environmental samples have shown that due to unfavorable burial conditions, "deamination" of bases on the side chains of DNA molecules often occurs, leading to distortion of sequencing information[7,8].

The conditions under which fossils are buried and the mechanisms of their formation are crucial for the preservation of oriDNA. Research has shown that certain fossilized volcanic tuffs, rapidly encased by volcanic ash and formed from living organisms, still contain intact nuclei of *in situ* organisms preserved at the outlines of their remains. For instance, intact nuclear structures are commonly found in fossils such as Jurassic plants and the Jehol Biota; nucleic acid stains reveal the presence of nucleic acid material in the nucleus location[9,10,11,12,13]. While these findings are fascinating, they also give rise to important questions. For instance, if the nucleus tests positive for nucleic acids, are these derived from oriDNA or eDNA? Moreover, if the nucleus still contains oriDNA, has it been cross-linked with other macromolecules or degraded to short-stranded DNA (< 40 bp), making it unusable as a valid amplification template? Or could it consist of long-stranded DNA sequences (>100 bp) that are still suitable as amplification templates? Addressing these questions is key to driving the most critical technological advancements in molecular paleontology[14]. *Lycoptera davidi* is an extinct ancient freshwater ray-finned fish belonging to the family Lycopteridae and the order Osteoglossiformes. This fish lived during the early Cretaceous period in present-day northeastern China. As prominent fossils of the Jehol Biota, it holds significant biostratigraphic importance for the early Cretaceous period (J3-K1). The fossil discussed in this study was discovered in the lower strata of the Yixian Formation in Shangyuan Town, Beipiao City, northeastern China, dating back approximately 120 million years ago. The fossil was preserved in sedimentary rock and formed under hydrostatic conditions when the fish body was quickly encapsulated by volcanic ash[15,16].

Fossil DNA extraction was conducted using nanoparticle affinity beads[17], leading to the establishment of a DNA library. Subsequently, second-generation sequencing yielded 1,258,901

DNA fragments (see Materials and Methods). To select oriDNA, it is necessary to remove eDNA contamination. This contamination can come from eDNA incorporated into fossils during the pre-lithification or post-lithification stages (paeDNA), and DNA from recent or present times (preDNA). Using the "Mega screening method", 243 oriDNA fragments were identified (Figure 1, Tables S3A and S4). These fragments are characterized by an average length exceeding 100 base pairs, rendering them suitable as templates for amplification. Further analysis revealed no identifiable "deamination" of the side-strand bases of these DNA sequences. Although the number of oriDNA obtained in this study is limited, it offers a rare glimpse into the deep-ancient fish genome. This includes insights into the genomic relationship between *Lycoptera* and local freshwater fishes, a transposase gene self-renewal mechanism known as "coding region sliding replication and recombination", and the mechanisms underlying the rapid diversification of Cretaceous fish species. With the improvement of aDNA extraction methods and the increase of fossil samples, we can obtain more oriDNA sequences, which are expected to bring us more knowledge about ancient life.

## Results and Discussion

### DNA information extraction using the "Mega screening method"

Utilizing the "Mega screen method," 1,258,901 DNA fragments of diverse origins spanning different periods were classified, and relevant biological information was extracted. This methodology involves two steps outlined in Materials and Methods and Figure S1.

### All sequences were divided into subsets based on lineage

Initially, as the Lycoptera genome data was unavailable, we conducted a BLAST search to align the fossil DNA sequences with all sequences in the NCBI database (Version 5, Nucleotide sequence Database) using the "minimum E-value mode". We processed data from the 1,258,901 reads, creating subsets based on genealogical lineages and determining the total number of sequences (TS) within each subset. Among these, 674,472 total sequences (TS) were classified as bacterial sequences, 131,312 TS as primate sequences, 30,211 TS as angiosperm sequences, and 26,852 TS as fungal sequences. Additionally, 19,747 TS were categorized in the arthropod sequence subset. With the 10,000 Fish Genomes Project (F10K) nearing completion, the genomes of representative species have all been sequenced, providing a more comprehensive reference database for fish sequence matching. Leveraging these advancements, we have identified 11,313 sequences, aligning with the subset of ray-finned fish sequences (Table S1).

### The ray-finned fish subset dominated by aDNA

To establish the sequence screening threshold, we initially set the E-value below 1E-07 and identified the sequences that met this criterion as qualified sequences (QS). Among the screened sequences, 693 met the threshold, while 10,620 did not. We hypothesize that the inability of certain *Lycoptera* sequences to meet the threshold could be attributed to the absence of *Lycoptera* genome data, inadequate matches, and limited similarity to contemporary fish genomes. Moreover, some unaligned sequences might have origins in non-fish species that have not been sequenced. Indeed, each species possesses its distinctive significance. Due to these limitations, we couldn't analyze these sequences in greater detail in this work.

To further assess the closeness of a subset to the respective modern genomes, we introduced the metrics, Affinity Index, and set the thresholds at 90% and 97.5%, respectively (see Materials and Methods). The percentage of the ray-finned fish subset passing the two threshold sequences was 8.51% and 4.91%, respectively. Compare this to other subsets; for example, the two percentage values for the human subset were 94.78% and 91.56%, respectively; while the percentages for amphibians and reptiles were 15.15%, and 12.12%, respectively. In addition, our analysis should introduce some other factors, such as geographical factors; the local geography is as follows: The cold climate and hilly topography of the region result in a sparse distribution of amphibians and reptiles, which limits their contribution to the local eDNA; therefore, the Affinity Index of this subset can be used as a background value to screen the corresponding lineages. From the above, it can be assumed that the ray-finned fish subset meets our criteria and most of the sequences in this subset do not come from the local environment, and they are aDNA or oriDNA. The Affinity Index for the human subset is high (Table S1), and most of the DNA fragments in this subset are recently contaminated preDNA sequences. As many extant species have yet to be sequenced, the values of the Affinity index for some subsets are correlational and may fluctuate in future studies.

**Some features of ray-finned fish DNA sequences**
When comparing the DNA sequences of 243 ray-finned fish (Tables S3A and S4) to the sequences in the human subset, it was observed that they are similar in length and GC content. However, there were notable differences in metrics such as Per Identity, Query Cover, average number of base deletions, and QS/TS (%). The human subset displayed a lower frequency of insertions/deletions when compared to that of ray-finned fish, due to the sequence differences between preDNA and aDNA. The DNA sequences within the human subset demonstrated aDNA characteristics overall, as depicted in Table S2, providing a sound basis for effectively distinguishing the ray-finned fish subset (Subset Affinity: 52.87%; see Materials and Methods) from the preDNA-dominated human subset (Subset Affinity: 97.15%).

**Identification of each sequence in the ray-finned fish subset**
In the subset of ray-finned fish, we utilized the "lowest E-value mode" to identify 693 high-quality matches. However, some of these matches had the same E-value when hitting genomes of different species, which made it difficult to determine their genealogical affiliation. The "MS mode" for analysis was used to enhance the hit reliability and the uniqueness of the search result (see Materials and Methods). As a result, we confirmed that 243 DNA sequences were specifically mapped to the ray-finned fish genome. Consequently, we concluded that the only genealogical origin of these sequences was the ray-finned fish genome.

It is reasonable to deduce that the sequences of the ray-finned fish subset can be divided into three parts: oriDNA sequences, paeDNA sequences, and preDNA sequences. By excluding paeDNA and preDNA sequences, oriDNA sequences can be recognized. These 243 sequences were then categorized into three groups: the "local fish group", consisting of 180 sequences (Table S4); and the "non-local fish group (including marine fish)," which contains 49 sequences. The third group, the "other ray-finned group," comprises 14 sequences that could not be classified at the order level, and its Affinity Index value is only 77.51%, indicating a notable distinction from the known genome.

**The Lycoptera oriDNA:** We further determined the source of the sequences of the ray-finned fish subset by combining the influence of geographical factors, geological changes, and evolutionary laws. Since there has been no marine transgression since the Cretaceous, the 49 sequences of the non-local fish group should be oriDNA. This result also shows a genomic connection between the *Lycoptera* and modern marine fish. In the local fish group, the 180 sequences mainly belong to Cypriniform genomes. The average values of the Affinity Index of the above two groups of sequences are 80.34% and 83.85%, respectively, and their conservation levels (average values) are almost equal. In addition, the fossil site is currently a hilly area without lakes and rivers. These conditions indicate that most of the sequences in the local fish group cannot be preDNA. At the same time, considering freshwater fish eventually become prey to other species, it is difficult for their DNA to enter the environment and become eDNA. Therefore, we can roughly infer that most of the 180 sequences in the local fish group and the 14 sequences in the non-order group could be oriDNA (Tables S3 and S4).

**Fossil DNA: The multiple resources hypothesis**
The Lycoptera fossil can be defined as a rock that contains internal voids. We have developed a method to quantify the percentage of these internal spaces about the total volume of the rock, which ranges from 10.8% to 11.3% (see Materials and Methods). Due to diffusion and osmosis, when the surrounding environment becomes arid, various molecules not embedded in the fossil may be lost as internal liquid water drains away. Subsequently, when water returns to the environment, DNA molecules may be transported into the dry voids within the fossil and become trapped. Furthermore, some molecules might be encased by minerals in the water and deposited onto the inner surfaces of the rock. However, if the environment dries out again, free molecules could escape.

Given that advanced primates exhibit the social behavior of burying their dead, DNA from the genomes of these primates, as well as from associated species such as scavenging animals, prey, and surrounding vegetation, can continuously be introduced into the environment of long-term settlements, contributing to environmental DNA (eDNA). Consequently, eDNA can infiltrate a fossil after its formation, and some of this DNA may also become buried in minerals within the water and permanently retained in the fossil. Thus, it is reasonable to conclude that eDNA from later periods can enter the internal voids of fossils formed during the Cretaceous period.

The mechanism of DNA preservation in Lycoptera fossils remains unclear. Water is the primary factor in DNA deamination reactions[18]. Fossils in sedimentary rocks are formed when organisms are rapidly covered by volcanic ash in a calm water environment (Figure 2). This forms a protective shell around the organism, which prevents the chemical groups on the DNA molecules from coming into contact with water, allowing nucleic acids to be preserved for a long time and possibly protecting them from external elements such as oxygen and chemicals. Although some oriDNA fragments may be lost or degraded due to geological processes, remnants may still exist. The eDNA in fossils comes from the environment, adsorbed on rock surfaces, cavities, and cracks; part of it has become permanent sediment[2,3,4,5].

Our proposal delineates three sources of fossil DNA, as illustrated in Figure 2: the first being oriDNA before fossil formation, including the fish DNA and DNA from *in situ* species, such as food found in the stomachs of fishes, small aquatic organisms, and fish parasites. The second source is paeDNA, and the third is preDNA. preDNA primarily originates from plants and animals associated with human activities and parasites, bacteria, and viruses. Historically, the Beipiao area was a center of agriculture and animal husbandry in Chaoyang City and likely served as the main source of preDNA due to human urban settlement, burial practices, and large-scale cultivation of angiosperm crops. Ancient eDNA has infiltrated fossils since the remains were buried in mud under hydrostatic conditions, particularly during periods of considerable lacunae before the fossilization of dense formations. Following full petrification, the impermeability of shale presents challenges for water-soluble compounds to permeate the pores. Despite these, the possibility remains that minuscule quantities of matter continue to seep into the fossils.

To elucidate the origin of DNA in fossils, we developed a model diagram based on our hypotheses and speculations (Figure 1, right): (1) Before the fossil forms a dense structure, the internal DNA fragments are primarily composed of oriDNA. At the same time, paeDNA molecules gradually enter the pores and remain inside. (2) As the fossilization process progresses, the interior becomes denser, causing the voids that can accommodate foreign molecules to decrease and sometimes disappear entirely. Concurrently, the ingress and egress of internal DNA molecules is obstructed, embedding the DNA within the rock formation and rendering it immobile. (3) Following this process, the surface area of the internal pores continuously decreases, reducing their capacity to accommodate foreign molecules. The remaining open gaps are relatively robust and resistant to collapse or blockage. Subsequent generations of external paeDNA can still enter and adsorb to the surface of the pores but are rarely embedded and remain predominantly mobile. (4) These mobile DNA fragments are displaced over time by later generations of external molecules. This replacement process persists over time, with older molecules being supplanted by newer ones up to the present day. (5) The oriDNA and paeDNA embedded within the fossils undergo continuous degradation. Consequently, this leads to a decrease in oriDNA and paeDNA and an increase in preDNA. In this study, a limited number of fish fragments (11,313 TS) were detected, whereas a larger quantity of human DNA fragments (127,977 TS) was identified, predominantly originating from preDNA (Table S1). This exemplifies the phenomenon of so-called "fewer old DNA molecules and more new ones" in fossils (Figure 1).

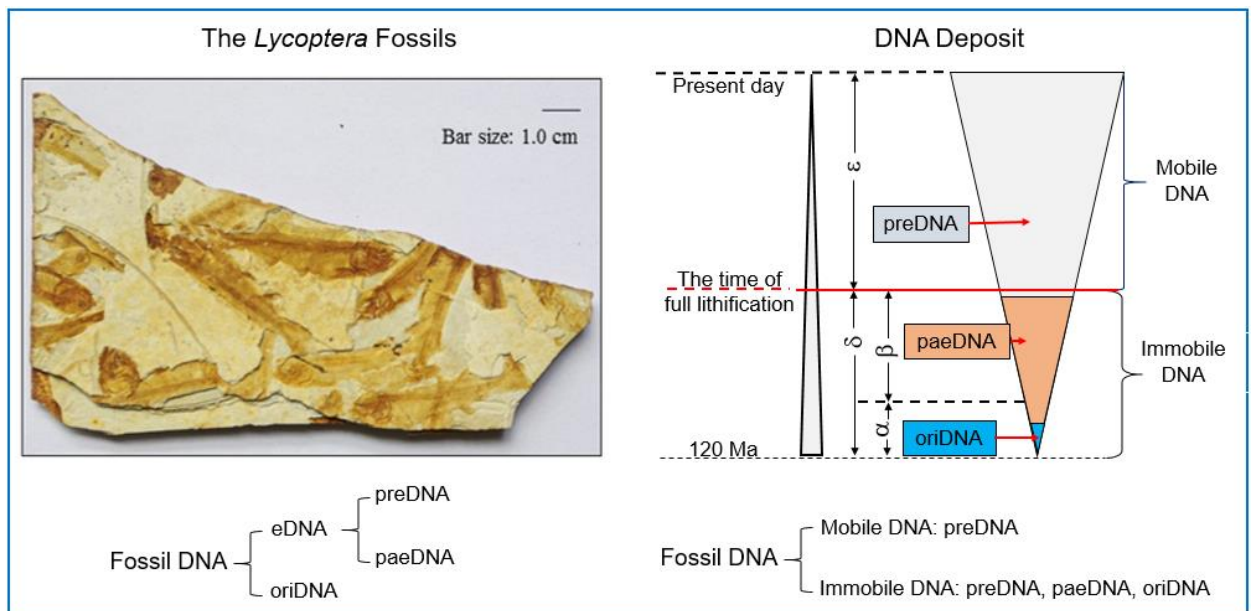

**Figure 1 The fossil formation process and the DNA contained therein**

**The *Lycoptera* fossils** were found in Beipiao City, Chaoyang City, Liaoning Province. The longitude is 120.84 and the latitude is 41.60. **The process of fossil formation and preservation of DNA:** During early lithification (α), a significant amount of volcanic ash descends into the water, rapidly forming a thick, viscous paste that envelops and buries the fish within a tranquil, undisturbed water environment. This results in enclosed inclusions isolated from their surroundings, allowing some paeDNA to become trapped within them. As lithification progresses (β), these inclusions gradually desiccated and were compressed into a thin, dense layer known as the "fish layer", each approximately 1 mm thick. The brown sections in the shape of fish within a plate represent the fish remains, with the surrounding whitened sections acting as the cofferdam area. Additionally, there is the "non-residual layer" without any fish remains. After complete lithification (δ), the "fish layer" and the "non-residual layer" are mixed and overlapped in multiple layers to form dense tuff plates and are further petrified into lithological fossils. After the soft tissue has decayed, the area between the skeletal remains becomes very porous, with many tiny cavities connected to surface cracks in the fossil rock. This allows for the transfer and exchange of internal and external molecules, providing a pathway for eDNA to enter the fossil.

**The "deamination" did not occur in Lycoptera oriDNA and some paeDNA sequences**

Although the "deamination" phenomenon is one of the ubiquitous damages of aDNA, can it be used as the primary condition for judging ancient DNA, that is, a fragment without "deamination" is not aDNA/oriDNA[7,8]? However, this phenomenon was not seen in the 243 oriDNA sequences evaluated in this study (Tables S3A and S4). This difference may be attributed to the closed structure of fossils after diagenesis, which may limit the exposure of organisms to external conditions, thereby protecting oriDNA from the influence of "deamination" factors. However, it is worth noting that some DNA fragments distributed on the surface of fossils, such as paeDNA fragments or preDNA fragments, may still be exposed to the environment without rock protection. These fragments may have undergone "deamination", which could further increase the gap between their sequences and modern genomes. The seriously distorted sequences have been removed by examining the MS pattern.

We identified numerous fragments that aligned with the Pan-genome in our DNA extracts. The affinity values for these fragments range from 38% to 90%, and the "Mega screen method" suggests a "unique" relationship between some sequences (IDs: 534-555) and the Pan-genome (Table S3B). Additionally, several facts support an opinion: these fragments likely originate from unsequenced genomes of Pan or other Hominoidea species, potentially representing extinct ancestral species; e.g. (1) The Pan lineage is not distributed throughout the East Asia region, and there are no fossil records of Pan species in the fossil production area, indicating these fragments are not a result of eDNA contamination; (2) our laboratory has never been exposed to samples containing Pan DNA. Furthermore, these fragments do not fully align with known Pan-genomes, and there is no existing sequencing record, ruling out experimental contamination; (3) Pan genomes have been extensively sequenced, and the observed affinity gaps between the known ape species and the sequenced genomes should not be so pronounced. (4) No discernible "deamination" phenomena were observed in these sequences.

**The "deamination" is not suitable as the primary criterion for identifying aDNA fragments**

In the method developed by the Pääbo team, "deamination" is the main criterion for identifying aDNA fragments[19,20,21,22,23,24]. However, several points challenge this perspective: (1) The assumption that the fossil oriDNA originates from a single individual of the corresponding species overlooks the potential for environmental eDNA contamination from either the same or different species over time following the formation of the fossils; (2) The selection process involves only a limited number of closely related genomes as BLAST references, failing to consider the broader influence of numerous environmental species and neglecting to utilize all available genome sequencing databases; (3) There is insufficient consideration of the unreliability of BLAST results, particularly when the E-value is excessively high, rendering them outside a credible interval; (4) The oriDNA fragments that matched the Pan-genomes we identified did not display

"deamination," suggesting that this characteristic is not a necessary condition for the identification of oriDNA (Table S3A and S3B). Consequently, the genomes of Neanderthals, Denisovans, and several species assembled by Pääbo's team likely represent "mixed genomes." The belief that "deamination" is necessary for identifying aDNA/oriDNA is incorrect which has hindered the search for long-chain aDNA or oriDNA, which could offer more comprehensive genetic information. As a result, several significant flaws in the literature concerning Pääbo's research on ancient species should be addressed, and these works should be re-evaluated.

**Genomic links between the *Lycoptera* and other fishes**

The relationship between the genomes of ray-finned fish and their morphological development is not fully understood. The classification of these fish is mostly based on their physical characteristics. A study by Zhang concluded that *Lycoptera* belongs to the order Osteoglossiformes[25]. Early researchers, such as Cockerell and Berg, suggested that species of the Cypriniformes order originated from Jurassic *Lycoptera* [26,27], while Rosen proposed that they may have originated from Gonorynchiformes[28]. The oldest known carp fossils date back about 60 million years ago. Carp fishes are found in freshwater environments worldwide, except Antarctica, South America, and Australia, indicating that their origins predate the breakup of the Pangaea continent. Molecular paleontological studies have provided valuable insights into the origins of carp fishes. Tao's research suggests that the genomes of carp fishes can be traced back to the early Jurassic period, around 193 million years ago[29]. Over 75% of the more than 180 sequences of *Lycoptera* oriDNA in the present study align with the carp genome, indicating a genomic connection between them. Furthermore, the genetic relationship of *Lycoptera* with carp fishes and other native fishes requires further investigation. Although the total length of the oriDNA obtained in this study is limited, further exploration is necessary for a comprehensive phylogenetic analysis.

**Decoding fossil DNA: Revealing parasites and prey**

In the fossil DNA, the ID: 282 sequence (Table S3) corresponds to the 28S rDNA sequence of the class *Ichthyosporea*. These are flagellated microorganisms that belong to the Opisthokonta (unranked). This small fish parasite's taxonomic position lies between animals and fungi[30]. Additionally, we have identified 4 DNA sequences corresponding to the *Macrobrachium nipponense* genome (IDs: 285-287) with a Subset Affinity value of 83.24%, of which 3 DNA sequences were highly divergent from the genomes of existing species. Two other sequences corresponding to *Penaeus vannamei* and *Rhynchothorax monnioti* (sea spider) genomes, both non-native species, were hypothesized to be derived from close relatives in their ancient decapod relatives (Table S3A, ID: 283, ID: 289). Upon using 3D X-ray computed tomography to scan the fossils, we found no remains of other animals. This result suggests that these fragments originated from the parasites and prey of the *Lycoptera* and that they are oriDNA rather than eDNA (Figure S1). It also brings new insights for studying the evolution of them.

**New mechanism for generating transposase-encoding sequences**

Transposons comprise over 40% of the fish genome. The transposition process relies on transposase as a crucial component[31,32]. According to Frith's study, the transposon sequences in modern genomes can be likened to "protein fossils"[33]. Their sequences exhibit significant similarity across various organisms, including prokaryotes, arthropods, and mammals. Transposons can introduce new DNA sequences into the genome. The discovery of 10 transposase gene fragments in the fossil DNA has uncovered a novel mechanism for coding region formation, termed "coding region sliding replication and recombination". This mechanism differs from known conventional mechanisms such as horizontal transfer, vertical inheritance, and gene and chromosome duplication[34].

This mechanism comprises three distinct forms of coding: (1) initial sliding replication in a segment of the coding region, where two distinct fragments are generated by shifting the reading frame and then combined into a new coding sequence with coding fragments from other transposase genes (Figure 2A); (2) sliding replication followed by the creation of distinct fragments, which are then assembled in tandem with each other, or with other transposase gene coding fragments, to form the coding sequence of the new enzyme (Figures 2B and 2C); (3) "Indel" events occurring at a single locus: when an insertion or deletion of a base in a gene's coding region results in a shift in the reading frame, directly impacting the transcription process by DNA. Figure 2D suggests a "deletion-correction mechanism" within the genome that can utilize the coding region where the "Indel" has occurred to generate a novel transposase coding sequence.

**A**

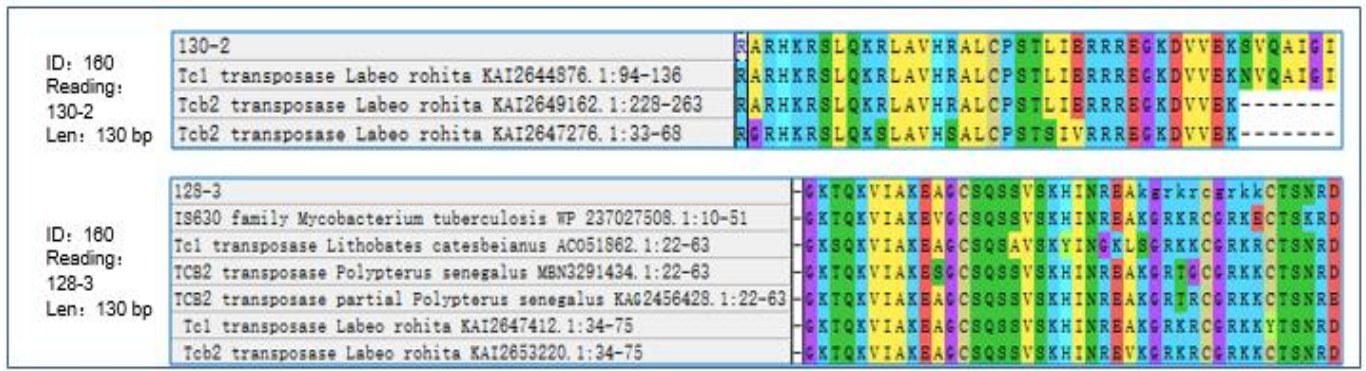

**B**

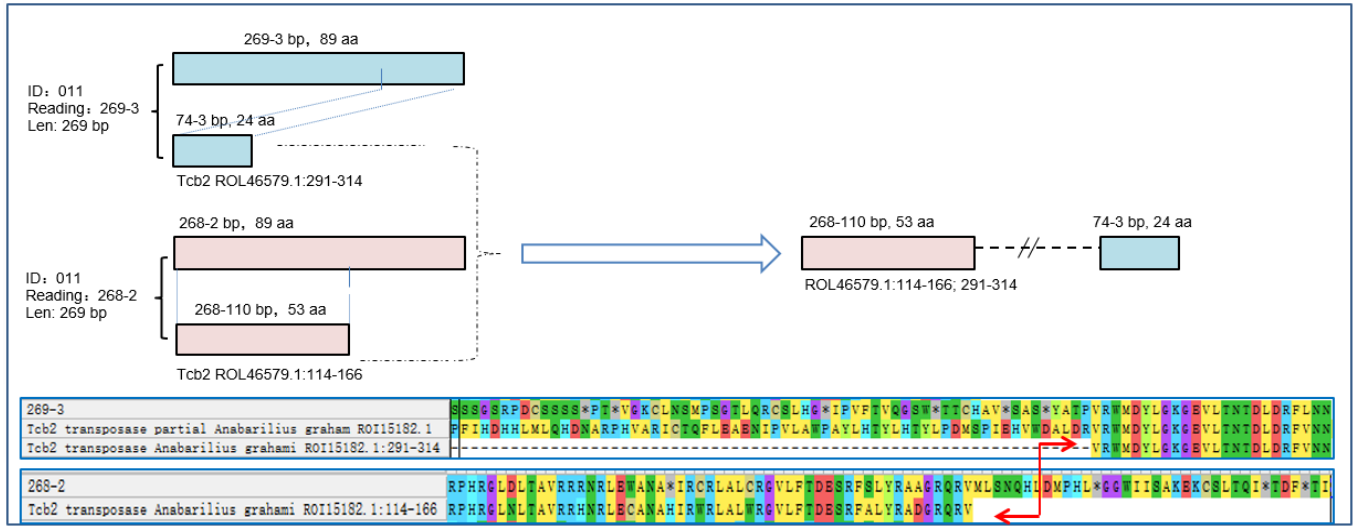

**C**

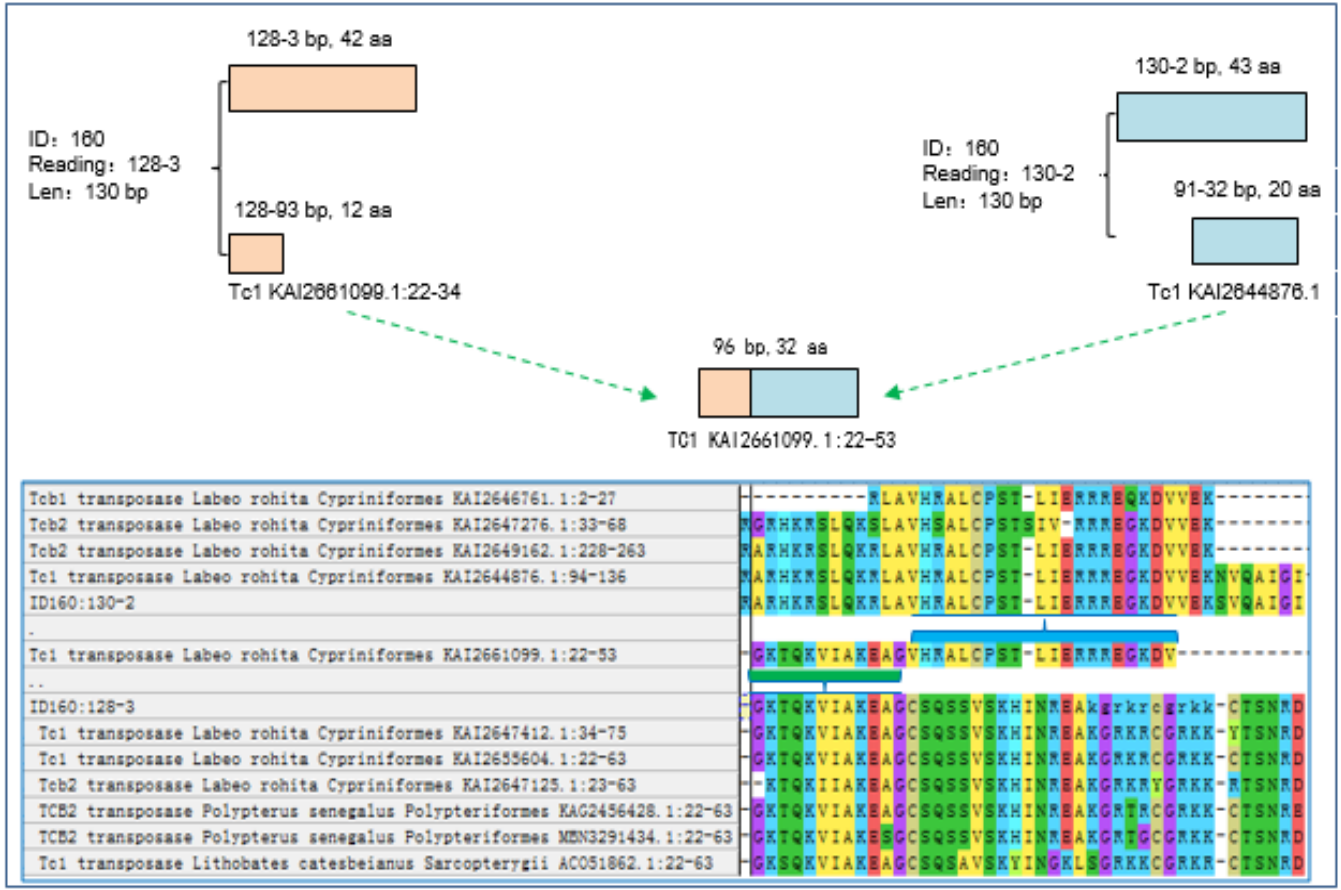

**D**

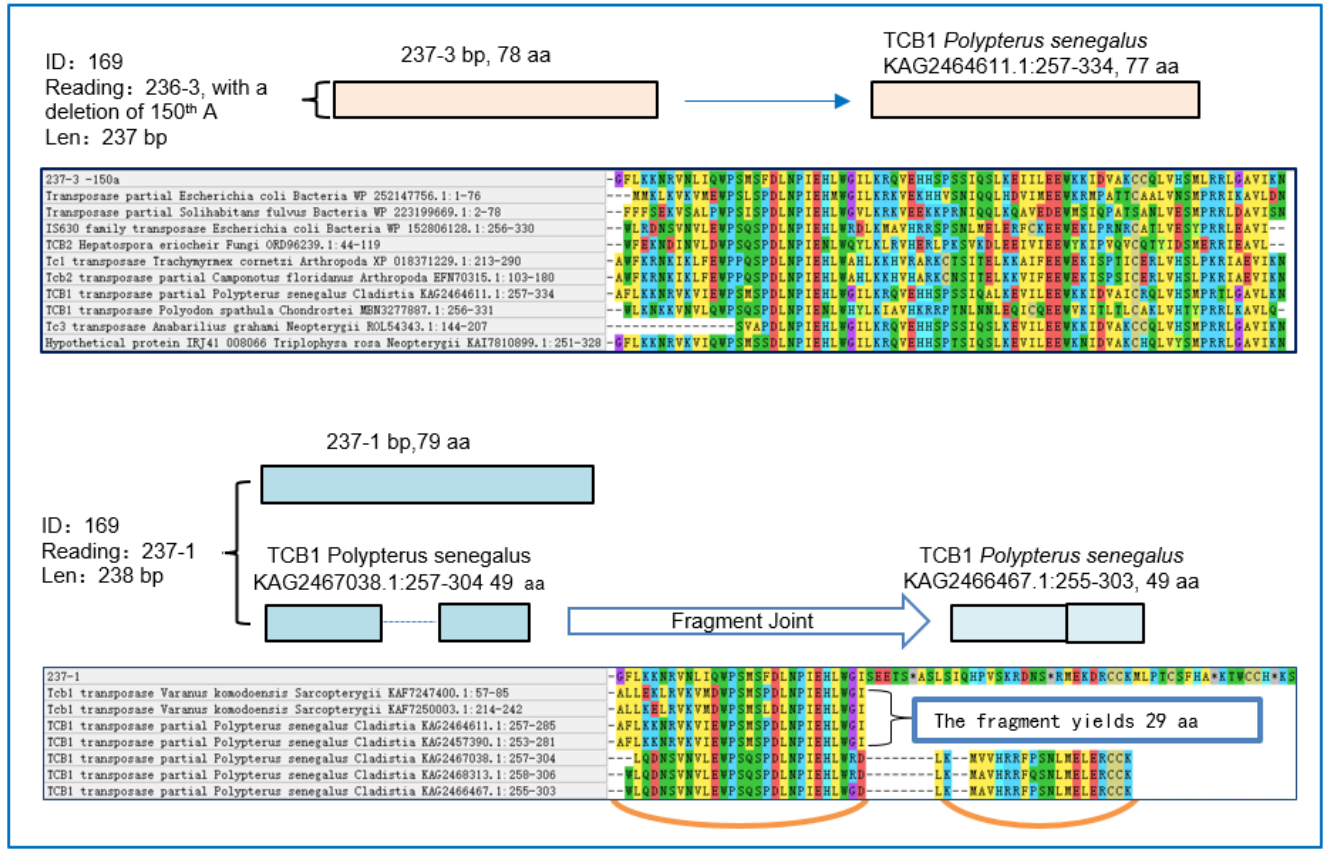

**Figure 2 Proposed model for New mechanism for generating transposase-encoding sequences.**

A, Multiple fragments produced by sliding replication are spliced and fused with other gene fragments separately to form a new enzyme. B, the two fragments generated by sliding duplication were spliced together with other gene fragments to constitute the coding region of a new enzyme. C, the two fragments generated by sliding duplication are joined together and spliced with other gene fragments to form the coding region of a new enzyme. D, A site in the coding region of the transposase undergoes "Indel", but still passages clips to constitute the coding region of the new enzyme.

During the Mesozoic period, a significant evolution of ray-finned fish occurred, leading to the emergence of numerous new species[35]. The genomes were able to generate a great amount of new transposase sequences through the process of "coding region sliding replication and recombination", without relying on external DNA input. This capability likely resulted in the rapid expansion and divergence of the genomes, contributing to the quick emergence of new species. Additionally, transposases can be integrated into the coding regions of functionally regulated proteins. For instance, the ID: 011 sequence, identical to Tcb2, was inserted into the growth inhibitory specific protein 2 (GAS2) gene of the ancestor of *Anabarilius grahami*. The N-terminus of Tcb2 and GAS2 share a common sequence of 52 amino acids and the C-terminus matches the conserved GAS2 sequence (Figure 3). These findings indicate that this translocation likely occurred before fossilization and is evidence of the early fish genome's evolution, and the pathway for expanding genome content. This also serves as molecular evidence for the early divergence of carp ancestral genomes.

**Figure 3 Transposase sequence fused in the GAS2 coding region**

We established a new protocol the "mega screen method" to select aDNA fragments. Using this method, we have successfully identified 243 oriDNA fragments from *Lycoptera davidi*, which date back to the early Cretaceous period, about 120 million years ago. The findings enhance our understanding of fossil DNA composition, set the groundwork for creating a DNA database of ancient species, and present potential avenues for bringing back ancient species. Additionally, they offer clear reference points for exploring the evolutionary connections between ancient and contemporary species. In addition, we have discovered a new mechanism for transposase gene formation, revealing the intrinsic dynamics of rapid genome expansion despite not relying on exogenous DNA input. We hypothesize that this mechanism is closely related to the swift emergence of new ray-finned fish species in the Cretaceous period.

Surprisingly, this research provides a new perspective on "Panspermia," the theory for exploring the origin of Earth life. Inspired by the study of extraterrestrial nucleobases in the Murchison meteorite[36], we propose the possibility that if rock fossils, such as the *Lycoptera* fossils, are transported into space, the DNA inside could potentially remain dormant, effectively and uniquely preserving life. Until these fossils land in a suitable extraterrestrial environment and release the contents, becoming the seeds of life, catalyzing the colonization of the planet.

### References and notes

1. Orlando, L. et al. Ancient DNA analysis. *Nat Rev Methods Primers* 1, 14 (2021).
2. Pedersen, M. W. et al. Ancient and modern environmental DNA. *Phil. Trans. R. Soc.* B370:20130383 (2015)
3. Foucher, A. et al. Persistence of environmental DNA in cultivated soils: implication of this memory effect for reconstructing the dynamics of land use and cover changes. *Sci Rep* 10, 10502 (2020).
4. Beng, K. C. & Corlett, R. T. Applications of environmental DNA (eDNA) in ecology and conservation: opportunities, challenges and prospects. *Biodiversity and Conservation* 29, 2089-2121 (2020).
5. Miao, B. et al. Assessment of contaminants associated with gold-standard ancient DNA protocols. *Science Bulletin* 68:5-9 (2023).
6. Liang, R. et al. Genome-centric resolution of novel microbial lineages in an excavated Centrosaurus dinosaur fossil bone from the Late Cretaceous of North America. *Environmental Microbiome* 15, 8 (2020).
7. Briggs, A. W. et al. Removal of deaminated cytosines and detection of *in vivo* methylation in ancient DNA. *Nucleic Acids Res*. 38(6):e87 (2010).

8.  Kjær, K.H. et al. A 2-million-year-old ecosystem in Greenland uncovered by environmental DNA. *Nature*. 612(7939):283-291 (2022).

9.  Pang, K. et al. The nature and origin of nucleus-like intracellular inclusions in Paleoproterozoic eukaryote microfossils. *Geobiology* 11, 499-510 (2013).

10. Bomfleur, B., McLoughlin, S. & Vajda, V. Fossilized nuclei and chromosomes reveal 180 million years of genomic stasis in royal ferns. *Science* 343, 1376-1377 (2014).

11. Yin, Z. et al. Nuclei and nucleoli in embryo-like fossils from the Ediacaran Wengan Biota. *Precambrian Research* 301, 145-151 (2017).

12. Bailleul, A. M. & Li, Z. DNA staining in fossil cells beyond the Quaternary: Reassessment of the evidence and prospects for an improved understanding of DNA preservation in deep time. *Earth Science-Reviews* 216 (2021).

13. Zheng, X. et al, Z. Nuclear preservation in the cartilage of the Jehol dinosaur *audipteryx*. *Commun Biol* 4, 1125 (2021).

14. Black, R. Possible Dinosaur DNA Has Been Found. *Scientific American* April 17 (2020).

15. Grabau, A. Stratigraphy of China. Part 2 Mesozoic. Beijing: Geological Survey of China 1-774 (1928).

16. Gu, Z. Jurassic and Cretaceous of China. Beijing: Science Press 1–84 (1962).

17. Su, T. et al. Development of nucleic acid isolation by non-silica-based nanoparticles and real-time PCR kit for edible vegetable oil traceability. *Food Chem* 300, 125205 (2019).

18. Lindalh, T. Instability and decay of the primary structure of DNA. *Nature* **362**, 709-715 (1993).

19. Fu, Q. et al., DNA analysis of an early modern human from Tianyuan Cave, China. P*roceedings of the National Academy of Sciences* 110, 2223-2227 (2013).

20. Wang, H. et al., 12 Human genetic history on the Tibetan Plateau in the past 5100 years. *Science Advances* 9, eadd5582 (2023).

21. Dabney, J. et al., Complete mitochondrial genome sequence of a Middle Pleistocene cave bear reconstructed from ultrashort DNA fragments. *Proceedings of the National Academy of Sciences* 110, 15758-15763 (2013).

22. Mao, X. et al., The deep population history of northern East Asia from the Late Pleistocene to the Holocene. *Cell* 184, 3256-3266.e3213 (2021).

23. Bai, F. et al., Ancient genomes revealed the complex human interactions of the ancient western Tibetans. *Current Biology* 34, 2594-3605.e2597 (2024).

24. Zhang, D. *et al.*, Denisovan DNA in Late Pleistocene sediments from Baishiya Karst Cave on the Tibetan Plateau *Science* **370**, 584-587 (2020).

25. Zhang, J.Y. Phylogeny of Osteoglossomorpha. *Vertebrata PalAsiatica*. 44(1): 43-59. (2006).

26. Cockerell, T. D. A. The Affinities of the Fish Lycoptera Middendorffi. *Bull Amer Mus Nat Hist* 51, 313–318 (1925).

27. Berg LS. Classification of fishes both recent and fossil. *Trudy Zoologischeskogo Instituta* 5, 85–517 (1940).

28. Rosen, D. G., Anderson, S. & Weitzman S. Origin of the Weberian apparatus and the relationships of the ostariophysan and gonorynchiform Fishes. *American Museum Novitates. Environmental Science* 2428, 1-25 (1970).

29. Tao, W. et al. Phylogenetic relationships of Cypriniformes and plasticity of pharyngeal teeth in the adaptive radiation of cyprinids. *Sci China Life Sci* 62, 553-565 (2019).

30. Mendoza, L., Taylor, J. W. & Ajello L. The class mesomycetozoea: a heterogeneous group of microorganisms at the animal-fungal boundary. *Annu Rev Microbiol*. 56:315-44 (2002).

31. Etchegaray, E. et al. Transposable element-derived sequences in vertebrate development. *Mob DNA* 12, 1 (2021).

32. Tafalla, C., Estepa, A. & Coll, J. M. Fish transposons and their potential use in aquaculture. *J Biotechnol* 123, 397-412 (2006).
33. Frith, M. C. Paleozoic Protein Fossils Illuminate the Evolution of Vertebrate Genomes and Transposable Elements. *Mol Biol Evol* 39 (2022).
34. Gilbert, C. & Feschotte, C. Horizontal acquisition of transposable elements and viral sequences: patterns and consequences. *Curr Opin Genet Dev* 49, 15-24 (2018).
35. Hughes, L.C. et al. Comprehensive phylogeny of ray-finned fishes (Actinopterygii) based on transcriptomic and genomic data. *Proc Natl Acad Sci USA*. 115:6249-6254 (2018).
36. Martins, Z. et al. Extraterrestrial nucleobases in the Murchison meteorite. *Earth and Planetary Science Letters.* 270(1–2):130-136 (2008).
37. Gilbert, M. T. P., et al, Assessing ancient DNA studies. *TRENDS in Ecology and Evolution*. 20(10):541-544 (2005).

**Acknowledgments:** We extend our gratitude to the "Chao-Xiang Talent Program" foundation of Haining City, the support that covers the costs of fossil collection, DNA extraction, and DNA sequencing from 2020 to 2021. We extend our heartfelt appreciation to the following individuals for their invaluable contributions to this project: Dr. H. J. Zheng from the Shanghai Institute for Biomedical and Pharmaceutical Technologies, Dr. J. Y. Ma from the Nanjing Institute of Geology and Paleontology, at the Chinese Academy of Sciences, Dr. A. M. Bailleul from the Institute of Vertebrate Paleontology and Paleoanthropology, at the Chinese Academy of Sciences, and Dr S. P. He from the Institute of Hydrobiology, also at Chinese Academy of Sciences. Their stimulating discussions regarding the manuscript have greatly enriched our work.

**Author contributions:** The study was conceptualized and designed by W.Q.Z. and Y.Q.G. The fossils were collected and overseen by F.L. and W.Q.Z. The DNA extraction from the fossil was carried out by W.Q.Z., C.Y.Z., and J.Y.Y. Alignment between the sequences and present genomes was completed by Z.Y.G. and M.J.C. DNA laboratory analysis, interpretations, taxonomic profiling, and annotation were conducted by Z.Y.G., Z.Y.T., L.J.Z., S.J.Z. and W.Q.Z. Statistical analyses were performed and completed by G.Q.C J.H.G, and X.G.Z. Phylogenetic analyses of mitogenomic DNA sequences were performed by Z.X.Q., W.Z., C.J.Z., M.F.T., and D. W. under the supervision of G.Q.C. and W.Q.Z. Figures were designed and finished by F.L., L.Y.Q., T.F.S., and M.R.S. Project coordination was managed by T.C.Q. W.Q.Z., Z.Y.T., M.J.L., and L.Y.Q wrote the manuscript.

**Competing interests:** The authors declare that they have no competing interests.

**Data and materials availability:** All data is available in the main text or the supplementary materials.

**Supplementary Materials**
Figures S1 to S2;
Tables S1 to S5

Supplementary Materials for

# Ancient DNA from 120-Million-Year-Old *Lycoptera* Fossils Reveals Evolutionary Insights

Wan-Qian Zhao[1,11]*，Zhan-Yong Guo[2]，Zeng-Yuan Tian[1]，Tong-Fu Su[3]，Gang-Qiang Cao[1]，Zi-Xin Qi[4]，Tian-Cang Qin[5], Wei Zhou[6], Jin-Yu Yang[7], Ming-Jie Chen[8], Xin-Ge Zhang[9], Chun-Yan Zhou[10,15], Chuan-Jia Zhu[11], Meng-Fei Tang[1], Di Wu[1], Mei-Rong Song[3], Yu-Qi Guo[11]*, Li-You Qiu[12]*，Fei Liang[13], Mei-Jun Li[14]*, Jun-Hui Geng[1], Li-Juan Zhao[11], Shu-Jie Zhang[11]

Corresponding author: wqzhao@zzu.edu.cn; guoyuqi@zzu.edu.cn; qliyou@henau.edu.cn; meijunli@cup.edu.cn

**The PDF file includes:**

Materials and Methods
Figs. S1 to S2
Tables S1 to S5

## Materials and Methods

### The experimental procedure for the wet lab

**X-ray 3D photography:** The fossils originated from Beipiao City, Chaoyang City, Liaoning Province. They underwent a cleaning process involving tap water, deionized water, and precise cutting. Subsequently, the complete fish fossil underwent 3D X-ray scanning using a Zeiss Xradia 620 Versa scanner (Carl Zeiss AG, Germany) to ensure the absence of co-deposited species. The initial scan was conducted at a resolution of 45 μm to capture the overall structure. Specific areas such as the head, thorax, tail, and periphery were scanned at a higher resolution of 6 μm for detailed examination. The ZEISS 3D Viewer software was used to examine and reconstruct 2D orthogonal slices of the scans, helping to identify any additional species present in the scans and reconstructions.

### Method for evaluating the internal volume ratio of fossils

After cleaning the fossils, dry them in a dryer (56°C, 72 hours) and weigh them (W1). Immerse the fossils in pure water for ultrasonic treatment (10 W, 2 hours). After taking them out, quickly wipe off the attached water and weigh them ($W_2$). Immediately put the wet fossils in a graduated container and immerse them in water. Calculate the total volume of the fossils ($V_2$) by reading the increased scale.

Internal volume of fossils ($V_1$) = ($W_2$ - $W_1$) × density of water

Internal volume ratio of fossils = $V_1$/ $V_2$ × 100%

The internal volume ratios of four Lycoptera fossil samples are 10.8%, 10.4%, 11.3%, and 10.8%.

### DNA Extraction, DNA Library Construction, and Sequencing

The procedure was conducted in a BSL-2 industrial laboratory (Biosafety Level 2), following rigorous ancient DNA extraction protocols[37]. The process involved UV sterilization of all items used and the utilization of Geobio® DNA Remover (Jiaxing Jiesai Biotechnology Co., Ltd., Jiaxing, China) on clean benches and solid equipment. It is important to note that this laboratory has never been exposed to fish, non-human primates, or plant samples before this work.

The stone was cleaned with tap water and deionized water, and the surface was clean. Immerse it in Geobio® DNA Remover for a few seconds and then quickly remove it and soak it 3 times in deionized water. The stone was washed with deionized water to remove additional detergent solution. After drying, the stone was cut into 10 × 10 × 5.0 (cm$^3$) raw materials, and the multilayered stone was divided into single layers approximately 0.8-1.2 mm thick. The textured portion of the fish was scraped off and the dropped particles were collected in a clean tube. Scrape off the contoured portion at least 1.0 cm from the textured portion and collect the dropped particles in a separate tube. Harvest the same weight of fine powder from the non-textured (coffered) portion of the fossil at least 1.0 cm from the textured portion. Use a set of mortar and pestle and a mortar and pestle to grind the small particles into a fine powder. Fossil DNA was extracted from the above fine powdered material using the Geobio® DNA (Solid State Matter) Extraction Kit (Jiaxing Jiesai Biotechnology Co., Ltd., Jiaxing, China)[17].

During long-term geological evolution, some *in situ* DNA molecules within the fossils undergo cross-linking reactions with other macromolecules (e.g., proteins and polysaccharides) as a result of the action of physicochemical and biological factors, to assess the concentration of the extracts, we examined the optical density (OD) values of the DNAmix, which contains both DNA-linking complexes and free DNA molecules, at the peak of the UV 260 nm. Five replicate extractions were performed on the textured portion of the sample and two replicate extractions were performed on the non-textured portion, so-called the cofferdam portion (Table <u>S3</u>).

The DNA extracted was repaired using NEB PreCR® Mix (M0309S). Next-generation sequencing (NGS) DNA libraries were prepared following the SSDLSP standard program (Sangon Biotech, CORP). Sequencing was conducted on an Illumina NovaSeq 6000 (S-4 kit) using the "Dual Label Sequencing" standard procedure. The reads met quality control standards and exhibited a satisfactory quality distribution. Occasionally, it may be necessary to manually remove excess adapters from both ends of the reads (sequences) to obtain accurate information. Combining forward reads with their corresponding reverse reads resulted in 1,259,658 paired sequences for the fossil textured portion (including 757 unnamed sequences) and 635 sequences for the fossil non-textured potions (cofferdam).

**The experimental procedure for the dry lab**

**The mega screen method** consists of two main steps. First, we employed the "minimum E-value mode" to group sequences by aligning fossil DNA with known sequences from the NCBI database (Version 5, Nucleotide Sequence Database). We applied an E-value cutoff below 1E-07 to filter for qualified sequences (QS) and identify the best match at a specific taxonomic level. In the second step, we used the "MS mode" by selecting the "exclude" option to conduct a search that omits the top result. If the E-value difference between two search results at the same taxonomic level (species, genus, family, order, and class) exceeds 1E-02, it suggests that the test sequence (query) belongs to the species identified in the first search result, indicating a unique origin. Conversely, if the difference does not exceed this threshold, the sequence is shared between two taxa and is not from a single lineage. In this study, when multiple hits involved both our target species (Actinopterygii) and environmentally dominant species (preDNA), and there was not a significant difference in E-values between them, it was impossible to confirm the true origin of the sequence. As a result, we had to abandon the test sequence without drawing any further conclusions.

In addition, the "mega screen method" can also be used to uncover relationships among similar sequences found across various genomic organisms, offering valuable insights into patterns of genome evolution.

The whole process involves the following steps:
1) Nucleotide Blast is employed to align sequences to the entire NCBI database without any limitation.
2) Creation of subsets for sorting the sequences: Using the "minimum E-value mode," results are matched and sequences are grouped into the appropriate subsets, determining the total number of sequences (TS). Subsequently, an E-value threshold of < 1E-07 is established to filter for qualified sequences (QS).
3) The subset primarily containing ancient DNA (aDNA) is selected.
4) Each sequence within this subset is screened individually using the "MS mode" to identify those originating from a single lineage, categorizing the "unique" sequences.

5) Considering various factors, including local climate shifting, geological changes, and evolutionary principles of the host species, we can ascertain if the sequence is oriDNA.

**Sequence affinity (Affinity):** This metric measures the similarity between subject sequences and hit genomes. It is calculated by multiplying the values of Identity and Cover obtained from NCBI Blast and then converting the result to a percentage (Identity × Cover × 100%). Mutations in the genomic and mitogenomic sequences occur continuously during the evolutionary process, leading to significant sequence differences between ancient and modern species. However, this variation is smaller in conserved sequences than in non-conserved sequences. A high-affinity value indicates that the sequence is highly coherent and related to the genome of the modern species, which is likely to be either a conserved sequence or a modern eDNA. A low-affinity value indicates a low similarity, and the sequence may be from either a distantly related species or other modern species that have not yet been sequenced. It is necessary to consider various factors, including the sequence composition and conservation, host species information, host genome sequencing data, etc., rather than relying on the Affinity value alone, when determining whether a sequence is an oriDNA.

**Subset affinity:** This metric refers to the average of the sequence affinities of all sequences in a subset; it indicates whether the subset consists primarily of aDNA sequences. High values (close to 100%) indicate that the DNA sequences in the subset consist mainly of preDNA.

**The percentage threshold for subset affinity (Affinity Index):** The number of sequences within a subset that exceed a certain threshold as a percentage of the total number of sequences. In this study, we set the thresholds at 90% (indicating limited similarity between the sequences and their hit genomes) and 97.5% (indicating high similarity between the sequences and their hit genomes), respectively. This metric reflects the closeness of the relationship between all sequences within a lineage subset and the modern genome. For a subset, a low value indicates the sequences are mostly ancient (aDNA), while a high value indicates that the sequences are mostly recent (preDNA).

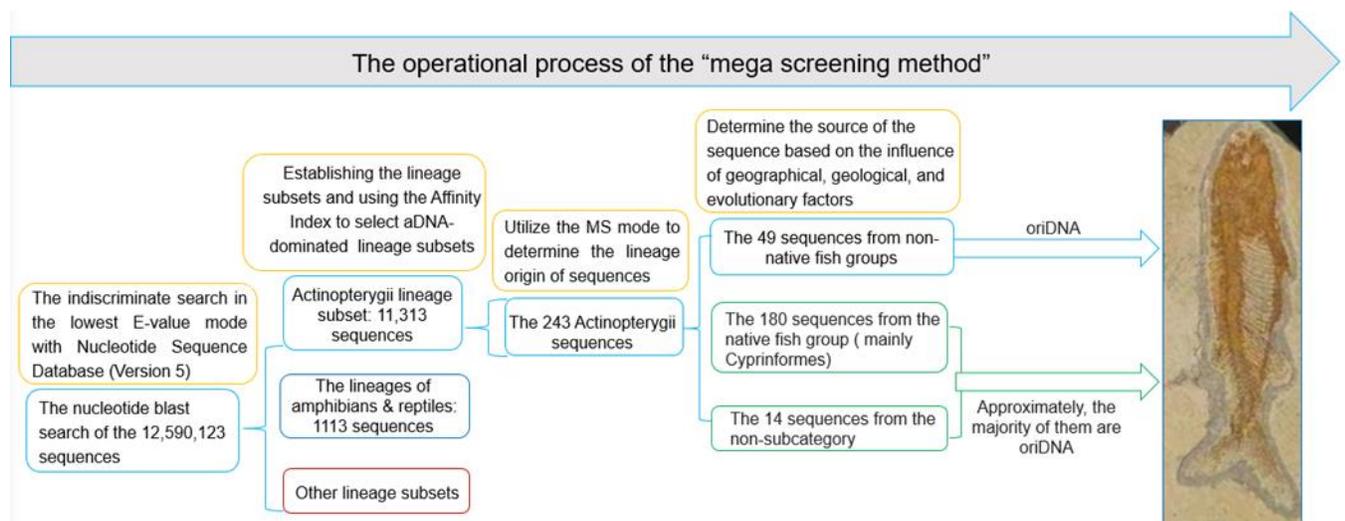

Fig. S1. Selecting for the fossil oriDNA using the "Mega screen method".

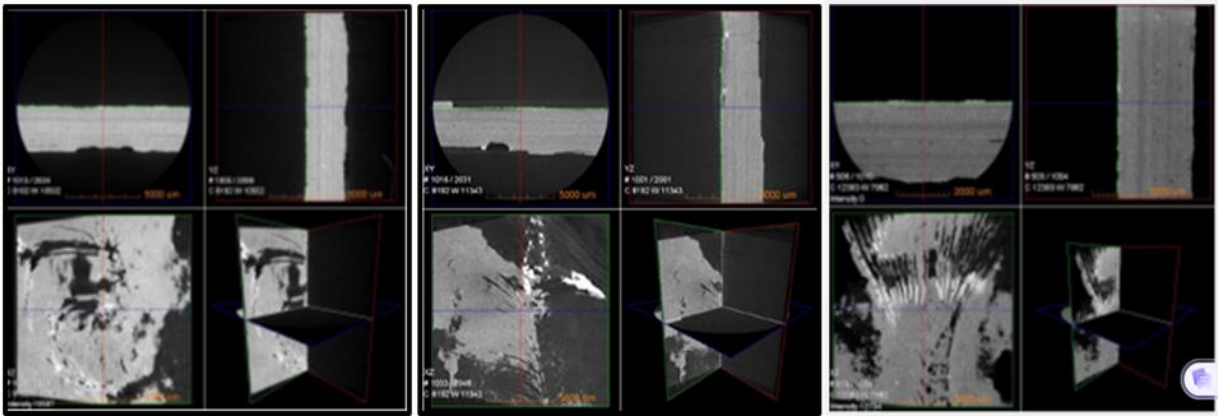

Fig. S2. 3D X-ray scanning the Fossil Fish.

Table S1. Sequencing reads and their matching lineage genomes

| Organisms | Textured parts | | | | Surrounding parts | | | |
|---|---|---|---|---|---|---|---|---|
| | TS | QS | Aff 90% | Aff 97% | TS | QS | Aff 90% | Aff 97% |
| Prokaryotes | 674472 | 597813 | 68.48% | 61.12% | 333 | 259 | 54.83% | 54.00% |
| Fungi | 26852 | 19372 | 60.91% | 51.96% | 15 | 9 | 77.78% | 66.67% |
| Algae | **1201** | **571** | **14.54%** | **8.41%** | | | | |
| Bryophytes | 366 | 186 | 90.32% | 84.41% | | | | |
| Gymnosperms | **513** | **442** | **11.54%** | **5.20%** | | | | |
| Angiosperms | 30211 | 20241 | 50.46% | 39.79% | 18 | 12 | 0.00% | 0.00% |
| Protozoa | 3913 | 2874 | 88.27% | 78.84% | | | | |
| Nematoda | 1512 | 837 | 18.52% | 8.12% | | | | |
| Mollusca | **1248** | **51** | **3.92%** | **1.96%** | | | | |
| Platyhelmintha | 881 | 406 | 79.56% | 73.40% | | | | |
| Arthropoda | 19747 | 2591 | 44.46% | 41.14% | **11** | **11** | **18.18%** | **9.09%** |
| ..Decapoda | 588 | 63 | 3.17% | 3.17% | | | | |
| **Ray-finned fishes** | **11313** | **693** | **8.51%** | **4.91%** | | | | |
| Amphibians & Reptiles | **1113** | **33** | **15.15%** | **12.12%** | | | | |
| ..Homo sapiens | 127977 | 127656 | 94.78% | 91.56% | 187 | 187 | 89.30% | 83.96% |
| ..Livestock | 2812 | 2167 | 59.07% | 45.82% | | | | |

TS, each subset's total number of sequences; QS, the number of qualified sequences.

Table S2. Some characteristics of DNA sequences in the ray-finned fish subset

| Markers | Homo sapiens | | oriDNA | |
|---|---|---|---|---|
| | Mean | SD | Mean | SD |
| Length | 135.65 | 23.8 | 139 | 21.28 |
| Per Identity | 98.32% | 6.46% | 68.65% | 21.05% |
| Query cover | 98.82% | 6.08% | 77.02% | 23.37% |
| Subset Affinity | 97.15% | 0.39% | 52.87% | 4.92% |
| Gap(s)/Sequence | 0.00 | 0.27 | 2.25 | 2.70 |
| GC% | 42.05% | 9.73% | 44.68% | 11.00% |
| QS/TS（%） | 99.70% | | 6.14% | |

**Table S3A. DNA sequences of the ray-finned fishes and the food web symbiotic specie**

| ID | Len | DNA Sequences |
|---|---|---|
| 1 | 101 | TGCATATTTCATTAAAGTCACCAGTGGAGGTGGGAAAGGAGTAGGTACAGCTCTCAGGGGAAGTAAAGGTTGTTTAGCTCAAGTGAATGTTACAGGAGCTG |
| 2 | 111 | GGGATCATTACAAGATTGTTCACTTAGCATAAGTTTTGGAGCTCATCTGAGAAGATGCCTCACTTTACCAAAGGTGTGTGACGGTAATAAAGGAAGGTGGCAAATGTAGCA |
| 3 | 149 | GACCAAGCAGAAAGAACCGACCAAAAACTGAACAGAACCGACCAAAACTGACCAAAACCGACAAAAAACGACCAAAACCGACCAAGAACCGACTAAGAACCGACTAAGAACTGACCGTATACAGAACCGACCAAAATCAACCAGAACCG |
| 4 | 133 | TTTTATTCCTCAAAAATGAGTCTTTGTAAAAAAAAAAAAAAAAACTTTACCAAGTATGATTTAGAAGCTTGTCCTTTGGCATGCAGAGAGGGATCTCCACAGAGCACACTGACAAAATTAGTTTCATTGTTTCTAG |
| 5 | 96 | ATTGTGAGTCCTTCAGCAGTTCACAGATATCACTTCTCGTTCCTTCTCCATTGGGCAACCTGGCAGCACTGTCATGGACTGTTTCGGAATCAGAAG |
| 7 | 74 | GATTTTTTGGGGTTCATCCGCAGGTTGGACATCGGGATTTTTCGTTTCTTCCAAAGCGTGCGTAAATTGACCGG |
| 8 | 97 | GCCGTGCGCGATCTGCTCGAATTCAAGCCGCTCACGTCCGTTCCTGTGAGCCCTGTGACCCCTGTAACCCCTGTGACCTCTGTGACCCCAGTGAGCA |
| 9 | 181 | GTCAACACTGAAACACAAATCAGCTCGTCAGTTACTACAGCTCGTGAGCTACTACAGCTCATCAGTTACTACATCTCTCTTCAGTTACTACAGCTCGTTTGTTACTACAGCTCGTCAGTTACTACAGCTCGTCAGTCACTACATCTCGTCAGTCACTACATCTCGTCTGTTACTACAGCTGGT |
| 10 | 125 | CAAGGTTCCATCTGAGCTAAACACAAAACATGTATTCAAAGAAGAAAAAAAGGAATTTCCCTGAGCTATTTTTGCACATTCAACAAAAAGATGATCACACGTGTCATGCAAAAGAAGCACAATAT |
| 11 | 269 | ATATTGTTCAAAAATCTGTCTAAATCTGTGTTAGTGAGCACTTCTCCTTTGCCGAGATAATCCATCCACCTCACAGGTGTGGCATATCAAGATGCTGATTAGACAGCATGACACGTTGTCTACCAGCTGCCCTGTACAGTGAAAACCGGGATTCATCCGTGAAGAGAACAACCTCTGCAAAGTGCCAGACGGCATCGAATTTAAGCATTTGCCCACTCAAGTCGGTTACGACGACGAACTGCAGTCAGGTCGAGACCCCGATGAGGACGA |
| 12 | 125 | TGAACTAGTGACCTCATGAAAACATAGAAATGCTTGGTGGAATGACTACAAGAACATTGGCTTGTTGGGAACATGTTCTCTTTAACAGCAACACCTATAAAAGCCTCCTCCTTCAAGCCATTTTG |
| 13 | 104 | ATCGGTGACTGTATATGAAGACGATCACTGACTGTATATAAAGACGATCACTGATTGTATATAAAGACGATCGGTGACTGTATATGAAGACGATCACTGACTGT |
| 14 | 68 | TGTTGAATAAAAGTTGCCATACTACCAAAATGCCTGGACGCAATTAGATAGACCTACATGCAGTGAGAAAAAGATGGTCCAACGAGTCAAAGCTGGATTTACAGGCCTGCATTGAGTGTTTTTGAAGTTGCAGCCACTGAC |
| 15 | 164 | CTGGACTGACACTGTGACTTCTTACATCAGTTTTTGTGAGGACGTGTGTGCTGACTAAAACCTTACGCACATACAACAACCATAAACCCTGCCTTCATCATTCCTCCTCTTCCTCTTCCCACACACGGTTTCAGCTGACAGTGGGGTGCTCTGGACTCCACTTACAACAGTGATCCCCAA |
| 16 | 163 | GCCTCTGTCGCGGCGGGTAAATACGAGAGTTGTTAATAATAATTCTGACCTTATGTTTTAATTTATTCCTCTATCGTGTGTGAGTGTGTGTGCCCTCATGCTTCACTCTCACCTCGTCTTCAGGGTGGTGGTTGGTGAGAATCAGATGC |
| 17 | 231 | AGTGATGCTGACTCGAGTATCACGTGATATACTTGCAGGTAGTTCGCAGCGGGATTCTCTAACAAGTGAGTAAAAAATTAGTAAAGCTTTAATTTGCTTTTCTGCAAATGGCTGAAAAGTGCAGCTCCCGACTCTCAATCAGTCCAAAGATCTGCTG |

18    72    ATACAGCATGACTGGGCGGTTCTTCTGATTTCCAGTGTATAGCTATACATTTCTTTGCTGCTATCAGAGCCA
AAAGACCACACATGTGCAGTATATGTTGTCCTTTTTCATACTATTACAGCAGTATGTAAAGGGACAAAAGAAA
19    150    ATAATCACGAGGATAGAAAGCAGCGACTGAAAAGAGCTTTTGCCGTAGATATTCTTGTTATAACATGTGTGGT
CTTT
GCCTTCAGATCATCTGTATATTGTTGGATCGACTGTTCTCATCTTTCTCTTGAAAATATCCCATAGATTCAGGT
CAGGCATATTGGCTGGCCAATAAAGCACAGTAATATCATGGTCAGCAAACCACTTGGAAGTGGTTTTTGCACT
GTGGGCAGGTGCTAAAGTCCTGCTGGAAAAGGAAATCAGCATCTCCATAAAGCTTGTCAGCAGATGGAAGCA
TAAAGTGCTTCAAAATCTCCTGGAAGATGGCTGCATTGACTTTG
21    96    GTGTGTGTGTGTGTGTGCTTCAGATATACCCTACATTATTGGGACAAAATATCCCTACTAATATCTGAAATTGT
TCACCTTGTGGGAACATTTTTT
GTTAACTCCACATCAGCTACATAAATTCATCAACTAACCATTCAGAAACATCCTGTTGCAATCTACATGTTGTC
ACTTCTTCTTAAATCTCTCCATCATCGTCCGATTCCGGTTTGAACATAAAAGGCTGAACAGTTCTGACATTTT
CAGTGAGCCTGTGTGAGGTAATCGGAGGTGCTAAAATGAGCTCTTGAAACTCTGTCCTCTTCTTGGGAGCAGC
AGCTCATTTGCATTTAAAGGGACACACAAAATCTGAGTGTTTTTGCTCACCCTCAAAAAGT
23    128    CTACTAAAGACATATTTAAAACAGTTCATGTGACTGTTATGAAGTGACAAGAAATACTTTTTGTGCGAAAAAAA
AAATGGCTTTATTCAACAATATCTAGTAATGGGCGATTTCAAAACACTGCTTCAT
AAAGGAAAAGAATTGAAAGTGGGGGGAAAAATCACATTATGAAATAAATGTTTTTCTCCAAAACACGTTGGC
CACAATTATTGGCACCCTTTTATTCAATACTTTTTGCAACCTCCTTTTGCCAAGTTAACAGATCTGAGTCTTCTC
CT
25    276    GCAATTATTTAAGACAATTCTCCTAACCTTTTTGGCATAAGGAGGACCTATTATGCCCCCCATCATATAAGTCT
CTGATGTCCCCAGAATGTGTCTGTGAAGTTTCAGCTCAAAATACCCCACAGATCATTTATTATAGCTTGTCAAA
TTTGCCCCTATTTGGATGTGAGCAAAAACACGGCGTTTTTGTGTGTGTCCCTTTAAATGCAAATGGCTGCTGC
TCCCGGCCCCCTTTCCAGAAGAGGGCGGAGCTTTAACAGCTCAACAACAACAAA
26    264    TGTTTGTGTACACATACACACATATATAAAAATTCCTGTTTGGATTTAAAATAGGCATATAATATATACTTAAGAGT
CTAAGACTGGATCATTATTGCAGTGATTATTATGTTTCTAGTATGTTATATGGTTGACAACAGTTATTCTAACC
CTGACTGATGGAGTGTGTAGCTTTTCATTTCTTTAACAACCATGTAGGAAGTCAAATTGTTGCCATATTCCAGG
ATGACAATGTCAAGATTCCACAAATAGTGAAAGAATTGTTGG
GAATATCTTCTCCTCCCTCTTGCAGCATCTCTTCTCTTCTCTGATGACGAGGGCGGGGCAACCTGTCACTCACA
TGAGATCCACCAATGGCAAACCACAAACATCCAGACATCCAATCAATTCCCCATGGACAAAATCAAGTCCTGC
CCTACATTTTTCTTGTTAGAGAAGGCATAGGGAAAAAAAGGCTATCACAACTTCGATTCCATGCCCACCTTTAA
AGGAACAACAGCATTTGTGG
28    103    TGTTCGAACCACCCACTCTCTAAGTGGGACTTGAACCCAGGTCCACCAGCATGGGAGTCAGGCACTCCAACAA
GGACTCTAAAGACCGCAGTCTTTACACGCA
29    133    GATGAAATGCCTTTCCGTACCAGCAGATGGCAGTAGCTGCACAGCCAATCAGAATGATCAGATGGCCCGATG
GACCGACGAGCTCCGACGCCAATTCAACATGTGGAATCGGCCGGATAAAAGCCGACAAGGA
CATCGACACAACAGCCAGTGGCACAGCTCCTCAACAAACCGTCCATACCGGTGTGATGAATACGATCCTCAAC
30    276    TGGACTGAACTGGAATAAATACTTTGAATGTTGCGATCCTATCAGACATAAGATAGCAACCTGAATCGTAACA
AAGCACTGTTCGCCAGAGGAGAACTGGCCCCCCGACTAAGCCTGTTTTCTCCCAAGGTTTTTTTCTCCATTTTA
ACACAGAAGTCTATTTGCCACCTGTTTGCCACCTGATGTCACCTGTTGGAGTTTGG
31    100    TGACGATTCATCTGACCCAGCTGCCCAGTCCATACTAAAGGACCCTCTTGGCGCAACCGAATTTCAGCATCCT
CCAGCCCAGCGTGGCACGAATACTGGA
CATCTTGGTTAAAGTTTTAGTATTTGTTTCCATGATTAATGTTTTTGACATTTTAATAGTTTCTTTTATACATAT
CGTCGCTGTCAAAAACCTCAATTTGAAGCTTGATTTTGGTAATAAGTGTTAGTTTCTGTCATTAATTCCTCTCCC
TCATGTCGTTCCACACCCGTAAGACCTT
ACATCCCATCCGCTGGGTTAAAACAATCCAATCCCTGGGTTAAAACAACTCTATCCCTGGGTTAAAACAACCC
ATTCGCTGGGTTAAAACAACCCATCCGCTGGGTTAAAACAATCCAATCCCTAGGTTAAACAACCCAATCGCTG
GGTTAAAACAACCCAACCACTGAGTTAAAACCACCCAATCGCT
33    189    CAGAAAAAGTAACTAACAGTTTACACTTTCTTCGTCTAACTTCAATATGGAAGGCGGTCTGGCGGAAGCTACATA
TTTTACTTCATAACTTGTTAAATATGGATATTTTACACAAACGCATTGCTTTGCTTCAGAAGGCCTTTATTAACC
CCCCGGAGCCGTGTGGAGCACATATTTATGATGGAT
34    184    TGTCCCAGGTGAAAAAAAAGTGCATTTCAATAATGTACTTAAAGTGCTCTATTTTCACACACTAATTTTGTACT
TAGATGCTCTTAAGATTATCTTAAGAAGTACTAAAGAAGAATTTTTAGTATATTAAGTACAAAATTAGTGTG
TCATGATATTACGTATTAGTATTGTGCTCTCATCTAACTTGAAGAAGGGGCTATTTTACAGTACTTTTCAGTTC
GTAATATTTAATGCTACAACATCTACGCACTGCACTATTCCAATTTTGCTTAGCTCAATAAACAAAAGAACATC
GTTGTAAAATTACATTTGAGA
ACCAGCCACTTACTTGCATATATTTCGGGAGAAACTGATGAATTCACATGCTGTAAATCCGACTGACAGGACT
CTGTGAACTGCAGCGCAAACTAAGATGGCGGCGCCCATCTCGCATTATAGATCAAGATAAAGATCGTTTATAA
AGGTTTTTAA
CATGCCGTTCCACACCCGTAAAACCTTCGTTAATCTTCGGAACACAAATTACGATATTTTGGTTGAAATCCAAT
GACTCAGTGAGGCCTCCATAGCCAGCAATGACATTTCCTCTCTCAAGATCCATTAATGTACTAAAAACATATTT
AAATCAGTTCATGTGAGTACAGTGGTTCAATATTAATATTATAAAG
GTTTCATCATATGCGTGCTCCTGGGAGCATTTTAGGCAGAATTGGTCACTTTTGCATGAAAAGTTACATTGCCA
CTAGAAATATGCTATTCTCACTCAGAAACGCTGTTCTGTGCTAGATATGCTTTAAAACAGCTAGTTTCATCATA
T
TTTCTTCACTCGAAAAGAAATTAAGGTTTTTGATGAAAACATTCCAGGGTTATTCTCCTTATAGTGGATTTCAA
TTGGCCCCAAACGGTTGAAGGTCAAAATTACAGTTTCAGTGCAGCTTCAAATGTATATGCTTTATAGTCACAA
ATGATCGCATCACAAGTGCTTCCACCAGACCCCGATTCTGTATTCTTCAAAAAGCTTACACTGAATGTCCTACA
CCTT

```
41   203   GTTACTCAATTATGTTGTTAGAGAGTCCTAAATAGAATCACTCCGGTTCTATTGTCCTGAAGTCAATGGGGTTT
           TTTTTACTGGGTTTTTGGTTAAAGGTGGGGTAAGCCATTTTTCAAAAATGTTTTTGAACACTGTTGATATTTGA
           AATCAACCCAAACAAATCCACCCCTCTTTTCTCATTGCTCTGCCTTCAAAACTCA
42   225   TAAATGTTCATTTATTAAACATATTAATGTTAATTTCCAACATATTTGGGTTCATTTTAAGCAACAATACAGTA
           ATTTTTAAACAATATTTGGGTTTTAAAAAAAACTACCCAGCACATTGGGGCAAATATTTAACCCAACCGCTGG
           GTTTGTCCATTTTCAACCAAACTTTTAGAGTGTACAAGTTCGATTAAACCATTGAAATATGAGGAAAATATTTT
           TATA
43   269   CTAGCCCGCAAGGTTATGATTATCGACACAATCCTCGTCAGAAAGGCAAAGGGGGAGGTGTTGCTGTAATTTA
           TAGTAATATTTACAGTATTAGTCAAAAGTCTTCCAAATATAATTCCTTCGAAGTGATGGTGCTTTACGTAACAT
           TATGTAAGTTGACATTTGTGCTGGCTACTGTATACAGGCCACCAGGACACCATACAGACTTCATCAAAGAATT
           TGCTGCTTTTCTTTCAGAGTTAGTACTAACTGCAGATAAAGTCCTTGTT
44   207   CCTTTCCCTTCTGCTGGCCTCCGTAACAGACAAAGAGCTGATCTGAGGTCCTGAAGCTTCGGGTTCTGTCCACG
           TACAGGTGCAGAGCGCGGACGGGACAGAGCAAAGCTAGGGCTGGGTCTGCCTCCTCCGAAGGCAGCGCTTGC
           AGGTTCACTACCTGATCTCGAAAGGGAGTAGTAGGAACCTTGGGCACATATCCAGGCCGGG
45   239   TTCTCAGGCACTCTTTACTTAAAGCTGCAGTCCGGAACTTTTTTTTGATTAAAAATTATCCAAAATCAATTTTTG
           AGCAAGTACATAACCAGCCAGTGTTCAAAACTATCGAATTATCTTAGCTCGATTCACAACGGTAAGCTTGTAA
           TAACGTTTTATAATAAGAGCGACCTGGTGGATTTCCGTGGGAAATTCAAGCATGCAGCAGTTCGTCTTTGTGTC
           ATTACATCACGTCTGTTT
46   216   ATACTAATAAGCAAAAATGACTTTAGATTAGATTTAGAAAGAAGAGTGTTGAACTTCCTTCTACTGTATCCTA
           ATCTTCTTTAATCCAGAATGGCAGCACAGCTGAGAGGTTTGTTTGAGCGGCGCCCTCTACTGAACAGGTGTAA
           ATATGCATTTCCTTCAGCCTGAGGATTATTCATTTCACTTTTGGTGTGAAAGGGCCTTTAAAGCAATTAA
           GGTTTGAGTTTTATGAACTTATTTCTTTTAATTTAGACAAACTTAAGTTTAACTGAAACTGGGTTGGGTTTTATA
47   171   TTTCCCAGCATGCTTTGCTCATGACACTCGAAAGGCCCCATGCAACTGATTGCCTCAATTTTTTTGAGTTTTGCC
           AACTTATTCAGGTTTACAGTGT
48   91    ATAATCATGGACATGATTCGAAAAGGTTAAATGGGCCTGAAAACAAGTTACACTTGTGCTCCTCCCTGTCAAA
           AATGGCACATTGGAAATG
49   184   GAAAAGGGGGTTTCGGGCCAGAAAAATTGTTTCAATATATCTCCAAAAAGGACAGATGAGAGAACCAGGAGT
           TTAAGCCCCGGGTATACTTCGGCCGTCTGCGTTTGTGAGTAATTTCGGGACCGCGCGGATGGTGCGCGTGCAA
           CAGCGCATGCGCAAGCAGGACAGAAGAGTACACTAGGTG
50   186   TGGAGGCCTCACAGAGCCATCGGATTTCATCAAAAATATCTTAATTTGTGTTCTGAAGATGAACGAAGGTCTT
           ATGGGTGTAGAACGACATGAGGGTGAGTAATAAATGACAGAATTTTCATTTTTGGGTGAACTAACTCTTTAAA
           ATTTTACTTGTCCAATGTTAAGAATGCATGCAGTTTTACT
51   131   CTGGAGTTGTGTTTACCTAACGTTAAAAATAAGCGGGGCAGCCAGAAATCTGCCCCAATGGCTCTCTTTGAGA
           GCGCGCTGCAGTTCCATTGTACACCACAGATTTACATAGGAAAGTTGTGGGGCGCTGT
52   176   ATGGCGTAATAATGTAACTTCTGTTGCTGTTCACGGTCAGGGACTAATTTTTTTCCGGCGGTAGGAATGTATTA
           GTGAAAATGTACTTCATGAAAGTTGCATTGATATATTTTTGGCTTTAATATTTGTATTGTGTGGTAACCGCTTC
           GCATCGTGCCTAACGACAGGGAACGACA
53   210   CGAGCAGAGAAAAACAGCATTTTTCATGGACAAAGGAAAGGTACTCCTCTCAGAAAACATACAAATAAAGAT
           ACTTTTAGATGTTTAATTGCTATTTGGAGCATCTGGATCTCTAGATGAGGGCTTAATGAACTAATTTTTGTGAA
           AACCTCAAATTCGACCAAAACTGGCATTTCTCACTTGTTTTGCCTGTAGCTCATAATTAAGTGT
54   157   CTCGTCAACATCACTGACTGACCTCACAAATGCGCTTCTAGAAGAATGGTCAAAAATTCCCATAAATACACTC
           CTAAACTTGTGGAAAGCCTTCCCAGAAGAGTTGAAGCTGTTATAGCTGCAAAGGGTGGGTCAACTCCATATT
           AAACCCTACGG
55   149   GGAGAAGAGTAAAATTTAGCTTCAAAATCAATTGAATGTTTTGTCGTGTGCGTGTGAGAGCACTGCTAGCATT
           CATTTCACTGTAAGACAGGAAAAGAGGAGCTGTCTGTGAGGATGTTGACTTTTTCCTTCGCTGATTTAATGACC
           TG
56   250   ATAGCAATGCAACTTATATATTTCAGGACATAATTAAAGCATCAAAACATTTAAATGCATTGAAAATCAGAGG
           GAACTCATATGTTAGAAAGTTTATGTGAATATGAGTAACTCAAGAGCTTAAGCTTTGGAGGCTTGGACGGTACT
           CATTGCTGTCCACAACAGCCTTCTGTGGTGTTAAAGAGTAAAAATACCACTGAACTCAAATCCCTCTCAGAAT
           TTCTGTATCCGGCATGTGTTTGATTTAACCA
57   140   CATATTTAATGTCCTGATGAACGCTTCCACCTGCTGACACAGGAAAAGTGACAGAACAATACACCTCCCAACC
           CCCTCCATCCCTCCGTCCAGCAGCTTCAAAGCCCGTCTGCCCGGAACATCGGCCATCATCAATCCAGA
58   143   TAATGACTTCACCCTTATCAGAGTTCAGTGAGAGTTCAGGGAATGCAGAGTTATTCCCTAACATTCATCCACA
           ACATAAAGCAGCATTCACAGCTTAATCATTCACTTTTGGTGCAGAGGAATCTTAAAATGCAATTTACAAT
59   133   TATTTCAGTGTCACATGATCCTTCAGAAATCATTCTAATATGCTGATTTGATACTCAGTTATTATCAGTTTTGGA
           AACAGTTTTGCTGCTTAATATTTTTTTGGAACCTGTGATATTTTTTTCAGGATTCTTT
60   154   CATGCTGATACGGATGAGGTGACTAAACACATACAGTACTTATGTAGAAACACAAATGATGCCCGTATAATCT
           CCAAACACCAGTCTTATTCAGTTCAGGGTGTATTATCAAAGAAAATACCACCACTTTTTCATATTTCATGAACT
           TTAGCAT
61   153   GGATTGAGATCTGGAGAATTTGGAGGCCAAGTCAACATCTCAAACTTGTTGTTATGCTCCTTAAACCATTCCTG
           AACCATTTTTTCTTTGTGGCAGGGCGCATTATCCTGCTGAAAGAGACCACAGCCACCAAGGAATACCATTTCC
           ATGAAA
62   93    ATTTAAATCAGAGTAGGTCACAAGGCCAAAGCCTCACCATCAGTTTGAAGAGCCCCTCTATTGATCACAGATA
           AAATCAAGGTAAAACCATTG
63   166   GCAGGCCCATGACAGTGAACTGGGGCACAAAGATGCACTGAAGATCCTCCCCAACCTGATCATTTGTTCCAGA
           AGTTTCCGTCTGGCAGATGTTTCAAGCAATCCAAACCAAGACCAACAGACAACCGCACTGCCAACACTGTACA
           CATTCACATACATATATTCA
64   78    AAACAGAGAACCTGCAGAATGGAAACGATCACGTAGTCTTCTTTCCTGGAGACATTCAGGTTGGTGCATGTCA
           TCAAT
```

```
65   177   TGTAATTCTTCTTGCTCTTTTCTCTGCATCTCAAGCACATTATTTCATTCAGTCTGTGCTCCATGTGCTCAGCTTT
           TGCCTTTTCAGCTTTCTGGCCACAAAGTCTCTGCTGTTGCTTCTCCGTTCTTTCTTTATTTCACAAATTCTGCCAT
           CTCATTCTCTTTGTCACTGAACCAA
66   150   CGACGCAAAATGCACTGCGCACCTAAAAATAAATATGTATAAAACGATGTTGATTTTTTTTTTTTTAAATAACG
           TAAATCTTTCATGGCTTTCTACTACATATTTCTGTAAGAGACATAATTTACGTTATATTAAGCCATACAAGTCT
           CA
67   165   ACACAACAGGAAATAGCTCCATGTTTTTCCTGGGCGAGAGGGGCCAATAACTGTCACTTTAGTTTAGATGGC
           TCGAGCGCAGTGGTCCAATGTCTGTTTAACAGCTTGGGTTGGAAACTTCTGCAATTAAAGAGTCCATGTGCAC
           ACGGAGATGCACTTCCCAG
68   171   AACCTATGTTTATAGACCCTTTTACAGCCCACGTAATCAAATAGTCACGCGTGCATTTTGGCAACGGAAGTGG
           TGTTGTTCCCCAGTGGTTCTAACGTATTCATAAACATCCAGTGATTTATATGCTCGTAATTTCTCTCGGGTGTAC
           ACGCTCGGTTTCTCAACTAAATA
69   237   GATGACCGAAAATCACCTGGTGAAGGACTTTTATTGACAACTTGCGAGCAGCCAGCATTCCATGGGATGAGGC
           TGAAGTCGCCTCTGTGGATCGAAACTGATGGAGAACACCTGTCCTCCCATGTGTAACAGGCGTATGCTAGTCA
           GAGCTGTGCGAGTAAACCTCACTCCCCGATCTCAAGAGATGCTCTAGCGACTGACGCTAGAGGTTGCAGCCTT
           TAGCCTCTGTGTTAGAGT
70   167   AGTCATTTAGTCCACAGCATGTTAGTTCATTAGGGAAATGAACCTAAGAGGGGAGCTAAATTTCACTAAATAT
           ATGCACTATATTAGCTTATATATTCATTATATTTACTTCACCTGTTAGTAATGTCTTAATTATTTAATATATGTT
           TAAATAAATTTGTCATAAA
71   181   TGTCATGATCAGACTATCACATTGTTAGCTTAGCAGCAAGCTAACACCGAATTGGTGCAATTTTTTTCCTAAAA
           AAGGCAAGCACAAATCCCACCTGCAGCTTATTTGTTTGGTCAACATGACCACTCAGTCACCTGTCAAGAAAAG
           TTCAGACCTCATGGATTTCAGAAATATAATGGAT
72   123   ATAAAGTAGTTTTCATTTGCGCATTTCGGCAAATTTAGCAGTTCAGTCCGTACATACAGAAAACTTGCCTTCTT
           CATTACTACTCTTTTTGTAAATCAGTAGGTGGTATGGTATTCTGAACAT
73   115   TGAAGAGTTCAGATGCAAAAGCCTCTAAGTGCCGTCTGAAATTTTCTTCTAAAATGTACATTTTTATCAAGCTT
           GTATGTTTATGTTAAGTTATTTAACTTTAATGGCAATTAAT
74   134   GACTATGTCTGCGCTTTCAAAAGTATGCAGTCTCAAAAATGAGGAGACCCCTCCTCGTTGCACTTTGGTGTGTG
           GACAGGAAGAGAGACAGCACAGCTGTCTTTACAAATATGTTCAGCTGAGTCAACACAGGA
75   260   ACTGTCCCTTAGAGCACACACACACACACACAACACACACAAAACCATATGGCCCCAGCATCAAGTGTCATGT
           TTACAAGCGTAGCTCATTGGCACTGCACTCGTCCTTCTTACCATTATCTTTCCATCTGAGTCCATCCATTAGTAA
           GTCCAGCGTAGAAACATTAATCCTCAAGTTATTCCAAAACAGTTTCTTGCCATCCTGCTCATCATTTTAGCGCT
           TTGAAATGTGGGCTAGTTACACAAAGCCCCATCAAGCCT
76   134   TCATGTTGTTCCACACCCATAAAACTTTCGTTCTTCTTCGGAAGACAAATTAAGATATTTTTGATGACAGAATG
           CTGTCAGATTTTCTTCCATTGACTGCCTTTGCAACTACCACTTTGACGCTTCAAGAACTT
77   274   GGCGTTTAACAGGGGCATGGGCAGCTCGGGAGCTCAGGGCTTCGACTACGGCATCTCCAACAACAAAGGTGA
           GAGTAAAGATCTGGGAGAGATAGAAATAAAATGTGATTTATTTATTGAGTGACGCTTACATCTCTGTTTATATG
           GTTACAGTCCACCTGCCAAATGAACATTTTTTTTCATGCCATGCTCCTGCTAAATATCTGTCATATTTAATCAG
           ATATATCTTGGTAGTTTATTGCCATATACATTCAATTGATTTCAATCCAATGGTG
78   218   GTTGCCTTTCACTGAAGGGTCAGTTGAAGTCAAATCCTGTGTTCTGCTTAAGGTCACACTGATAGAAGAGACT
           CATTGTGCTCAAATAGAGAAATACTGAGAAGAACTGTCACTCACTCGACTAATAAAAAGCTCTCCCTTAAATA
           AAAACAAGTGCTTCACAAAACCTGTTTATTTCCTCACGCTTTAAGCAGACTGGTCAGGATGCTGTGGCTCAT
           ATGAGCCACAGCATCCTGACCAGTCTGCTTAAAGCGTGAGAAATAAACAGGTTTTGTGAAGACTTGTTTTT
79   217   ATTTAAGGGAGAGCTTTTTATTAGTCGAGTGAGTGACAGTTCTTCTCAGTATTTCTCTATTTGAGCACAATGAG
           TCTCTTCTATCAGTGTGACCTTAAGCAGAACACAGGATTTGACTTCAACTGACCCTTCAGTGAAAGGCAA
           CTGAGGCTGGGGTCAAGGCATCAAGAGCCACCACACACAGACGTGTCAAGGAATTTGGCTACAGTTGTGGTA
           TTCCTCTTGTTAAGCCACTCCTGAACCACAGACAAATGTCAGAGGCGTCTTACCTGGGCTAAGGAGAAGAATAA
           CTGGACTGTTGCCCAGTGGTCCAAAATCCTCTTTTCAGATGAGATCAAGTTTCGGAAACCAAGGTCCTAGAGT
           CTGGAGGAAGGGTGGAGAAGTTCATCACCCA
80   249   TGAGTTATAAAGTCAGAATTCAAAGAAAAAACAGTCTTTTTTTTCCTCACAACTGGACATAACACGCAATTGC
           AAGTTTATATCTCACAATTCTGACTTTATAACTCGCAATTGTGATTTTATATCTTGAAATTTTGAGAAAAAAAG
           TCAGAATTGTGAGATA
81   163   GAAGCCCACTAACTCTCATTGATTTTGCTTCAGTGTCTCTTTATTGAAACTTTATCAGGTGGATCTGAGCGCAT
           GGAGAGGGAATCCGGAACGAGTGATGAACTTTTCTGTCTCTCCTCCTCATGCGTACACCCCGCTACATCG
           TGGAATCGAGATGAGGGCATTTCAGGGTACATGAAGAAAAAGCTTTTCTCTTGCAGGGCTTTCTAAGGCAAAA
82   144   CAGCACTTCATAAATTAAAATTACTTGTCCACAAGCAACACTGGATCTTGAGTTCAGTTCCTGAGGCAATGTTT
           TAGACATTAATATAACAAAACCATTTGGGGATAAAGTGTAGTTCAGCATGTGTGCGCGATCCTCGGATTCCA
83   218   GAGTAATTGATTTAGAGAATGTTAAAATCAATTGGCAGACAGCTGGGGTCTGGGAGATTAGAAGTGAAGTAT
           AGGTGGAGTGGAGAGCAGCAGACGGCCCCAGAGCAAACAGTAATCTCAGCATTATAGACCAGGGGGAGATG
           TACCTGTTTTCTCTTATTTCACTTTCACTCTCTCTTTGCTCCTTTTCTCCTGTTGGGGACGGTCCATAGTCTCTT
           ACACCAACCTTGTCTCCATCT
84   240   TTAATCATTTTCATTTATGTCACTGCTATTGCTAGACGGCATAGCGGCAAAATGTCCGAAGGAAATTATAGTG
           GAGCACAAAAAACGGAAAATACAAAGGTTAAAAGAACAACGTGAACAATATATACTGAAAAAATACCCAA
           ATTGACTTGATACTTCAAGGCATTGCATACATCAAGACAGAAGTGGCG
85   192   CCTATGCCAACATGTCCAGGTCACAGTCCAACGGGCACCATACCAACATTCCCTCACTGACACATTAACACAA
           TGAAACAGGTTGCAACCATTGTTAACAGCCTGCAAA
86   109   GGAATTTGTCTAAAATTAGCCCAAGAAGAAAGAAAGAGCTAATTATTGTCAACTTCACATGTCTTATCCCACG
           CAAAGATGGTGAAACACTCACCCAACGTAAAGAAAATTCTGATACGTCTACCACCAATAACGCTTCCCTCTATG
           TGTTTCACAA
87   156   CTTAAATAACTTTTTAATTTTTTTTTTTTTTTTGAGTGATTCCAATATTCAGCAATAATTTGCCAAATGTCTTCTTTC
           AAAATAGACCTTAGATCTGTATTCCAAAGCATTCATGAATTACAACCATTTATTTCGCATGTCGT
88   141   
```

```
89   276   AGCACTTACCGAATACAACAGACGAGTCAGTGCCATTGTGAGGGTACCGATAGTGACAGTAGTGGCAGCTGA
           AAATCTTGTCCAATGTATAAAATGAAACTCGGTGGACTGGACGAGATCATTTGTAGTCAGAGATGTTGCCTGT
           ATTTAGCAAGAGCTGGTCTGAGGCATGTGGTTTACAAGCACCCAATTTAAGGGTGTCCATGCCCCCTTTGCTAT
           CCTATGGTCGTTGACACTCTACTTTTCTGTGAAGCCCATCAAGTTTTGATTGCCGTG
90   205   AACGATGGCAGGGGAAAAACGAACCGTTTTGAAAGAGTAATTACGTTGTAGCCTACTAGAATTGATGCCGAC
           CATGTTTATCATAAGTGCAACAAGTCATTTGCATGTGAAAGTAGAAACATGCAATTCAGACGAAACTTCTAAA
           ACGTGCATATTAACTTTCCTCATTAAATGTGTCTGTGTGTCCGACACAGCCTCCACGACG
91   159   GATCGGAAGAGATGAAGAATATACGTCATAGATTGCTTTGGAAGTCAGGCATTTTGGTACATTATCATGTGCC
           AGTTTGGTACATCCATACTACAACCAGAGGTCAGGAAAGATGCCACTATTGGCCAAAAATGATGACTAAACA
           GGTCAAAATTATGG
92   138   TAGAGTAAAGAGACGCAATTAAGCAGGTGGTTTTCCCCTCAATCTGGCGCCCCTGTGCGGGTGGCTGTAGGGG
           AGGGTTGATGTCGGAGCACTAGCCCCTCTCATCCTCAAGCACATGGTCCACATTAGTGTGGAATG
93   218   CAAGCCAAAAACTGTTTTTAAAGTAAACTGGTGCCTTCCACTCCTCCCATGTGCTTCTCCTTTTCACAGTCTGG
           CAAAAATAATGGGTAGTTCCAAAATATGTGGTCACATGGCAATAAAAGGGTACGGTTGCTAAGATACAGGGG
           GTGGGTCAACATACCCACTGGCTCATCTTACCCCACTCTCCCCTACACATAGAAAACTCATAGGCCCGGTTT
           GTGGACTATTTTAACAATGTCTTAGAGTCAGACAGCTCTCGGATTTCATCAAAAATATTTTAATTAATTTGTGT
94   178   TCCGAAGATGAACGAAGGTCTTTTAGGTGTGGAACGACATGAGGGTGAGTAATTAATGACAGAATTTTCATTT
           TTGGGTGAACTAACCCTTTACGTTGCTTTGT
95   108   AGTGAATTGAATTTCACTATGTGTCTGTCTATGTGTCTCTGTGTGTGTGTGTGTCTGTCTATGTGTCTGTGTG
           TGTGTAGAGAAGTATTGGATATAGTGGTGGAAT
96   141   GCCACAAGCTGTTTTGTTCCTTGCCTTCAGGCAAGCGTTATTGCACTGTCAAAGCAAAGACTAGCAGACTAAA
           TATTCCAACCCACTGCAGTTCATAGACTAAATAACACTTTTATAGGACATGTTCAAGGTTTTTTGCTG
97   173   GCGCTTATCCGGATTGTTCAAACCGACTTAAAACTAAAAGATTATGTTCTCTTGCGCGAACTCATTCAGGAGT
           GGTGACTTTCCACGTATTCCCTATTAACTTCTGAGCTGCTCAGCTGAAGTTATGGTTGATTGATCTTCGGCTGG
           ATATCAACACACCCATTAATGTACTA
98   287   GAAAAAGCCGAAGGGAAGCTGGAGGGGGAAAAAAAGTTCAGATTAGTTAAATAATGATTAAAATTATTCAGA
           AAAATATATACTGATTATGTCTTACTTTCCCAACTGCTTTTGGAGGTCAACCATGTACTGGTAACACTGGGCGT
           AATACGTCATCTGAGCCTCCACAAAATCATTTAGACAGCGGAGATGATGGGCCTGGGAGAGAAATGATATGG
           TTTGTTATAGGTCAGACGTTGAGTAAATCGAGGGGGTTGTGAGATGTGGATTGGCAGGCGGCCTAGTGG
           GACAAGTTGAAGAAAAAATGGAAGAGAAAAGATTATTAGATATCATTGGGTTTTTTCAACAGACATATAT
           ACAACATCAGTCCTTTAATCTGACTGGTGATA
99   104   CTTCACATTTGAGAGTCACAGATGTGCTGGAGGTCCATGAGACCACAATTTCTGTTGCAATTGGTCTGTCTTAA
           AATGAAATGTGTTAATGAGCAAACAGATAAAATAGATGAATTGTATATGAATAAAAATGACTCTGATATGTCT
100  209   TGACTTACATTGAATAATTAATGTAAAGTTTTGGGATTTTATAGTTCCATTTTTTCTAACTA
           TCTCAAAATCTTACAGTCACTGCTGGAAAGGGTTCAAATATGCAGAAGATGCTGGAAAACTGAAGAATCTGC
101  113   AGGACCTGGAGGATTTTTCTGAAGAACAGAGCTCAATTTAA
102  146   CATTGTTATGATTCCCTAGTCTTTACAGTAAAAGGACCTGGAGCCCTAGGGGTCAGTAGATGTTTTACAGGAC
           TTCCTGTGTTGTGCGAAAGGGGGCTCTGGGGTCAGTACACATTATGCAACAAGGGGCCCGCAGGCCTAATTCT
           ACTAGCGTTTCTGCTGAAGTTTTGCTTGTAAAAAAAGTTCATTGTCTATTGTCAATAGTGTAGTGATGTATGAA
103  213   GCACAGCTCACACAGAAGCCACTTTCAAGCAGCTTTTTGTTGGTCAATGCAATGTGAAATCTTTCCCTT
           GTCTGAAATTCCCAATTGTGTGGATATCTATTGGAATTACTTGATTTTGCTTGTGAAAATTCATT
104  76    GCGCCAGATCGTGCCCACAGGACGTCATCGACCGGCATCAGCCCAAACTCTTCCTCTTGTCCCCCAAACATGA
           GCT
105  216   CTTATCCAAACGCTGCAGTGACTTGACTGAACTGTTTTCTTTCATGTAACTGTCGCGTCCTTACAAAACTTGC
           AGAAACCGATGTTAATATATCACCACCTTCAGTATTTTTCAGAATTTACTTGGAATGCACGTTCATCAGCGACA
           GCAGCAGTGTGTCTTTGTGTGGGACGTCCTGCAAAAGAACCCCACAAACTCCGTAATAAGAAACGCG
106  165   TGAGAGCCATTTCCCCAAACACATTTACATCAAATTAAACATATATGCCAACATTTAAATCACGTACTGTGAA
           TTTCTCTAATGGAAATGTTATTTGACATGGCTAATTTATTTTATTAAAAAAAATTAGACAAAAGGGTAACTTTTT
           AAACTTAGAAAAACCCAAA
107  239   GCTGTGTGCCTGGAGCTGAGGGGCTGAGGTGAAGAGCATCCAGTGGCGTCTCACAGTTCCGCTAGGCAGCAC
           ACTAATTGAGAAGTCGCTCAGGGGTTGAGCGCAGGGGGCTCGCTGCACTGTGCTCCCTATCCAAATCAAGCCC
           CTGTAAGTGGGGCTGACATGGACCAATCACCATCTGGAGGTTCCTCCACTAAAACTCCACCATTCCTGACTAC
           CATTAAAGGATCTGCTGCTAT
108  121   CACACGGACAGGTACTATTTTATGCACTGTGGCAAAATACAGTGTATTAAAAGGAATGAGTCACTGCTGACCT
           TTCAGCAAGTTGTACTGCCATGTTTCCTTAAACATAGATTTGCTAAGA
109  136   TTTGGGTGGTTTCTTGGTGGGGGCTTATTGGCCCAAATGTTGTGAAATTTACAGGACCTTTATACAGTGCATGG
           CATTGTAAAGACCTTAAGTCTATCCTTCACAGCCTCTCTTCATCTGTGTACAATTCAGTATG
110  251   AAAAATTGGATGACCAGTAGTTTCCTACTTCTAATTTCAGAAAAAAACAGAGATTCTAATTTTTGGACCAAAA
           ACTTCACGTAATAGCCTAGAATACTGTCTAACACTTGATGGCTGCTCTGTTAAGTCTTCGTCGTCAGTTAGGAA
           CCTGGGTGTGCTCTTTGATACCAATCTTTCATTTGAAGGCCATGTTACTAGCATCTGTAAAACCGCATTCTTCC
           ATCTTAAAAATATATCTAAACTACGACATA
111  135   GCATGGTTATACTTTTCTCTCTAAATATTGTCAATCTTGCCAGTAAAGAAGTTCATAAAGTCATTACTGTTGTGC
           TGTTTGGAAACATCAGAAGTTGAAGCTTTATTTCTTGTTAATTTTGTTGTTTAGATTTGTG
112  148   TTACCCTAACCAAAAGGACAAACTGTTGAGAAGGACAAAAAGGAGACTTTTTATCCAAAAAAGTGACCAGTT
           TCTGTTTTTTTTTTTCTTTTCCAAAGAAGAATGACCCTCATTGGACTGAATGTAAGAAATACTGTCTGAATCGGGC
           AG
113  97    TGTCTCTTTTGTCCACTTTCAACCACTTTTGTCCTGATTTCTTCGAGGGGATGGTAAATATATTGGCTTTTTCTG
           TTCTTTTGATCTAATGGACGAA
114  131   CACAGCGATCAGGTTATTGCCATGTCCACCTGCAAGAAAAACATAACAGTTGCATGAAAATCCAGTGTTGGGA
           ACAGAGCAATCTAGCAGAAAGGTTAATTATTTATTATTGTAAGCAATCCCTGACATAT
```

115 133 CACTTTCTCTAATAAAAAACATCAGTATGGTGAGGATGACGTTGCTGTTTTATCCACCGTTTGTGGCAGAATTG
TGGTCAGGGGCTTGAAGGCTGACAGTGAGCCGTTAGTGTATATGTGCACTGAAACACGG

116 133 ACAACAAAATGGAAATATTTGGATACTTTCTAACATTTACAGATGGAGAATATACCTGTGAAATGTAAGTTAT
GTACTAAGGAAAATGCCATATTTAAAATACATTAAACAGCGCAAGTATCTGTAAGAGCAT
AGGAAATGATATAGATAGAAAATAAAAACATGGAATATAAATCTAGTATGGATGATGCTAGGATAATAAAAA

117 193 ACGTGCTGTGGATCCACTATTGCTCAAACAAGACTTCAGATGAAACTTTAGCAAGTTTTCTAGCAATCCATTAT
TAGCTTATACACCAATTCTGCAATGTTTCAGCACCATCTCACAATCA

118 221 GTTACATATGATTTCTATTTCAAATAAATGCTGTTCTTTTGAACTTTCTGTTCAGTAAAGTTTCCTGAAAAAAA
AAAAAGTCACTGTTTCCACAAAGATTTTAAGCAGCACGACTGTTTTGCAATCTTGATAATAATAACCATTATTA
ATAGGTGAGTACCAATATTAATTGATGATAATTGAGCACCAAATCAGCATATTAAAATGATTTCTGAAGGATC
A

119 189 CATATAGGTCCTATACCTATAAAACACACAAAAAAAACATTTGAATGAAAACTTGAAATGTAATAGTTTGGACA
CTGTTGCAATAAATCTTGCCAAGAATTTCCTAACTTTCAACGTTTCGTTTGAATTGTGAAATACATGGGACTGA
GACAATGAGACAATCAATCAATACAATCACTACACAAAACTA

120 126 GCCAATTGGGATATACTGTAGATAAATGAAATGGGGAGGATAGGCTTATATAAGACTGTATAATTTGGTGTTTC
GGTGCCAACTTGAGCCATGATGGTTTCAAGCCTTTGGACAAGTAGTAATTTAT

121 113 CATGGGTACCATGCAACGACATTCTGGACCCCTAATCTACTGGACACCTTCCATTCTGCTCATCCCAACAGACCT
GCCCCTCGTGGAAGAGGACAACCACCACGATGTCGGGGT
GTGTTGTGCCGAATTCTGACCCCTGTCAGATTATTAACCCCTAGTATTGGTAGGTTTAGGATTAGATGTGGGG

122 167 GAGAGGGGTTAGGATTAGGGGTTGGGTTAGGATTAGGCAATCAGGTAGCGACTTCAACGAGGGGGGTAATAA
TTTGGCAGCGAGTCGAAATTG

123 142 CTTTTTTAAAACCAGTTGGCAACATAATGACAGAATTTTCATTTTTGGGTGAACTGTCCCTTTAAATAAATGTG
GTGTAATGTTCTTAACTATGATTCAAGTGGTCTGTTGTGAATTGGGTCTTTGATTTGGATCTTTTATA
AAAGTATCATTCTGACTGGATTATAAGAGCAGACTGATCTCTTAGAATTCTGCTCAGAAACACTGCCCATGAAT

124 121 AAAGAATGGTATAAAAACTTCCTGCAAGAGCAGCTTCTCCCAACGATA
ACTAGTGTGCGTTACTGTACAAGTTACGGGTTTACTACAGAAACGGAGAAGGTGAATTCAGGGCAGGATAAG

125 278 GAAAAAAATAAATAAAAAAAAAATTGTGGTTTGGGAATTTGGATACAGGAGAGATGAAGTATCACAGAGCCAGG
TAATTTGCAAGACCTGTCAAACTGTCGTTGCGGCAAAACAGAGTAGGGGCCATTCACACAGAATGTGTTTTTC
TGGTTAAAAACATGAGACGCAGGTCTAGAGACTTGTTTTTTTAAAAAATGTAACTTATTTTA
ATCTCTTTCAGTAGCTTAGTCGGTATAGGGTCTAACATACATTGTTGATTTTGATGATTTAACAAGTTTAGA

126 171 CAATTCTTCCTCTCCTATAGCAGAAAAGGAATGGAATTGTTCCTCAGGAGATCTGTAGCACGACATCTGTGAT
ACTGTAGTAAATGGTTGCATGGTT
CCCCATGATCCTTTTCACAGGGAAGTTGTTTGGCATCACAGAATGGGTACAGGAGGCTCAGAGTTCGATTTTA

127 224 CCTGTAAATGTATGTATAGTATTTTCTATACTACTGTAAACCTTTATCAACTAGATTTGGTACATTATTTTGGT
AAAAACGACATTGTACTTCAACAAAAATACTTTACATTTACCTTTATCAGTCCATTCAAATGTTTCAAATACAT
AG

128 229 TTTTTCTTTTGTTGCTTGATTAACATTAATGACAGACAGCAGCAGGTTTATTAGGCTGCTGTCACTTTAAGGAC
AAATGCACAGACCGAATATACTTATGGTTATTCAAAACTTTTTAACTTATTTAAAACTGTGTTCACGAGGTCAC
TGGCCAAAAAAAGCATTTTGACATAATTTTGTGTTTTTGTCCATTAAAGTGCACCAGACAGGCGCTAAACAG
ACAGTGTG

129 167 CACAATATTCTATTTCGTTCCCTCGATTTGCTAAATCGCGTACACAATTTACTATTTCGTTCCCTCGATTAGCTA
AATTGCACGCATTATTTACTATTTCGTTCCCTCGATTTGCTAAATCTTGACGACGATTTACTATTTCGCTCCCTC
GATTTGCTAAATCTTGA

130 140 TTGAGCTTCAAGAAAAGACGTGCTGGAATTATAGTAAACAAAATAATAACATGAAAACGCTTTGCTGAAGTTC
TGCTTGTCAGTGATGAAGAAATAAAAAATACTGCTCGCAGCCTCTTTCTTGTTCTGCCCCACATTTT

131 101 ACTGTACTGCTGTTATTAGGCATTTCACATCAAGCACGATAACTATAACATAAAGTTTAATAAACATT
GTTTCTTTGAGAATAGGGAAGTCCACA

132 141 CCCTCAACTGTATCGTAAAAATGTTCATTACACTAATGCTGTGTTCGAAATCGCCCCCTATACCCTCATTCACT
ATTCCCTACATTACTCCACTAATATAGTCCACTTGGAGTGAATGAAACGAGTGAGTGAATTCGGACA

133 90 CTATCCTGACAAAAAAAAAAATTATCCTAGTTTCCACAACTGTTTTCAACACTGATAATAATCAGAAATGTTTC
TTGAGCAGCAAATCAT

134 105 ATATAAGACTGGTTTTGTGCTTCAGGGTCACATATAACATTAATTTACTTTGTCACTTGTGTTTTAGTCTCGCCC
ACAGGTGTTTGCTGGCATGTCTGATGGTCT

135 105 ATATAAGACTGGTTTTGTGCTCCAGGGTCACATATAACATTAATTTACTTTGTCACTTGTGTTTTAGTCTCGCCC
ACAGGTGTTTGCTGGCATGTCTGATGGTCT

136 112 TGTGCGATTTGGTTGATATCCCACAATTAAAACTTATTTAATACAGACTTCAGCACATCGTTTGACATGTTTTTG
CATTAGGATTTGAATGAATCGATTACCACTTGGCTGTT

137 178 TTGACTTCTGAGAGACACTGCCACTCTGAGAGGCCCCTTTTATACCCAACCATGTTGCCAATTGACCTAATAAG
TTGCAAATTGGTCCTCCAGCTGTACCTTATATGTACATTTAACTTTTCCGGCCTCTTATTGCTACCTGTCCCAAC
TTGTTTGGGATGTGACTAT CATGAAAT

138 111 TTTTTTTCTTGTTCAAGTTAAGCATAAAAGGCAGTGCTAAATCTTTTGTCCAGTGCAGATGTTAAATTTAGGTA
GGAATATGATGGGTAATTTAACACATTATTGCAGGAA
TAATTTGGACTTCTCTGTGCATGCTCTTTGAGGTGGTGACCATAGACCTGGATTATATGAGTCACAGACAAGA

139 152 CAGTTTGACGTAAAAATATAAAAATGGAAACTACTGCGGAAACAAAGACATCTACATCTTGGATGCCCTTG
AGGTAAG

140 116 TAAAATGGTGTAAGATTTAGGAAATTTGATAATGTAATTATCATTTTGTTGCGAATGCAGTGCACTTGCAGAG
AGAGACCGCCCACACACTCTTCTGATTGGCTGTGATTTTTGAT

141 249 TTTGCGTTCTCTCACAAAACTTTTGTGTTCCGCTAAGAAACTTTGGTGTTCCCTCGCAAATCTTTTGCATTTCCCC
CGACAAACTTGATTTGCAGGCAACTAAAAAGTTTTGCAAGCGAACGCAGTTTCTCAGGGGAACGCAAACATTT

```
         TGTGAGTGAACACAAAGTTTCCTGGGGAATGCAAAAAAAAATTGCGAAAGAATGCAAAAGTATTAACATATA
         ATCTTTGCGTTCTCTCGCAAATTTTTTTG
142  233 AAAATCTTGTAATAAAAGATGATGTTGTATTGGCTATGACTAGACAGCAAATACAGACTCTCCATGTTTTTG
         AGTATTATTAAAGTACAGGCAAAAAAAAAAGTAAAAAGTTATTAACCAGTATTTTTTCTAATCAAATCGGTT
         TCAGATTGGTAATAATACAGAAGGAACACATGAACAGAAACGTTAAAATACTGATTCTGCTCAGAGTGAATTT
         TTAAGCCCTGATGG
143  172 ATATGCCAGACCCAACAATGCAGAAGAGCTGAAGGCCACTATCAGAGCAACCTGGGCTCTTATAACACCTGA
         GCAGTGCCACAGACTGATAGACTCCATGCCACGCCGCATTGCTGCAGTAATTCAGGCAAAAGGAGCCCCAAC
         TAAGTATTGAGTTCTGTACATGCTCATA
144  250 CCAGGTGAGAAAAAGTAACACAACATAACATTTCCATAAAAAGTAACTATGTAACGCAGTTACTTTTTTAGGG
         AGTAACACAATGTTGTAATGTATTACTTTTAAAAGTAACTTTCCCCAAAACGGGTTAAAAATAAACTCATTTA
         CAGCCACTTTTTCCAAATCCATTGTTTAGTTATACCGTTACTGTTAGTTGCTGTACCAGTTAGTATTGTTCTTGT
         CTTTGACTGTTAATTGCCAAGTTTATGTT
145  102 TATTGCAGGCAGATAATATTGGATATGTTCAGCTATTAGGATATAAAGCTTACAAAATGTTTATCATAAACTA
         GCACTAAGACCCAGAATCTGCCTGCAATA
146  119 AAGCAGAATATGGCTGTTTCTAAAGCAGTGCCATCTGCTGTTAAAAACTAAGCTCAGAATCGGTTCAAGAGAG
         AATCACGATGCATTTGGAAAATCTCAGAATCGATCAGGAATCATTT
         GGTGCATCACTAATAAAAAAAAATATTTAAAAAAGCATTTTAAAAATCTAACAGTTTTTCACTGTAAAAACAAA
147  247 TGCAGTTTTAAGGAGCAGTGTACTAGTTTTCTTTTTTGTTTAGATGGTTAATATTAATGCAACTATCAGCGCAC
         TTTTTTTTTTTTAACAATCCGGCCTCAACCTAAAATGAGTTTGACACCCCTGATTTAAACTGTTCTGACCAGCA
         GGGGTCACCAGCGTGTGGTGTTCCG
148  210 CATACTGCCCTAGGTTAACAAGACCTTTAATGGAAGGGTTACCGTAATCTCCTCCTCCCTTCTTTTTATTTGTAT
         ATGAGTCTGGGAATAACTATGTCTTGCTATTGTCTAGTAGTGCCCTCACTGTCATCTGATTCAGGGTCTCCAGA
         GAGCAGTTTCCTGTCTGCTGCATCAGCACATCGTGCTCACATTACCTTCATAGTCAGGAGG
149  288 TTTACTGTTTAAGAAGAGTATCTTCTCACCTTGACCTTCCTTGTGTTGTAGGTAGTTAGGCGACGGCTGTATG
         TATGGCAAAGGGACCCGATTAACCTGTGGGTTGAAGAATCATGGGTAACCTCATTAGCAGACCTAGCTGTTTG
         GGACAAAAGTCCAAACAAGTTAAGTCAGATGAGAGCGTCCTGAAAGAGTGCTACCAGCAACGGAGAGAATG
         GCCGCTCCCGGAGCCTCCTAGAGAAGAGAAAAAGAAAAGAGGATGTGGAAGATGTTCTTCTTAAACAGTAA
         CCTGTTCGGGCGTAGGGGCCGTCACTTTGTTGGAGGCCTCAGTCTGCTGTGACTCTCTGCAAGAGCGCTTTTCA
150  248 AACCTGGAGAGCCTCTGCGTGTTTTGTGAACCTCCCTTACTCAACACACCTGATTCAGATCATCAGCTCTTTGG
         GAAAGAGCGCCAGTGAAGTCAGTTGAGTGCGCCTGATACGAGAAACGTACATACAAAATGTCCAGTGCTGTG
         GGTCCACGGGCGTAGAGACGTGGCAGCA
151  122 CGAATTACTTTTTTTGGCTATTATTTGGCTATTTAAGCTGCTTTTTTGCTACTCACTGCACAGAACAGGCGCAG
         AGAGAGGTGATCCTGTGCTGGGGAGTCAGACCTCACGCTTGCGGGGCA
         AGGGCAGACAGAAAATCCCAAATTCAGTATCTCAGAAAATTAGAATATTGTGAAAAGGTTCAATATTGAAGA
152  229 CACCTGGTGCCACACTCTAATCAGCTAATTAACTCAAAACACCTGCAAAGGCCTTTAAATGGTCTCTCAGTCT
         AGTTCTGTAGGCTACACAATCATGGGGAAGACTGCTGACTTGACAGTTGCCCAAAAGACGACCACTGACACCT
         TGCACAAGGAG
153  77  TTTTAACAGACGACATACTGAACCGGTATGAAATTGGCAGAAAGCTTGGTGAGGGAGGATGCGGGTCCGTCTT
         TGAA
         TCAACGCAGTTTTATTTTCATAAAAAACAAAACAAAAGTAAAACATAAGAACAAATAAAAAAAAATTGCAAAC
154  258 AACAGCACAAATAAATGCAATAGAATAGAGAGCTATGAAAAAGTTTTTCAGGCTAAGTAGGTTTTTATTCAGG
         TAAGAAATACTACAGAAATTAAAGTAATCAAATGTAAAATAACACATCCAGTGTTCCCCATATGCGTCTTTGTC
         TATATTATTATTATTAAATATTAATAAGACGAACGACACA
155  113 ATCCACTCAAAAAAAAATATATTTTTGATATATTTACAGTAGGAAATTTACAAAATATCTTCATGGAACATGAT
         CTTTACTTAATATCCTAATGATTTTTTGGCATGAAAGAAAA
         AGTCCGTACAGCAAACAAATCCAAATCGTAATCCAACAAACAAGGTCAAGAACAGGCAAAGAACATACACG
156  249 AGGCAATCAATGTGGCAATGAGAAAACGCTTGGTAGAGACAGGGGTTAGCTGGCAATACTTCGCAATCTGTT
         GTACAACTGAGCTGGATTTAAACAGGAAGCTACAGGAAGTGGAGGAAGAAGAAGAGCACAGTCAGTAATGA
         GGAGATGGCGCCCTCTGGCGGCTTGATGAATGGAT
157  121 CGAATGTCCAATGTCAAGCCAACTTTATGCCTGTTTTTGTTTTGTTTTTTACTTTATTTTTCTTTTCTCTGTGCCG
         GATTGCTGATGTCTTTTACCCGGGCATAGATGTCCATCCATCAAT
158  93  CAGCTCACACTCCCAGATTCACTGTTTGATTGGTTACATGATCTGGACTGAAACACACCAGTTTCACTTCTGCT
         GTAGTCAAGAGATAATTCA
         GACCTCGGCCTCACTGCGCAGTGACTCGTGGCTCTGTTCCACCACATTATGGCCAATCACGGTTAATGACTAGT
159  192 GGTCACGGCAAATCAGAGTTAAGTGGGAGAACTGGCCATTCTCTGATCGTGCCGCACAGCGCTTGGCTGTCTC
         TGATAGGTTAAATTGTTTGATTCTATTATCAGTGAAAACTCTGAG
160  130 TTATCCCTATTGCTTGTACACTTTTTTCTACCACATCTTTTCCTTCCCTTCGCCTCTCTATTAATGTGCTTGGACA
         CAGAGCTCTGTGAACAGCCAGCCTCTTTGCAATGACCTTTTGTGTCTTGCCCT
161  121 CTTAGATCCGGTAGTCTTACTGTTCCCAGAGCCAAATTAAAACAAAGAGGTGAATGTGCCTTTGCTGTTGCAG
         GCCCTAAACTGTTTACCAGTTCACAGGTCTGCACCTACATTACAGATT
         GCTTTCTCTCTACCTATATCTCTTTATGTCTTTTTCCTTCCTTCACCCTCTCTCTCTCCTCCCCCATCACTCTCTC
162  164 TCTCTTTGTGACACTATCTACCTCCTTCTCCCCTTCTCCTCCCCCCTCTCCGTGCTCTATCTCACTCAATCTAATT
         TTCTCTTTCCT
163  52  AGCGAAGGCTGACCCGTGTGGTTCGATCCAACAGAAGAGCTACTGTAGCTCA
164  141 TAAACATGTTTAAACATGTTTAAACATGTTTAGCGATGTTTAGCATGCTTAAACATGTTTAGCGATGTTTAGCA
         TGCTTAAACATGTTTAAACATGTTTAAACATGTTTAGCGATGTTTAGTGTGCTTAAGCATGCTTAAA
165  133 GCTACAGGCGTTGGTGGAATGCTACAGGCGTTGGTGGAATGCTACAGGCGTTGGTGGAATGCTACAGGCGTTG
         GTGGAATGCTACAGGCGTTGGTGGAATGCTACAGGCGTTGGTGGAATGCTACAGGCGTTG
```

166 165 CAACGCCTGTAGCATTCCCCCAACGCCTGTAGCATTCCACCAACGCCTGTAGCATTCCACCAACGCCTGTAGC
ATTCCACCAACGCCTGTAGCATTCCACCAACGCCTGTAGCATTCCACCAACGCCTGTAGCATTCCACCACTAA
AGGAATTGCGAATAATAAT

167 84 ACGTAAGTGCAGCAGAAATGTTTTTGTAACCTTGGCCAGATCTGTGCCTTGCCACAATTCTGTCTCTGAGCTCT
TCAGGCAGTT

168 161 AGCAAGCCCAAAAACACCGCCAAACATCAGCAAGCCCACAAAAATCAGCAAGCCCAGAAACACCGCCAAAC
ATCAGCAAGCCCACAAACATCAGCAAGCCCACAAATATCAGCAAGACCACAAACACCGCCCCCAAACAGCAA
GACCACAAACACCGCCAA

169 238 TGATTTTTAATGACAGCACCAAGTCTTCTAAGCATGGAATGAACAAGTTGGCAACATTTTGCAACATCTATCTT
TTTCCATTCTTCAAGAATTATCTCTTTTAGAGACTGGATGCTGGATGGAGAGTGATGCTCAACTTGTCTCTTCA
GAAATTCCCCATAGGTGTTCGATTGGGTTCAGATCAAAAGACATACTTGGCCACTGAATCAAGTTCACCCTGT
TCTTCTTCAGAAAACCA

170 175 TGACTTGAACCCCATAGAGAATCTATGGGGTATTGTCAAGAGGAAGATGAGAGACACCAGACCCAACAATGC
AGATGAGCTGAAGGCCGCTATCAAAGAAACCTGGGCTTCCATTACACCTGAGCAGTGCCACAGGCTGATCTTC
TCCATGCCACGCTGCATTGATGCTGTAATT

171 176 GAATTACAGCATCAATGCGACGTGGCATGGAGGCAATGAGCCTGTGGCACTGCTCAGGTGTAATGGAAGCCC
AGGTTGCTTTGATAGCGGCCTTCAGGTCATCTGCATTGTTGGGTCTGGTGTCTCTCATCTTCCTCTTGACAATAC
CCCATAGATTCTCTATGGGGTTCAAGTCA

172 167 ACCCATGCATGCATTCCTCTGCAGTGTACACCGTATTGTGTCACGGGAAATAGTCACCCCAGTTTGGCTTTCT
ACTTCTTTAGATAGCTGCAGTGAACTTGCTTGCCAATTTTCTTCAACCCTTCTCATCAGAAGACGCTCCTGTCA
AGGTGTTAACTTCTGTGGA

173 149 CAGGTAAGAGACACACACGCACACGGCCATCCACATGATGTAAAAGAAAACATGATTCATCAGACCAGGCCA
TCTTCTTCCATTACTCCAGTTCTGATGCTCACGTGTCCACTGTTGGCTCTTTCGGCGGTGGACAGGGGTCAGCA
TGG

174 145 TCCCGTGGATGCGGATATCCATCCATCCATCCCGTGGAGGCAGATATTTATCCATCCATCCGTCCGTCCATCCA
TCCTCCTGTGGAGGGGGATATTTATCCATCCATCCATCCCGTAGAGGGGGATATTTATCCATCCATCCAT
TGGGGAGGCCTGCGTTTCGGCATCCTCATCCTCTCAGTCTCAGATGCAACCATTATCTGAGCCAGGGACTCAG

175 164 ACCAGGAGACATCATGCCTGCATTTCACACACACACGCACACACACACACGTACGTACTGACACAC
ACGGAAACATCCGTGTGTG
ATCTCCAGTGCAGACAGAGCCTCATGACACAGCCTTCAGGGGAGAGAGAGAAAGAAAAAGAGAGAGAGAGG
GAGGGGAGGAGGGGGAGAGAGACAGAGAGTGAGAGACAGAGAGGGAGAGAGAGGAAGAGAGAGAAAGA
GA

176 142 GCTCGGAGGCAGAAAACTCAGGAGACTGGGGCTGCACGGAGGCAGAAAACTCAGGAGACTGGGGCTGCTCG
GAGGCAGAAAACTCAGGAGACTGGGGCTGCACGGAGGCAGAAAACTCAGGAGACTGGGGCTGCACGGAGGC
AGGAG

177 147 TATCACTAGCCCAAACCTAAACCTAAACCTAACCCTAACCCTAGCTCTAACCCTAAACCAAGAACTAAACCCTA
AACCTAGAACTAACTGTAACCATGAACTTAGCACTAACACTAACATTAGTCCTAACCCTAACACAAATCCTAA
CCCTAA

178 152 AACCACCCCCCACCACCGCCTCCCCACTACTCCTCGCCTTGCAGAAAAGACAATGATTCTAATACTGTCACGA
CCGCTGAGATATACTTCATGTAACACATGTTCCCTCGACCTTTACGGCACGTGCTAAAGAATTAATTCAGATAT
GTTTTGGGAATACATGTAAATGTGCCTGTCG

179 178 CACCTAGCCTCACCCCACCTCGCCCAGCATCACATGCCACGCCTCGCCTCGCCTTATCTCGCCCCATCTCGCTC
CACCTCGTCCCGCCTCGCTTCCCCTCGCCCCACCTCGTTCCGCCTCGCCCCGCCTCACCCCGCCTCGCCTTACCC
CCTAGCCT

180 159 TAACTCCGCTTCTCCGTGATCCTGCAGATCTTCCTCCACTTGGCGCTGCATCTGTGTGCAGTTGTTTTTGCCATC
GCAGCCATTGTGCTGTA

181 92 GTGAGGACAGGAACATAGATTGACCAGTAAATTGAGAGCATTGTCTTTCGGCTCCCTCACCACAACAGACTGA
TAGAATGACCACATCACTGTAGACACCGCACTGATTCACCTCTCAATCTCACACTCTCCCTCACTCCTGATCAA
GACCCTGAGATATTGAGCACCTTTTTCCGGT

182 178 GTCCAACTAAATGCAAACTGGATGGGGATGGCATGTCGCTGCAGGATGCTGTGGTAGCCATGCTGGTTCAGTGT
GCCTTCAATTTTGAATGAATCCCCAACAGTGTCACCAGCAAAGCAACCCCCACACCATCACACCTCCTCCTCCAT
GCTTCACGGTGGGAACCATGCATGTAGAGACCATCCGTTCAACCCAAACCAAAGATCTCAAATTTGGACTCAT
CAGACCAAAG

183 230 TGGTATTACTTGCTTCCCCCCGGGGAAATTAGGTCAAAAATTCCCCTTTCCCCGTTTCCCTGTATCCCTGCCTGCC
TCTATCCCTGTCTATCTGTCTCTCTATCTGTCTCT

184 110 ATCAAATCCCAAGATTTCAGGAGAGAGCTTGTGAGTGAGGAAGTGTGTGTGTGTGTGCATGTAAATATGTG
TGGTGTGTGTAGTATGTGGGGGGTGTGGTGTGTGAGTGTGCGTGTGGTGTGTAGTGTATGTGGTGTATGTAT
GTGTGGTGTGTGTGTG

185 163 CAGCGGATGGATGGATGGATGGACGGACGGACGGATGGATGGACGGACGAATTGCTAGATATTGATGCATTG
ATAGATATTTGTTAATTATTGCCTAACACAAATAATAAATAATTCCAGATTAATGGAATGCTATTCGATATGGG
ATACTTTATTATATATGCTTGGACTGGGTAGGAGTGAACGAACGAACGAACGAACGAACGATCGAATGAATG
GATGGATGGATGGATGGATGGATGAACAAACGGATGAATGAA

186 259 GTCGATGGTGCATCATAAGAGGGGTGCCACGGGAATCTTGGAGTGGGGAGGGAATGGGGACTCTTCAAATGG
AGAGGGAAACTCATGCAGGTCTACGGGTTAGGGTACAACATCTCTCCCTTATATCACTCTCTCTCTCTCTCTCT
CTCTCTCTCTAGGAGGTGTTAGCGTCGTTTGATAGGAGGAAGATGGGATTGATGCCACAAGCCCTGCTTCTCG
TGTTTTTTGAGTTCGGCTCTTGGGAGGTAT

188 249 GTCTCCTCAATTTATTTGGTCGTCTCTAGGCTGAGTCACTCACCCAGAGTTGACTCACTCACTGATTTACTCACT
CAGCCATCCAGCCACCCAGAGTTAGATCACTCATTCACTGATTTACTCGCCCGCTCACTCACTCACTCACTCAC
AGATCGGAAG

189 159

190   144   GGTAGGAAAGGTTCCACTGCAACTTTCACTTACATAACCTACTATACTTCTGTCAAAACATGTCTGCATGCACA
ACAGCACTGTGACTGGATTCACCCGGCGTAGACAGTGACAGAGGACTGCAGCTGCTGCTGACGTGGCGCT

191   117   GTAACCAAAACATAAGAGCCATGAAGATAGTTAACTTCACTTGAATGTACGGTTGTTAACATCAGCTAACATT
GGCCACCAATATCCACTGGTCATACCAGACCACATCAGACTGCA

            GTCAATGTTGCCATTGCCGCCATCAAGACCAAGCGCAGCATCCAGTTTGTGGACTGGTGCCCCACTGGTTTCA
192   247   AGGTTGGCATCAACTACCAGCCACCCACTGTGGTTCCTGGTGGAGACCTGGCCAAGGTCCAGAGGGCCGTGTG
CATGCTGAGCAACACCACTGCTATCGCAGAGGCCTGGGCTCGGCTGGACCACAAGTTTGACCTGATGTACGCT
AAGCGTGCCTTTGTTCACTGGTATGTGG

            ATATTTGAAAAAAGGATCACAAGAAGGAAGGTATGGTTACTATAGTGGACGTTTAAAATATTTCTGCAGAATG
193   170   CACTCCAACATCCATTATGATTTCTGTGTATTCACTGTTTTCTTTTGTTGACCTTTCTCTTAAAGCAAGTTGTGT
GGCATTGAAGTGAAGTCGGAGG

            ATTTTTCGGGCACACAGCTCATTTTTCAGGCACACAACTCACTTTTCGGGCACACAGCTCATTTTTCGGGCACA
194   206   CAGCTCATTCTTCAGGCACGCAGCTCATTTCAGGGCACACAGCCCACTTTTCGGGCACACAGCTCATTTTTCGG
GCACACAGCTCATTCTTCGGGCACGCAGCTCATTTCAGGGCACACAACTCATTCTTCG

            ATTTTGGTCAGAAATCAGCATTTTTAGTTACACCCACCCATGTTCTCCAGACAATTGCTAAGGAATGGAAAGA
195   111   GGTACAAACATTTTTTTTTTCTGATGATAAAGGAAAGT

            CTTTTAGGTGACATCCACGGCCAGTACACGGACTTACTGAGGCTGTTCGAGTACGGGGGATTTCCGCCCGAGT
196   118   CCAACTACCTGTTTCTGGGGGACTACGTGGATAGGGGCAAACAGT

            CTGCTGTGACAGGTTCACACACACACATACACAATCATAGACATGCACACAAACACATACACACTGAT
197   131   GCACAGGTGTGCATACACACTCACATACAGAAGCACAGACATGCAGACACATAGACACA

            AGGCCCTCCTCTCCTCTCTCGCTACCAACCTCTCAGTTGTCTCTTTCTCTATCCCTCTCTCTCTCTCAGCTCTCTC
198   173   TCTCTCTCTCTATCCCTCCTCTCTAGCTGGAGGGAGATCCTCAACTCCATGACGTCTTGGGGGAGGTCATCATA
TGAACGACGCCACAGCTCCTCCT

            CACACACATTCCTAGCAGTATTGTACACACATTCGTAGCAGGTCCACACACATTGTAGCAGGTCCACACACAT
199   181   TGTAGCAGGTCAACAAACCTAGACGCGCGAAGCGCGGCTAGCTCAGCCGCGCAGCGCTAGACGCGCGAAGCG
CGGCTTTAACTCATAAAATGACCGATTTCCGTTGTA

            TCACTAAATGAGCCGATAAAGTGACGTTGAAATGATAAAGGTCAGAATTTAGTTCACAGATAAGATCTTCCAG
200   118   TGTAAATCATTGTTGTCAAATGGAGTGGAATGAAATGGCTGAGGA

            ATCCATCCATCCATCCATCCATCCATCCATCTATCCATCCGTCCGTCCATCCATCAGACCGAAAATGGATACTG
201   248   TTTGAGGAATTGCTTGAGACTTGGAATGAGACACAGTCTGTCGACCTTCTGACCTCTGAGGTGACCTTGAATG
GTCAGCCTGGCCAGGTGGCTGAGCCGTGGAACGCCGAGCAGGAGGGGACGATTCAGGACAACCGGGGTGTCG
CACCGACCTGAACTTTCGACCAGCACGTT

            TGGTAAGCAGAACTGGCGCTGCGGGATGAACCGAACGCCGGGTTAAGGCGCCCGATGCCGACGCTCATCAGA
202   155   CCCCAGAAAAGGTGTTGGTTGATATAGACAGCAGGACGGTGGCCATGGAAGTCGGAATCCGCTAAGGAGTGT
GTAACAACTCA

203   50    ATCCGCTAAGGAGTGTGTAACAACTCACCTGCCGAATCAACTAGCCCTGA

204   82    GGTTGATATAGACAGCAGGACGGTGGCCATGGAAGTCGGAATCCGCTAAGGAGTGTGTAACAACTCACCTGC
CGAATCAACT

            GCTTCATACTTAAGATAATTAGCATGTATTAGCTTCATACTTAAGATAATTAGCATGTATTAGCTTTATACCCA
205   222   GAATAATTAGCATGTATTAGCAAGTATTAGCAAGTATTAGTCTCATACCTAAGATAATTAGTATGTGTTAGCTT
AAATCTAGGATGATTAGCATGTATTAGCTTCATACCTAAGATAATTAGCATGTATTAGCCTTATACCCAGAAT
A

            GGTTGGTTGGTTGGTTGGTTGGTTGGTTGGTTGGTTGGTTGGTTGGTTGGTTGGTTGGTCGATTGATTGATTCTGGAT
206   109   GATCTTCTGGGATTGACTGTTTATCACTGAAGTGG

            ATTTCTACCAATTTCACCTGTGAAATTGCTGGTTCTCACATCAAAACATCCAATACTGTTAAAAATTTAGGTGT
207   98    AATCATTGACTCCACCCTTTCCTT

            TGTTAACACAAATAGCTTATCAGCCAATCACATGGCAGCAACTCTGCCGAAGTTCAAACAGAGCATCAGAATG
208   160   GGGAAGAAAGGGGATTTTAAGTGACTTTGAAAGTGGCATGGTTTGGTGCCAGACGGGCTGGTCTGAGTATT
TCAGAAACTGCTGA

            CGTTCTGTGGCAGTTTGCCTCCGTTTCTAAACACCTCTGAAACATTTATGTTAAAAAAAAAATATTGTCTGTGG
209   263   GAGATGGGAGCAGCTGTGACATGTTTGTCAGCTGACCCACTTTTTCCTGATTATGAATTAAAATTGAATGATG
GCGTCTGAGATTGCCAATTAAAAATGCATGCAGTGCCACGGTCAGAGCTGAGAGGCCAAACGCAGTGTCGCC
TCCGGCCTCCTACAGGAAACCATAATTCACTACTGGGCCTTTTA

            GAGGCGTTCTAGCAACTGATGCTAGAGACTGCGGTCTTTAGCCTCCTTGTTAGAGGGCCCGATTCCCATGCCA
210   171   GCGGACCCGGGTTTGAGTCCCAGTTAGAACAGGACGTGTCACTTTGGTGCCATGACCTGGATGGGAATGAGGT
TTAGGGGGTAAGTGTAATGTACGTA

            CTTCATCCTCCGTCCTCCTCATCTTCATCTTCGTCATCAAGCATCGGCTCCTCATCATCAAAATCCTCCTCGTCCT
211   126   CATCATCATCTTCCTCGAGGGTGTCATTCAGCGTTTTGAGAGTTTCGGTCT

            GGCGACAGTCGCTCAAAGAAGCGTTCGATTTCAGTACTACTCAGTACTACGGTACTACTCTCTCCCACCACCC
212   252   ACCCATCCATCCATCATTTTGTTCTTCTTTCAATCTGGTCCTTTCGTCCATCTATCCATCCATCCGTCCGTCAGT
CCGTCTGTCCATCACGGCCGCTAATTGTTCCGCTCGACCGATCAGGTGAGGAGTTAGTTTCGCACTCAGTCCCT
TATCTCGTCTGAATTTTTGTTCTTCTTTCA

            ATGCCAACACACACATACACACAGACAGACAGACATGCACACACAGAACTTCACATGACATCAATCTGTGTT
213   210   GACTGTGTAGTGATCTCATTACAGACAACTTGCTGTTGATTAAAGCCTACTGTGTCTCTCTCATACACACACAC
ACACATTCTCTTTTCTCTTTTCCTCCTTTAGCCTGGGTCTCTCCATTCTGCGGTTTCTCCT

            AAAATAAATAAAGATCTAGACTTTAATAATAAATGATAAATATACTAAGATATTAAGAAAATTAAAATATAGA
214   127   TTAATACTTAGTATATTTATCATTATTATTAAAGTCTAGATCTTTATTTATTTT

            CATTAAAATCATACCATTGACCAAGCACTTTTTGCCTCTGTGACTGTGGTTGACCAGTGGCATTCATTTATGAG
215   122   TTCACAACACATGAGAATCACCATGACTCTGTCGTACGTTTATGATGT

```
216  187  GCTCTTATTGGCCTACGTCCGTTACACTCACCCCCCTAAACCTCACTCCCATCCGGGTCATGGCACCAATGTAA
          CCCCTCCTCTTCGAACCTTGGTGGCCAGGTCAGGGGATCAGGTGAGTTAGCCTCCATTACACTGGCAACCACC
          CACCTCACCAAGCATTGCGGTGGCACGTTTTGCATGGGAA

217  264  ATGGAAAAGATACTCTTAAGATGAGACGGAAAAGTGTTTGGCCCAGTTTGTGACTACAGTTGGCTCTTAAATA
          ATGAGTCTTGTCACAGGCATGTAGTTAGTACCCACAAGACCTGATCCCCTGTAATTTAGAATCAGTATTCCAA
          GGATAATGTCAGTTAGGATGTACACAGTAAACCTCAAAGGATATCCCCCGGTTATGCCTGAATATAACCTCTG
          CTATGCTTTGAATCTGCCTGTGATCAGCATGAGTATCTTTTCCAT

218  168  GATAGATAGACAGACTGATTGATTGATTGATTGATTGATTGACAGGTAGACAGACTGAATGCATATAGATTGT
          TAGATAGTTAGATAGATAGATAGATAGACAGACTGATTGATTGATTGATTGACAGGCAGGTAGACAGACAGA
          AGGCAGATAGATAGATAGAGACAGA

219  199  GTCAAGAAGGAGGTCAAGGAGGAGGTGAGCGTTTAAAATTGCATTCTTCTCATTTCACCGCAGTGTATTGTTTT
          GTCTAAAATCTCAATCTCAAATCCTTAGTATATTGGCTCACTGACTAAATCTTATTTTTAATTTACTCTTCCTAG
          GATAAGGAACTGCGTGATGTTGGGGACTGGCGTAAGAATATTGAAGACAAG

220  157  GACGTCTCAATGGTTGATTCGTTTGAACTTTGACCTACGTTGCATTGAACGGTTGAATGATTGAAAGCGAAAC
          TGATTGATGGACAGGCAATAGGGTGAGGTAGAGGAGGAGGAGGAGGAGGAGGAGGGGGAGGAACAGGGTGCT
          TTGGGGAGAGACT

221  276  GGCTATGGTTGCAGATATGAGATGATACATATCGTATCATATCACATCCATCAAGTCATTCGTTCATTCGTTCA
          TTCATTCATTCATTCATCCATCCATCCATCCGTCCGTCCGTCCGTCCGTCCGTCCGTCGCGAATATCAATACAGTCATCC
          TTCAATCCATCCATTCATCCGTCCAATCATTCATTTTTCTTAAAATCTCAAAGCAGTTTTACAGTCAGTTCAGA
          GTTTAAAATATGAAATTATTTCATTTGCGACGTATTCTTGATGTAATTTCAAG

222  188  CCTGGAGAGGGCCTCCGTCACCGGTGCCTCTGGTCGCCCGGTCGACCTCGAGGCGCTGCGCCCCGCCGCCGAG
          GACAGCGCTGCACCCGAGGAGAGCGCTGCACCCGAGGAGAGCGCTGCACCCGAGGAGGGCGGCCGGGACGA
          CGCCTGACGTCGTACGCCGTCGGCGAACCTCGTCGGCGCGTCGT

223  62   GTTCTCCTGCTGCACGGTATGGTTCTCCTGCTACACGGTATGGTTCTCCTGCTGCACGGTAT

225  105  TGTCGTGTGTTTGCAGAGGAGGCGGTGACTCGTGTGATCGATTTTAAAGTCTGATATTCGTCAAAATATTGTGT
          TCAGTTCATATTTGTTGGCGTTTAATAACT

226  90   GTGGTTTACATGTTTTAGTCTACTTGAAAACAAATTTCTTATTGCAAACCTTTGTACAAGGAATGTAAACTTTC
          TCTTGCATATCAGTAA

228  129  CATGTGTGGAGGATCTGAGATTGCATCATCTATTTGTTGCCTGTAGATACGAAGCTCTGCGGTAGAAATGAGT
          TCCTCCTCAGGGATGCTGCTGAGATTGAAGACGAAGCGCAGCGGAGTCTCCTGTGA

229  266  ATCCTGAAAAAATTAATCACTGTTTCCACAAAAATATGAAGCAGCACAACTGTTTTCAACATTGAATGTAGCC
          TTGGTGAGTATACAATACTTCTTTCAAAAACATTAAAAACATTTGCGGACTCCGAACTATTGAATGGTAGTGT
          ATATTAGGGTTGGTAACCGAGAACCGGTTCTCATCTGGAACCGGTAGTGTTTTTTGAAAAGAACCGGAACCGT
          GCAAGATTTCTAAGTTGCGGTTTCGAAAACGGTTCTGGTGCGATGGG

230  236  GAGCTGCCTGAACCGAAGAAACACATTGCATTACTTAAAAATACTTACATTTTCTTCATCTGCAGCTTTGCCTC
          GCTTTTGAAAACAAAATGGCAGCCCTGGGGGTGGAAATGTGCAGATTAAGGGGTGTTTAAATATTAATAATAAG
          ATCCCCTTTTTAAATCACTAGGGGAGCGAAATCGGAACGGCTCGTTTTTTCACATGCTTGTAGAGAAAGGCTT
          GCCAAAACAAAGGTACT

231  225  GCCGTGGGCAGTGGCGGCGTCTTGCATATCAATAGTCGGGGACCCCTGGGCCGTTTCCCCCCCACGGTAGTGC
          CTGGGGGTGGTTAGGGAGTTACGTGGGTGGAATGTAAGCGGAGCGGTGCGTACAGCTGTGCAGCAGTGTGG
          ATGTTATGAAGTGAGGGAGCGCCCAGTCACGGCGCTGCGCTGTGTGTATGTGTGTGTGTGTGTGTGTGTGTGT
          GTGTGCG

232  202  AGATCCAGAGCACTCCGGAGCAGGTTTTCATTAAGGATGTCTCTTGACATTGCTTCATTCATCTTTCCCTCAAT
          CCTGACTAGTCTCCCAGTTCTCGCCACTGAAAAACATTCCCACAGCATGATGCTGCGACCACCATGCTTCACTG
          TAGGGATGGTATTGGACAGGTGATGAGCAGTGCCTGGTTTTCTCCACACATATA

233  246  AACCTGCTACCAAGAGTCGGTCTGGCACCTGACGACGCATGATAACCTCTCAGCTCTCTGCCGATACTCAGTG
          ACAGACTGGAGGCTATCTGCGAGTGTGTTCGTGTGTGTGTGTGTGTGTGTGTGTGTGTGTGTGTGTGTGTATGTGT
          GTGTCAGCACGCTGACTGACAGATTAAAACAGGGAACAACAAAATCAAACGGTGTCCCGCCCTACGGCTCTG
          ATTTCCAATGACAAAGCAAGAAGTCTA

234  118  CGTCGAATTCTCGTTCTGCCTCGTAGAGAGGCTCGAGATTCCGTCAAAACCGACGTCTCCGGTGGGACTCGAA
          CCCACAACCTCCGAATTGCTTTCGGATCAACCGCTAGAAGTCCAA

235  152  GATAGATAGACAGACTGATTGATTGATTGATTGATTGACAGGTAGACAGACAGAAGGCAGATAGATAG
          ATAGATAGATAGACAGACTGATTGATTGATTGATTGACAGGCAGGTAGACAGACAGAAGGCAGATAGATAGA
          TAGACAGA

236  193  ACGATAAAGAAGTTTCTAAGGGAGGAGGAGGAATCACAGGACTGGAGACACACCTGTCTCTGGTCCAGGTAAC
          CTTATTTCTGATGCCATAAATTACGCTTTTAATAAGAATAACTCCAGTATACTGCTGCAAGCCATTTGGACCAT
          CCCCCTTTTTACTAACTAGAGATTTTCCACACAGCACAGCTCCACCA

237  235  ATGCTTAGCAAAAGAGTATTTTGACATTTTTGTGTGCTAGTTGACCGTTTAGGTGTGAATCCGATGCCCATAGT
          ATACTTTGCTTATGCACATTCTGCTCCGTTTCCGAGCTCACACACAAAAATCTTGTTTTCATAATCGTCGGAAG
          CCGAAATCAAAAATCAATTACGATTAATCGGACAGCCCTAATCAGCACAACAATGATTGTGTGGTGTTCTTGT
          CCAATCAATTGCTG

238  235  CAGCAATTGATTGGACAAGAACACCACAACAATCATTGTTGTGCTGATTAGGGCTGTCGATTAATCGTAATTG
          ATTTTTGATTTCGGCTTCCGACGATTATGAAAACAAGATTTTTGTGTGTGAGCTCGGAAACGGAGCAGAATGT
          GCATAAGCAAAGTATACTATGGGCATCGGATTCACACCTAAACGGTCAACTAGCACACAAAAATGTCAAAAT
          ACTCTTTTGCTAAGCAT

239  136  ATCCGAACTCTAACTCTTACCCTGACCCTAACCCTAACCCTAAACATAAGCCTAAATTAAACCCTAAACATAA
          CCCTACCCTAACCCTACCACAACACTAACCGTAACCGTAACCCTAAACCTAACCCCAACCTA

241  129  AGCGTGACCTGCGAAGCAAAGAGACCCCAGAGAAGACGAACGTTGTGAAATCACCTGGACAGGCGGACCTGC
          ATCCAGAGGGAGAAGACAGCAGAGAAACCGAAGAGCTTCCATACATGACAACAACCG
```

242 116 CCTAAATAGCCAACCCAAGACTGACGTGGGTCATAGTCCACCTGTCGTATCTGTTTTTACACTAAATATCTGTA
AACATATGGTATTTCAATCAGGACCACTTCCGTCAAGTATTA

GCTCGCTCGGCTCCACCAGAGACCATCATCCTTATGGCTCCAACAGGCTCCCTTGTCCCTCCGGCTACGCCTTG
243 271 GTCAGACGTCACTCTGCCTGTGCCATGGACTTCTGAGCTGTCTGCTGCACTCCGTCTCTCCACCCCTTCAGGTC
TGTCTGGTTCCGCCTCGGATCTGACAGTCACTGCGGCTTCGCCTCTCAGCCTCCAGGACCTTCGGTGTTGACCAGT
CGTATCTGCTGTCCGTCTGCACATAGGGCTCCACCTCCATCAGTTTCAT

AACTAACAATCCAACTACAAGTGTCAAACCATCTGAGACAACTAACAATCCAACTACCAGTTTCAAACCATCT
244 209 GTGACAACTAACAATCTACTCACCAGTTTCAAACCAACTGTGACTATAAACAATCCAACTACCAGTTGGATTG
TTAGTTGAACTGGTAGTTGGATTGTTTATAGTCACAGTTGGTTTGAACTGGTGAGTAGATTG

CCTGAATCATAAGCATCGAATATTGAATATTGCATTCTCATATTTAACAGTTCAGTAGCGGTGAGATGTGATA
245 126 CCTGTGGGGGTTTGTGGAGGTCTCTGAGCCTCTTGTCCTTTGCCTGCGTTAAA

CTCTGATGTTGACCCTGCCAACATAGCGTGCTTCAAAGCTGCATTCACAAAGGACCTAAACAGGCAAAAGGA
246 206 GAACTCAAACCTATGGTGGTTAAAGGTGGCAACCGCTCTAGATCCTAAGTTCAAAAACCTTAGGTGCTTGCCC
AGGGCAGAGATGGGAGAGGTATGGCAAAAGCTGAGTGAGATGCTGAAGGATAGAGAACCTG

CGATAGGTAACATTGGTTTTGCTTTGTTTCACAGAATATATGCTGTTAAAACACACAGTAATACCAAAACACA
247 220 CTTCACTTATTGATGACATCCTACCCTCGGCCGCCTGGCACATCAGATCCTGCCATAATCATTGCTGGGGTTGC
GTACGTATGTGTGGGGTGGCGCTATCAAAATAGGGACGAGACCCTTTGGGGTAGGGGCGTGTTTGTTTTGGT
TCCGCGAAAAGGCGCTGTGGGTTGGAAGGCGCTGTGGTTGGAAGGCGCTGTGGTTGGAAGGCGCTGTGGTTGG

AAGGCGCTGTGGTTGGAAGGCGCTGTGGCGGAGAGGCTCAGTTGCCCTGGCCGACGATGTACAGATATCCGG
248 148 CGAG

CTGAGGCCGGGGAATGGCGTGAACCCGGGAGGCGGAGCTTGCAGTGAGCCGAGATGGCGCCACTGCACTCCA
249 85 GCCTGGGCGACAG

ACAGTCACATAAAATCTAAAGTTAATAACTGAAGGGTCAAGATCCCAACTAAAGCAAACACTGAACACTATA
250 92 ATGGTGATACACAAATTGTG

CTTATTCTTGTCTTGGATAACCAGGCGCTTTCGGGCAAAGTAATCAGTTTTTCCCTCTCTCCTTCTCCTGAACTT
251 172 CACCTGGTACCTCTTGAAGTAGGCCTTGTTCTTCACTACTTTAACGAATCCCATTTTGTAGACTTTTAGTCTCC
TAAGCCGGAGCCTGGAGTCAAA

GGCTACGTAGCTACCGCCGGCTACGGCTACGCCATGCAGCAGCCGCTGGCCACCGCCGCTCCAGGGACCGCA
252 128 GCCGCAGCCGCCGCCTTCGGTCAGTACCAGCCCCAGCAGCTGCAAGCCGAACGGAT

GGAGGGTCAGCATACGGAGGGTCAGCATACGGAAGGTCAGCATACGGAAGGTCAGCATACGGAGGGTCAGC
253 103 ATACGGAGGGTCAGCATACGGAAGGTCAGCAT

CCCCCTACACCCAGATTTTTATTAGACATTAGTCTGATTCCTATTCCACAGGCGTTTCCCCTTCCACTTTAACAC
254 144 TCAGCCACACCAGAGCCAATTTTAAAAAGGCTCTTCTGCCTAGCCTGGAAAAAAAAAAAAAAACCCTTGG

CTAGCAACTAGCCATAATGCCACTCATAGCTAGCAACTAGCCATAATGCCACTCATAGCTAGCAACTAGCCAT
255 138 AATTCCACTCATAGCTAGCAACTAGCCATAATGCCACTCATAGCTAGCAACTAGCCATAATGCCA

GTCGAGAATGCCGAGCTCGCTCTCTGCCTCCGTCGACGCAGAAGACAAAGTAAAAGGTCCGTAGTGGCGGTA
256 148 GATGATTCGTAAAGGTTCTTGCGCGGTGTCGGGGTCCAGATCAAGTGCGCGACCCTTGCCCTTGAGCTGCTGC
TGG

GACAAGATGCACAGCATAGCCATATTCTAGGTAGACATGAGTTTTTTCTTTAACTATATATATATATATATATA
257 93 TATAGTCGTAGTGGTCTTG

TGATGTTGATCTGAGAAACCCTGTTTTATTATAAAAAATAAAGCATTTTTATATATAGGGTGTGTTCACCCACC
258 271 CACCACCATCCAACAAGCCATCTGATCACACATCTCCTTTTACTATTCATCCATCCATCCATCCATCCATCCATC
CATCTCAAATACACCTGAACAAAGCCAACAATAAGAGAAGTTCAACAAGTGTGCAGGCGTCAGTCTTTAGCTCC
TTCTTCCTCTCACTGTATGCATGTTTAAAAGAGCGTCAGTATATCATCTTAA

TTTGTCGCACAGCAGGTAGATTCTTCACCACCGGTAACGCAACCAGCGGTACGGTCCATACGAACGATTTTC
259 148 AGGTTAGAAGCGTTCGGAGCTTTAGAGTCGTAGATAGCGTCAGAAACAACCGGTTCCAGACGACGGGTGAAA
GAA

GTGGTGATGGGGGTGGTGGTGGTGATGGTGGTGGAGGTGTGGGTGATGTGTGGGCAGAGGTGGAGACACGGG
260 82 AGACAAAGGT

CTTGTAATTCTTGGCAAACATGAACCCACCCTCATTGATTCCTGGTGTGTTTGTGTTGTAGTTATTCAACTTTTA
261 177 CACCACCACCCTGCAGAATTCTTTGTGTAGACATCGTTTCATCTGTGATTTGTCCTTAAAAACAATTCTACATT
TTGAAAGCTGAATACAGATGTTAAAATA

TCTTAGCGTAGCCTGAACAAAACCCTGAAAATTAATTCTTTCCTTCACCCCCCATCATCCCGTCACCTCCTCCC
262 86 TCTTTATCTTCT

CATGGATTATCGGATTGTGCAAATTTAGCACCACAACCACCACAACCACCGCAACCTCCACAACCACTACTAC
263 137 TTATAGAGCCTCCACAGCCTCCACAACCACCGCAACCTCCACATCCACCACAGCCACCACTTGT

CATCTTCTTGCTTTTAGGCTAAAACAGAAGCACCCACAGCCTCCACAAAACCTAGAGGAAGCAATTTTTTAGGT
264 122 TGAATGTTGACAGATGCATATTGAACTTCATCTTCCTGCAGCACATTCT

CCTGCTCAATCTTTCCGCCTCCATACGGATGTCAATCATCTGCTGGCGGTCCTTGAAGCTTCTCGATAAAAAAC
265 205 GTCCCTCTGGGGTGTAGAGGGAGCGAGTAGGAGGTGAGGCACCGGATGTTGCGGACCTCCTCTGAGTCACTTTC
GCCACCGACGGGACCCTCACGTTTGCGATGCGGCGGGAGCCGAGAGCAGTCATCGTCG

AATATGCAATCAGCTTGATGGTTCAGAACATAAAACACTCTAAAAGGAATGTAAAAGTGTTTATTCATTCCCC
266 248 ATTTCAAGATCGCTCATGCTTGTCTTTATCTTCCTGTAGGAGCTAGAAATGATGAGAAGGAGCGGCAAGTGGA
AGGTATGGAGAGGAACAAGTCTCTCCAAGATTACAACGAGTATTACGATTACTCAATGCAAATGCACCGCAA
ACTGTTCTCTGGCCAGAAAATTGGCAGAAT

TTAAATTAGAAAACAACCAAAAATTCTGTTTAAAATTAAGACTTGGTTGTTAAAGTGTTAAATGGACCATGTCC
267 185 GTAGTTAATGTACTACACCTGTTGTTAGGACACTTCTGTAAACTTGAGGGGAATCAACTTCATACAGCTACTGT
TACACCATTTGAATAAAAATAACAGCAAAAATGACTGA

CTGCGATCAATCAGAGAGCCGCGAGTCAGTGGAATTACTGGATCGATCACTGCGATCAGTGTTTATCAATCAC
268 134 CCACAACAGGCCATATGGACCGGCCGGTACGTCACAAAGTCCCCGGATTTGTTGCAGGAGA

| ID | Len | DNA Sequences |
|---|---|---|
| 269 | 148 | CGTCTGCAGACGGGCTACAGAAGACGGGTAGTATAACCGGACCGAACCCAGGTCCCCAGCTCTCATTACACC CGGATCAACGCATCGCTTACAGTCTGTCGGTCGGAACGAGCTCGGTTATGGGATATTAACTCAAAAGAGGATC GTG |
| 270 | 180 | GATGTGTTAAAAACGGACGCATGCAAGAAGACTCGTCTGGAGTTTTACATCATCAATCTGTGACACGTTGGGT GACTGTGAGCTGCATTTACATGTGGTGACTAGAAAACCAAATTTCTAGTGTTAACAGGGTGTGAGAAAAAAA GTGAAGCGGGACTCGATTGAAATGTGAAAAAATAG |
| 271 | 110 | TAAACCCCCCCAAGGGGGTTTGACTGGGGGTTTGACTGGGGATTGACTGAGGGTTTGTCTGAGGGTTGACCGA GGGTCTGACTGGTGTTTGACTAAGGGTTGACTGAGGG |
| 272 | 108 | ATCACCCTCACTCTCCTGTCCTCTGTGCTGTCCCTGCTTTAGGTCACGTGCCTTCAGGACTTCTTTGGGGACGAT GACATCTTCCTCGCATGTGGCTCTGAGAGGTTT |
| 273 | 237 | CAGCAAAATAAAGTATGGCATCGCAATAAACACAATCAAACCGGACAAAATATAACCATCTAGCCATTAAAA ACACGAAAAATCAAGGAAAAAACGTCAGACAGGCAAATCTCGTGGTCTCGCGAATCCAGCAGCTTCGCTTCT GAAATGCTTCTGGTGTGAGTTGACACCGCACAGAGGACGGAACAAAACCTTGTCGGAGCCGTAACACACCGC AGACGCATGCAGTCTAAACAA |
| 274 | 184 | GATTTGTTAACCGTCGTAATGCATTGGTTTTGTTATTCTCGTAGTCATCATCACCATCATCATCATCATCATCAT CATTACTGTAACACGTGTTCCAACTCTTCCAAATGTAAGGCACATAAACTCTTCCAAGGCCACGTGACTCCAG CAGCCTTGCCTTTATCATTATCATCATCATTATCGT |
| 275 | 167 | AGTTTTACGTTAAAGATGATAGATGTGTTATTAAAGTCTCAATTTTTTCCTTCAGGTCGCTAATCTGATCTTGC AAAGCTTCTGTCTGTGTTTACTGGATGCTTCCAGAATCAGCCCTCCTAACAGGTACTGAAAACCTCTAAATGTA ACTTTGGTTTCCTGAGATG |
| 276 | 113 | CACAGAACTTATTTTACTCTTAAATTCAAGGAACTCATTAAACCCCATTAAGAAAATCTTGAGGGAACCAGGG GTGGGGTTTTGACATCAGACTTCCTCCACTCTATTTCTTA |
| 277 | 65 | AGTACTAACCAGGCCCGACTCTGCTTAGCTTCCGAGATCAGACGAGATCGGGCGTGCTCAGAGTG |
| 278 | 117 | TAGATGGTTCGATTAGTCTTTCGCCCCTATACCCAGGTCGGACGACCGATTTGCACGTCAGGACCGCTACGGA CCTCCACCAGAGTTTCCTCTGGCTTCGCCCTGCCCAGGCATAGT |
| 279 | 113 | CTGCTGTCTATATCAACCAACACCTTTTCTGGGGTCTGATGAGCGTCGGCATCGGGCGCCTTAACCCGGCGTTC GGTTCATCCCGCAGCGCCAGTTCTGCTTACCAAAAGTGG |
| 280 | 100 | CATGCGGACCAAGAAATAATCAAAATCTCTTCAGGTGAAAATGTTAACACAGAGTATGTATTGCTACATGATC ATGAGGGGGTCTCAATGTTTCCTCCAA |
| 281 | 73 | TGTAATAAAATACCTCACTCACCGGTCAAATAATAAAAACTCCATCATCACCGCATCACTTTTACAACATTTT |
| 282 | 149 | ATTCGGCAAATTGGATCCGTAACTTCGGGATAAGGATTGGCTCTAAGGGTTGGGTACATCGGGCCCTGGTTGG AAGCCGCTGATGCTGGCTTGGACTGCTGCGTGGGAACATGCGGTGGACCGAGCCGGCGTCGGCGTGTGGACG GCCA |
| 283 | 150 | TTGTGCAACAGCGATAAAGTCGGTAGGAACAAGACGGGTGCCCTGATGCAGTAGCTGGTAGAAAGCGTGCTG ACCATTAGTTCCAGGTTCACCCCAATAGATTTCTCCGGTGGAGGTAGTAACGGGGCTGCCATCGACACGAACA GACTT |
| 284 | 121 | ACGCAAGGCCTGAATTAGTGACACCATTTTGTTGTGCAGTCCAGGAACAGGTCGTCAAGACAAGGAAGGTTTG TCTGTGACGAAGCAGAAGACGGAGGACACACAGTGCACGTCACACCTTC |
| 285 | 150 | TTGTTGGTAAAGGCTCTTTTTTTAAATAAGTCCCAAATGGGTTTGTGATCGGTATGCACATTAACCTTATGACC TAAAAGTATGTGTTTAAATTTCTTTAGACCCCCATATAAAGCTAAACACTCTCTTTCTGTTGTACTATAGTTTCT T |
| 286 | 100 | GTGAAAAAGCATTCAGTTTAGAGATATCAGCTACTTAGGATATGAATAGTCAAACAGAAGAACGGATTTCAG TACCTGGTATTTCTAATATTCATGACTT |
| 287 | 150 | ACGCAAGGCCTGAATTAGTGACACCATTTTGTTGTGCAGTCCAGGAACAGGTCGTCAAGACAAGGAAGGTTTG TCTGTGACGAAGCAGAAGACGGAGGACACACGATGACGTCACACCTACAAAACGAAAAGCCCCCCGCAGGA CCACAC |
| 288 | 150 | ATACAACACAATTTAGATATATTTTTTTATTTTTTATGATCTTTATATGTATGTAAAAATAACTATGCACAAACA TTCTACATGCATGCATCTACTTATCAATATATATATATATATATATATATATATATATATAATATTTTAGATA T |
| 289 | 109 | CAGCGGCTGCCCGGGCGGCGCAAGGAGGCCTGCCGACGCAGCGGCCTCGTCCAACTCCTTCCCGTTGTATCCC ACGGCTCCTTCGACCTCGGACTCCGCCCCGGCGCCC |

## Table S3B. Alignment of DNA sequences to Pan-genomes

| ID | Len | DNA Sequences |
|---|---|---|
| 290 | 150 | GCCCACTTCCGCGGCCACCGGTGCAGGGCATGCCCCCGGGCCGTGGAGGTGAGGGAAACAGGCGCGAGTCGC GGTTTAACAACCCCCCCCCTCCCCCAAGGCCCCGCCGCGCAGCCAGGGCGGCAAACTTTAGCGGCCGGTAAGGT ACAGA |
| 291 | 150 | ATCTTAGAGACCAATGACATATAGGATCTCAGCATTTTGTAATATTGAAAGGAGCAGGTAATTTTTCAACTAA ATCAAAATACATAAAAGCTTGCTCATGAAATACCACCATCTTAGACAACACAGTTCTCTTGGGTGTCTTAAAT ATGA |
| 292 | 150 | AGTCGCTCCAGCTCTGAGACTTCATCTCCCTTCTGTAATGTAAGGGATGGATCATGCTTGTCCTCACAGGATTACAAA GTCAAGTACAGACATACTGCTTTTTGTTTCCATAATGGTCTCCTGGAAACCACATTTTGGGTCATATTAATGAT TT |
| 293 | 150 | TAACCGAAGAATCACCAAAAGAAGTGAAAATGGCCTGTTCCTGCCTTAACTGATGACATTACCTTGTGAAATTC CTTCTCCTGGCTCATCCTGGCTCAAAAGCTCCCCCCCTGAGCCCCTTGTGACCCCCGCCCCCCCCCCCCCCCCCGA CCA |

```
294  149  GAGCTGAGATCACACCATTGCACTCCAGCTTGGGCCACCGAGTGAGACTCCGTCTCAAAAAAAAAAAAAAAA
          AAAGAAAGGAAGGGAATTTGCCAAACCATTAATAATAATGCATTAATGCATTCTTTGTAGAATCAGGGCAGT
          GCACA
295  109  CCAGGAAGTTTAGAAAGTTGCTTTAACCAGGGCACATGCTCTGGAGACTGTGTGGTTAATAATGACATTTAGT
          ACTTTCTTATTCTTTACACCCTCCTCACGGGGCATT
296  150  AGAAAGCACACTGTACAACTTCAACAAGGACAGGTCAGTTTTCTTGTTTAAAACCACAGTTGGTGCCGGGCAT
          GGTGGCTCACGCCTCTAATCCTAGCACTTTGGGAGGCCGAGATCGGAGGGATCACAAGGCAGAGAGCAAGTCA
          CGACA
297  150  AATTATCCAATGTAATAAAAAAATACTAAAAACTGAAGCTCAACAAACTCCATCTTGGATAAAAATAACAAT
          AACAAGACACAATCAAAGCAAAGATTTTAAATTCAAATAAAAGAAAAGAATCTTCTTATTATTTAATTATTTT
          TTTTA
298  74   GTCTCACCTGTCACCCAGGCTGGAGGGCAGTGGTGTGATCTCGGCTCACCCCAACTTCTGCCTCCCAGGCTCA
          A
299  119  ATAAATATTTGAGGTGATGAGAACATTTGAAATCCTCTCTTTTAGCTGTTTTGAAATATACATTATTTTTAAGT
          ATAATCATTATGCTGTGTCATTGCACAGCATGGTGACTAATAGCAA
300  72   GGTTAGATCACTCCCACTGCACCAACGAGCACAGGGTAGGTCTCAGAGCAAGTCAGGGGAAGTGATCTAACC
          GTCTCTAGCACGTCTTTATCTCCAACCCTCAGGGACTCTCCTGTGTGTCCCAGCTACTCTCCAACCACGCCCAC
301  150  AATTCAGCGGGGGTCAGATCCAGGCACCAATCAAAAAACAACCAGATCCCACAATACACACTTACTCTCC
          AGCA
302  150  CGATTCCATTCCATTCCATTCCATTCCATTCCATTGCAATCGAGTTGATTCCATTCCATTCCATTCCATTCCGTT
          CCGGATGATTCCATTCCATTCCATTCCATTCCATTCCATTCCATTCCATTCCCTGCAGTCGGGTTGATTCCATT
          CCTGTAATGCCAGCTATTCAGGAGGCTGAAGCAGGAGAATCACTTGAACCCAGGAGGTGGAGGTTGCAGTGA
303  131  GCCGAGACTGAGCAAGACTCTATCTCAAAAAAAAAAAAAAAAAAAAAAAAAAAAAAAAGACACGC
          CTTTTTTTTTTCCTCTCCGCGCCCCCGCGGGCCCCTCCCTGCTCTGCGGGCGCAGAAATTCCATCTCCAATTCCT
304  148  GTTCCCAACATCCCCCAACCCCCCCACCCCCACGGTAGTACAAACCCCAAACCGGTAGCAAAGATCGTGGGGT
          ATTTGTATTCTTCCAAAATTCCTATGTTGAAACCCTAATCCCCAGTGGGATGCTGTTTGGAGATGGGGCCTTTG
305  87   GGAAGAATACAAAT
          GGCAACAAGAGTGAAACTCTGTCTCAAAAAAAAAAAAAAAAAAAAAAAAAAAAAAACAAAAAAAAGATACCCAGT
306  150  GGTTGGGGTGGGGTCTAAAGGGAGAAAGATAAACCAAATGGCAGATCCATAATGAAAGGCGTTGTGTTTTCC
          TCACATT
          ACTCTCCAAATGAAGTTTCCACCTATCAATCCCTCTCCACTCAAGGCAGATGGTTCTTAGACCAAAAAAAAAA
307  150  AAAAAAAAAAATGTCCCACCACCCACACAAGCCCCATAAATTATGTTGTGAATAAAAAACAATTACAAGTA
          GGTTA
308  146  TCATAGTCAAACTGCTAAAATGCAATGATAATGAGAACATCTTGACAACCAGAGAAAAATAACTCATTTCACA
          CACAGGAACAATGATATGAATAACAGCTGCCTTCCCTCTCACCTCAGCCGCCCCGGAAGGTGGGTCGAGGGTC
          GATATATACAAGAATATTCATAGCAGCTTATTTCACAACAGCTCCAAACTGGAAACAACTCAAATGTCCATCC
309  149  ATATGGTAACAAGATCAACAAACTGTGGTATATTAACACTGGTATACTAACACAGTGGAATACCATACAGCA
          GTAA
310  150  AAAACTTTTCCTATTTGAGCCATAAACCCTTCCTTTGTTCTTCGGTGCACACACTTTCATTTGTGCTTCCCCAGT
          CTGCGGGAGTTTTTTTGTCTTTTTTTTTTTTTTTATGTTTTATATATAAAAAAACTCTTAAATTTTGATATTA
          TTACCGGCGTGAGCCACTGCACCCGGCCTGAGTCTTTTTGTTTTGTTTTTTTTTTTTTTTGAGTCAGAGTCTC
311  150  TCTCAGTGGCCCGAGCGGGAGTGGGGTCGGGGCAATCTCACGGCAAAATAAGCCACCGGGGGGTCAAGA
          AA
          GAATAATCAATCTTTTGTTTACGAAGCACTTCCACAATTATTACTTTTTTTTTTTTTTAACAAATCCTCCTGTC
312  150  CATTTCACTATGTAGGCAAGGCAAGTATTGAAATTATAATTATAATTTCATTGATGAGGAAATATCGTAAGTG
          CA
          CACATTTTCTCTAAACTCATATAATATTTTACTTCTTTTTCTCTGTTCAAGCCAATTATTAAATACATAGCATAG
313  150  GCGATACCTGGATGCCCAAAGATGGATGCCACGCTCCCTGCCGTGGAGGGGGTCTCAGTCTATGCTATGTATT
          TA
          CCAGGATGGTCTCGATATCCTGACCTCGTGATCCGCCTGCCTCGGCCTCCCAAAGTGCTGGGATTACAGGCGT
314  149  GAGCCACCGCGCCCTGCCCCTCATCCCGTTTCTTTTCCTCCCGCCCGCCCCCACAACCCCGGTCGCTCGCTGA
          GC
          GGCTGGGCATGATGGCTCATGCCTGTAATACCAACACTTTAGAAGGCCAAGGCAAGCGGATCACTTGAGGTCT
315  147  GGAGTTTGAGACCAGCCTCATCAACATGGTGAAACCCCGTCTCTACTAAAAATACAAAAAAGCATTAGCTGG
          GC
316  89   CGGTCCCTGTCCTCTGTTGCGATTTGGCAGGCAGCTCCAAGGATGTACGGAGGCACCCCTCCCTCCCCCCCAAA
          CCCCACCCCCGGAGG
317  94   AGCAAGTTTATAACAATAAACTTGACAACTTTGGTGAAATACATCAATTCCTTCAAAAACAAATTACCAAATC
          CAACTAGTAGTTGGATTTGGT
318  150  AAACTCCGTCTCAGAAAAAAAAAAAAAAAAAAAAAAGAATGTATTTTGCGGGGAAGGGGTAATTTGAAC
          ATAGAGTAACAAGGCAAAAATACAAGCAAATGCAAGGAAGCTGGAAAGGATACACACCGGAGTTGCGGGAA
          ATACATGA
319  150  ACTGTATTTTCCAGGGAAAAGAATGTCTAGCATCTCCAAAATCATCTCTCTGAGCTAAAATGGCATCATGCAA
          AAGGAAACTTACTCATTCCCTGGGAAATCCAAGACAGAGACCTTTCTGAGAGCGACCTGAAGTGAGGTTTGAC
          AAGA
320  150  GGATGTGGGCATTTCAGTTTCCCAGGCCCCTCTATTTAGGAAGAAAACTCCAGGCTCCACCCAGATAAATACA
          GACTCCTCCTTCCTACCTAATCCAGGCTCAGGATGGGAGGCCCAGCCTCCACATCCAGAACACCACACTACGT
          TCAC
321  112  TTTGGCAGTTTCTAATTTTCGAAGAATTGGTCCATTTCATCCAAGTTGTTGAATTTACCATATGTATATACAAA
          GTATATCTTTAATATCCTTTCACTGGGAAATTTGAAAA
```

```
322   62    AAATGAAGAATGTTAGATGTCTCCACTGACCTACTGAGCCTACATCTCAACATTCTTCATTT
                  CTGGGGATATTTGTTTTTCACTATAGGCCTCAGTGAGCTCAAAAATGTCCATTTGCAGATTCTTCAAAAATAGA
323   150   GTATCCACAAAGCGGAAACAGACGAAAGCTATAACTATGACAGATGAAACCAAACATAAAAAAGAAGTTTCA
                  CAGA
                  GAGACTCCATCTTAAAAAAAAAAAAAAAAAAAAAAAAGGATACAAAGAGAAAAAAGGACCAGCAAAAAAG
324   149   AAAAAAAACCAATGATCTAAAAGATAACACAGGAATATTCAACAAACAACTTCAGTACAAAACAAAAAAGC
                  AGGTAAAA
                  AATGTTTCTTTTTATTTATTTCTTTTTTTTTTTTTTTTTTAGACAAGGTATCACAAATTCCCCCAGGCAGTAGTGAA
325   141   GTTGCACAATAAAGGCTCAAAACAACCATTACCAACTGCCAAAGCCACCCAACTAGCTGTGACTAC
                  TAGGTTAAATTGTGTCACCACCACCCCCACCCCACACAAGCACAAAAAAAATTTAAATACAAAGCCCAAGTA
326   150   CAAGTGAAATTAAAATAAATTTTAAAATAGGCTTTTCAGATGTAATCAAAATTATCAGAGATAAGAATGGATA
                  AGGGT
                  TAATGCATCACTGAATCAGGAAATGTTAGGAGCTTTGCTCACACATAGGTAAACCGTGTAACTTCTGGAGTCA
327   150   TTGAAAAATAATATTTCTTAATATCTGACAAGTCGTCAGCTATGCACTTAGGACAAGCCTCAAAAACACTTGT
                  TTGA
                  GGAAAATGCAGTCTCTGGGGTAATTTTATGAGAAATACAGCTTCCTTGAGTTAACAATGAAGAGCTTTAAAAT
328   150   CAAATCTCATGTTGATTTCATCTTTTTTTTTTTTTTTTTTTTTTTTTTTTTTGATATTGAGTATTGATATGGTTCC
                  CTTTGATCTTCTTCCCCTTCTTTTCTTCCCCTTCCCCTGCACCCAGAATAGGGCCTGGCAGAGGCTGGCAGGAG
329   150   CCAGCGCTCTGTATCCCCTCTTACCTTGGTAATGATGTCCTTGAACTGACCCTCCTTCCAGGCCTCAGTGGGAG
                  AT
                  ACAGGAAGGGGAACATCACACACTGGGGACTGTTGTGGGGTGGGGAGAGGGGGGAGGGGATAGCATTAGAAG
330   150   ATATACCTAATGCTAAAGGAGAGTTAATTGGTTCAGCAAAACAAAATTGCATAAGTAAAAATAAGAAACGAA
                  TAGCAAA
                  GATTCAGAGGAGATGCTTCCCAGAGGGCTCAGTCCTGGAGTCAGAGATGCTTCCTGGAGGGCTCAGTCCTGGA
331   150   GTCAGAGATGCTTCCTGGAGGGCTCAGTCCTGGAGTCAGAGAAGATGCTTGGCGGGCGGCGAGGTCGTGTAG
                  GCGGA
                  ACACTATGGGGACTCCAACAGAGCCATACCTTCCTGTCTACGGCGGTTGGACCTCCTGGCTCCGGGCACCAGC
332   79    TTCTCC
                  GTGTCTAGGGCTTCACCAATTGTATGTTCATTCATTCAATCTTGATGGCTTGCTTTCCTCCCTTCCTTCTTTCTTT
333   145   CGTTCCTTCTAGCCTGCCTTCCTGCTGTCCTTCTGTCCTGCCTACGGACCCCGCGTCGCCCGACCCCCGC
                  ACTTAACTATGGAAAAGATATTGTTGGTGGCATTAACGAAGACACTGAAGAATATCACCTAAGTGATGATGT
334   148   TGTACAGTATGGAAAAAATGAAGTGATTGAAATCGTGACTGACCAACAAATATCTTGTCGATAGTTAAGTAGA
                  GC
                  GAATGGAATGGAATGGAATCAAACCGAGTGGAATGGAATGGAATGGAATGGAATGGATTGGAATGGAATGG
335   149   AATCAAACCGAGTGGAATGGAATGGAATGGAATGGAATGGAATGGAATGGAATGGCATCGGATGGACACAA
                  CACGGCA
                  GTTACGTGTAAGGGTGATCAAAGATTTAAACTGTCATGTCGACAGTAAGCCGAGGAACAGGCTGTGCAGGGG
336   149   CCTTAATAAATCTGCCCTTGACTTATGGCCGAACCTCATTCAAGATCGGAAGGAGCACACGTCTGAACTCCAG
                  TCAC
                  AGGGAGTAATTAAACCTTTTTTCCCACATGATAACAAGTATGATCGTTCACACTTGATATCATGTGGGAAAAA
337   150   AGGTTTAATTACTCCCTAATTGTCTGGGAAACAAAAGATTTCAAAATACGGAGCTCCTGGGGGAAGGAGAGTT
                  AATA
                  TAACAGATTCAGGATTGAATAAAAAAGAAGTTATGGGCATAAGTAGTGCAACATGAAAAGGTACATTACAAA
338   150   CAGTGTCACGGGTGGAAAAGGGGCAATGCATGGTAGCTCCTAGTTTCCAAACGGGTTTCAGAAAGTATGTCAC
                  ACTTT
                  CAAAGCATAAAAAAAATAAAAAAGGTAAAGAGTACTAAGATCCAGAAAATTAACAAGTAGTAAAATTTTTGT
339   150   TTCAGTCTGTCAAGATTATCATTCCATTAAGCATCCAAAAATTTTACTACTTGTTAATTTTCTGGAGCTTAGTA
                  ATCT
                  TAGTCTTGATAGTTTGTTTTTTTATGAAAATTTGTGCATTTCATTTAGGCTAATTTATTGGCATGTAATTGTTTA
340   96    TATAATTCCCTTATGTATACC
                  TTATTCAATTCAAAATTGATGACCTTCTACTCTGTAGTCCCTCCTTTGAGTCTTCTCAAGCAGATACCCTCCTGC
341   125   TCCTTCAACATTTATACACTGGAGAATAAAAGTTGAAGGAGCAGGAGGT
                  AACTTCAAAGCAAAATGAAGCTCTTTTGGTTGCTTTTCACCATTGGGTTCTGCTGGGCTCTAAGATCTCAGGCC
342   84    AGATTCTCCT
                  ATAACTGTTTTGTCTTTCCCTCCCCCGCTTCGTTCCCAGGTTTCGGTCTAAGTACCACCCAGATGAGGTGGGGA
343   143   AGCGTCGGCAGGAGGCCCGGGGGGCCCTCGAGGCCCAGGGGGACCAGGAGGTCCTGGAGGGCCTGGGAT
                  TATCAAGGAGTGACTGATAGTAATATTCTTTTTTTTTTTTTTTTTTTTTGGAGATAGCATCGCCACTATGTTGCTC
344   150   TGGCTGGAGTGCAGTGGCTTTATCTCGGGTCACAGCAAGAAAGGAAGAGCACACCTAGGAAAAACAGTAAAG
                  ACA
                  CTTGCTTGGTTTCTTTCAGAAATTGTTGCCAAGGTCAGAGAAGTCTCCCAGCCCGACTGGACGCCCCACCAGA
345   150   AGTCACGCTTGTGTTGACCAAAGAGAACTTTGACATCATTATGAATCTGTTTTACTCTTCCCAGTGCCTTGAAA
                  AAA
                  ATACATGGGGCCTTCAGAGTTCATGGAAAAAAGATTATTATGAAAAAAGTATGCATGGATTTCAATTTTTTTG
346   105   CACCAAAACAAATATATCTCATTTCCGTTTTC
                  TGCATCTTTTATGGCAGGGAGTCGTCGCATGGTGGATGTCATGGACGTGAACACACAGAAAGGCATTGAAATG
347   148   ACCATGGCTCAGTGGACACGCTACTATGAGAGGGGTGGCCTGGGCGGCTCCTGCAGCCTGGAACATGGCTC
                  GTT
                  GACAGAGCGAGACTCCATCTCAAAAAAAAAAAAAAAAAAAAAAAAAAAAGATATAAAAGCAAAATTAAGGGGGC
348   150   AGAGGGGTATTAAAAGGTGAAATATGGGGAAGAAGATGGACCAAAATGCGTTGGTTTGATATACAATAACCA
                  CAAAACA
```

349 118 AAGTGTCTGAAATGCCGTGGCAGTGGAGACACTTGATTTGAGATCTCCAGTAGCCTCCAAATATTTGGTGGAT
GAGGGTGGTAGAAGCAGCAGAAGGGGATGGCCTTGGCAAGGTGGT

350 150 AGAAAGCACACTGTACAACTTCAACAAGGACAGGTCAGTTTTCTTGTTTAAAACCACAGTTGGAGCCCGGCAT
GGTGGCAAAACCCTCTAAACCCACCACATTGGGAGGCCCAGGAGCGAAGAACACCACGTCTGAAGTAAGTGA
CCAAA

351 150 CCCCATTCTTCCCTCTCTCTGTCCTCAGAACACTGCCTCATATCCTTCCCTGGTCCCTGGCTCTCTGAGTCCCGC
GGTTTTTTTTTTTTTTTTTTTTTGTTTAGAAGAGATAGGAAGAGCACACCGATGAACACCACTCACGACCAGAG
A

352 150 CTCACACCTATAATCCCAGCACTTTGGGGGGATCGCAAGCACTTGCTTCCAAACTTGGATCAATGAATGAGCA
CTGCTTTTTGCAGACATAGAGCAGGGCTCTAGATAGAGTCTACTCAGTCCTACTCTGGAGTAGCCCATAGTCCT
ATA

353 108 TTTTTAGTGACAAAACATAAAGGTACTGAGAAAAGAGAATCCCCTTCACCAGCACCGAAGCCTAGAAAGTAG
AGACTAATGTTGGAATCTTCAAATTCTAATTCCTTT

354 149 TCTTAAAAACATTTTTTTTTTTTTTTTTTTTTGGCCAGGCATGGTGGCAAACCCCAGTAAAACCAGCAATTTGG
GAGGCCGAGGCAGGCAGAAAAGCTAAGGTAAGGAGAATAAAGGAAGAGCACACGGATGAACACCAGGCAC
GACAA

355 148 TCCAATCTGCTATCTGCTTGTGCCTGAGTTTGTGCAATATCCTCCTCATTGAGCTTCCTGCCTCTTCTCTTGCCT
ACCCCCACGACCCGCCCCCATGTTTTTTATTCCGCACGGGGGGGGTCGTGTGTTTTATGGGGGGGAAGGTGTA
AGGAAGTGCTGGTGGATTATTACATCGACCCGGCCGATGCAAGCCCTGACCAAGAGATCAGCAAGATTAGCC
TGCAAGCCAAGAGCCAGCTC

356 92 CTTAGTTTTCTATATACTCATTGGCCAATTAAGCAACCACACAAATCAATGCTTCACTACAAATATCTTAGGCCT
AGACTTGAAGTTAGGATTTCAATTTT

357 100 CAATCCTTCACTCGGCTTTCTTTTTCCTTCATGGTACTTAAAGCTGATATTAGACGGTCATTACTTGTCTGTAAG
TTCCCGCAGTGTCTTCACCTACGAGATCGGAAAGAGCACACGTCTGAACTCCAGTCACGACAAGAGATCTCGT
AT

358 150 GAATCTCTTCATATAAAATCAAGACAGAAGCATTCTCGGAAACATCTCTGTGATGTTTGCATTCAAGTCACAG
TATTGAACAAAAAAATTCATAGGGCAGGTTTGGAAAAAACGTTCGTTCGATCGGAGGAGGACGCGAGTTCAC
TCCA

359 149 TCTCATAAACTGATGAGGCTGTGATATCAGGCTCACTTGACAGTCGTCCTCAGCTGAGAAGGAAGCCAGCACC
ACTTCTGTGCGTCCGTGAATCTCACATTGTGGGTGCTGGCT

360 114 AGAGTGAAGAGGCTTGGAGGATCTGATCATCACATCACAAGGGAATTCATTTTCTGCCAGTTACGTGGTCCCC
ACAGGATGCCGAGTTGCAAAGCCCACAGATTAATCCGGAACAAATAGCGCCACAAAAAATAATTTCACCAGG
TCAT

361 150 TTCCTTCTAACAGACAGGACTCTCAGATGCAGGTATGTAGGAGTTTTCTAGAGGTCAAATCAAGAACCTGTTT
GCCTGGGTATCAGCAGCGGTGTCTGAAGAAAATAGGAGATCGGAAGATCACACGAATTAACTCAATTCACGA
CACGA

362 150 AGGCCTTTTAAATATACCTATAGTTTGGATATCCTCCATCTTTAATTAAGCTGACTTTTAACCGTCACATTTCCT
TTTTTTTTTTTTTTTTTTTTTTTTTAGAAAGATTATTTCTTTGTTGCCAAGGCTGGATTGCAGTGGCTTGA

363 150 AGGCTGGAGTGTAGTGGTACAATCTCGGCTCACTGAAATCTCTGCCACCCGGGTTCAAGGGATTATCCTGCCG
CAGGAGGCTGAGGCAGGAGAATACCGTGAACCCGGGGGGCAGAGA

364 118 AGACCAGCCTGGGCAACATGGCAAAACCCTACCTCTACAAAAAATACAAAAATTAGCTGAGGGTGATGGCAC
ATGCCATTCCCCAGCTGGGTGACCTTGGGTTAGGGCCCTCTCCTGTCTCAAAGAGCAGACACGCCACATGCTG
GGTCT

365 150 CGGCGGGGCCAAGGGGAGCATGAACCGGCACGTGGCGGCCATCGGGCCCCGCTTCAAGTGAGGGCCCTCTTC
CTGGGGAGCACAGGGCCCCGTTCTTGCCGATGCCCATGTTCTGGGACACAGCGACGATGCAGTTTAGCGAACC
AACCA

366 150 AGAATGTAGCGCCTGCTTCCCCCAACCCCCAACAGATTTCCTGTTTTTTTTTTTGTTTCTTTCTTTATTTATTTTT
TTTATAAAAGCATACAAAATAGAATATAGGAAGAGCACACGTCTGAACAACACTCACGACAAGAGATAGCGT
AA

367 150 TGTCCCTTATTTGCCCTGGCCCTTCCCTGGCTCTGCCATACCCCTTCTCTGGATTTGGATGTGTCCTGTCCCTTA
TTTGCCCTGGC

368 86 TCCACTTGAGGTGATTCCATTCCATTCCATTCCATTCCATTCCATTCCATTCCATTCCATTCCACTCTGTTTGATT
CAGTTCCATTCCATTCCATTCCACACGATTTGATTCCATTCCATTCTATTCCGTTTCATTCCATTCCACGCCAT

369 150 GGGATTACGGGCGCCCCCCACCTCGCCTGGCTAATTTTTTGTGTGTATTTTTACTAGAGACAGAGTTTCACCAT
CTTGGCCAGGCTGGGGTGGGGGTGGGGGGTGGAGCGCGCCCAGGGGGGGCCACACCACCACCCACCCACAACGGAGGGGG
GACCA

370 150 TGACGTTTCTGTGTCCAAGGCCTGAACTGAGGTGTCCGCGCCTTGTAAGCGTGCTGGCTCTGTTTCCAGTTGGA
AGATATAGCAGACAAGATCGGGAAGAGCACACGTCTGAACTCCAGTCACGACAAGAGATCTCGTATGCCGTC
TTCT

371 150 CCTCATTCTTTTCTAAGACATTACCCTGCCACAGGGAAAATTAAGAAAGAAATCATCCTAACTCAGTGAGGAT
GATTTCTTTCTTAATTTTCCCTGTGGCAGGGTAATGTCTTAGAAAAGAATGA

372 125 GGAGAGATGGACCTGGGTTGGATTTTATCCATTATCAGTCTTTCGATAGTTTGACTTTTTTTAACCATGTGCAC
ATTTTGATTTGAGGTTTTTATATGTTTATCAAACCAATATGTGCACATGGTTAAAAAAAAGTCAAACTATCGAAA
GA

373 150 GGATCTCATTATATTGCTCAGGCTGGTCAAGAACTCCTGGCCTCAAATGATCCTCCCACCTTGGCTTCCCAGAG
TGCTAGGAGTTCTTGACCAGCCTGAGCAATATAATGAGATCCCATCTCTACAAAAAATTTTTTTTAAAGTAGCT
GA

374 150 ACAACCTCATCTCAAGACCCTTAACTAATTACATCTGCATTCACCCTATTTCTTTTTCTTTTTTTTTTTTTTTTTTT
TTTTGTTTTGGTTTTTGTTAGGTAGAGGACAAATGTGAACATACAGAACGCAAAGAGATCGGGGTGTGCGGAG
A

375 150

| | | |
|---|---|---|
| 376 | 150 | TTGTGCCAGCCACAGCGCTAGACATGGGCATACGGCAGCTTTACTGCTGCTCAGCTCCTGTGTGCCCATCCTG<br>GAGCAATTGTTCTTTTTTTTTTTCTTATTATAGATATGTTTTTAAAAATTTCAATAGCTTTTGGAGTACAAGTGGT<br>TT |
| 377 | 150 | AAAATCTATAGAGACAAAAAAATTAGTGATTGCCTAGGGCTGGGAATTTTTTTTTTTTTTTTTTTTTTTTTTTT<br>TTTTTAAGGAGTAAAGGTGTGTAGCCCAAGGATGAAAGAAAATGTGAAATAAACGCACAAGACAACAAACGC<br>CCC |
| 378 | 148 | GATCGGAATAGCACACGACTGAACACCATACAAGACAGTACATCTCCAATGACAGCGTCTGGTGTGAAGGGT<br>GCGGCGGCAGAGGGCGGCGGGCTGGCGAGTGGAGTCATGGAGAGAACGCGGGACTCGTCGCCTGGGGCCGG<br>GCGGC |
| 379 | 78 | AGGAGCTGCTAAGAGAAAAACAGAAGGAGCAGCAGCAAATGATGGAGGCTCAAGAGAGAAGCTTCCAGGAC<br>CTGAAAC |
| 380 | 116 | TGGAGCTCAGCGAGGGGAAGGTGCTTTCCTTACCTCTAAACAGCAGTGCAGTCGTGAACTGCTCCGTGCACGG<br>CCTGCCCACCCCTGCCCTACGCTGGCCCAAGGGGTCCAGATCT |
| 381 | 132 | CAGGACTTTATATTGCCTAAGGTTAGGAGAACGGAAGCTCTCTGGTTCCAAGCTGGGCTCTGAGCGGTAGCTG<br>GGGTGACGGCTGCTCTCTGCAATTGAGGTTGGCTCCTTCAAGCTGTCCCCACGACTCAA |
| 382 | 150 | CTTTCTTTCTTTTTTTTTTTTTTTGTCTTTTTATTATATATTGGGTTTAACAGTGTTTGCCAGGCTGGTCTCGAA<br>CTCATGACCTTATTTGATCCACCCACCTTGGCCACACCGAGAGCTGGGATTAATCCTTGAGCCACCGCACTCA<br>A |
| 383 | 150 | TGAAAAACTGGTCCAGCGGGCGGCGGACAAAAGTTTCCGAGCGCGCTCCACAGCTGGGCGGCGAGGACCCC<br>CCGGCGGCCGGCCCCGCCTCGGATGCTCCCGGCCTCCTTCCTGTGCTCGTGACGGCGCCGAGGGCGGCCCCCC<br>CCCCC |
| 384 | 131 | AACTTAATATAACAAACAAATTGAAATAGCTTATTAAAAGTGTCCGTATATAAAAATGGCCTTGTGATGTACA<br>GTAAAATGCAATAAAAATTATCCTCCTTTGTTCCGGCAGCCCACCCCCCGGACCGACA<br>ATTACAGGCGTGTGCCACCACGCCCCGCCAAATCAGCAGTTTATGTAAAAATTCTTTAACCAGGGAGGTGGAT |
| 385 | 150 | GTTGCAGTGATCTGGGATCGCACCACTGCACTCCAGCCAGGGTGAGTAGGGCAAAACTCATGAGAAACTGCT<br>CCAGC<br>GACCGCCAGCAGCCCTGTACTCTCAGCCAGACCCCCACGTACCAGGCGGGAGATGAAGATTCCTAGAACCCG |
| 386 | 150 | CTCCAGGCCCGGCTCAGCCACCCGCACGCACCTGCAATCACGCATGAACAAGCCGAGGTGGCGGTCTCAAGC<br>ATGCTT |
| 387 | 79 | GCACCAACGTCCGGAGAAACCTGGCCTTCCACACACTCAGCCAAGAAGTCCTGCTCAAGGAGTTCTCCACCAC<br>CTTCTC |
| 388 | 150 | AGGAGCTGGACAGCCTTCTCCACCTCCTCATCTTCTCCTGGAACTCCTTTGTCTCCTGGAAACCCATTTTCCCCT<br>GGAACCCCCATTTCCCCAGGGCCCATAGTCCCCAACACATCACCCCCAAGGCAACCCCACCCCCCCCCCGACA<br>CC |
| 389 | 74 | GTCTCACCTGTCACCCAGGCTGGAGGGCAGTGGTGTGATCTCGGCTCACACCAACTTCTGCCTCCCAGGCTCA<br>A |
| 390 | 95 | TCTTCAACTGATGCAGCTACATCTGAGTCCAAGGAGACCCTTGGCACTCTGCAATCCTCACAACAGCAACCAA<br>CACTCCCAACACGTCTAGACCT |
| 391 | 120 | GGCTGCGGCCCGGGCGGCGGCGATGCTGTGGCCGCGGCTGGCGGCGGCCGAGTGGGCGGCGCTGGCCTGGGA<br>GCTGCTGGGCGCCTCGGTGCTGCTGATCTGCTGGTACTCGCAGTCCTT |
| 392 | 150 | CTGAGACCAGGTCACACAGTCCTCCCATCTTCACTTCTTTGATCACCACCACAGAGACCACCTCACACAGTACT<br>CCCATATTCACTTCTTCGATCACCACCACCGAGACCACCTCACACAGTACTCCAAGCTTCACTTAATCGATCAC<br>CA |
| 393 | 89 | GTGTACGGCACCCTTGAAAACAAGCAGGCGCTAAGTTTGAATCTTGAAACTTCCTAGCTTTCAGCCACTTCCTC<br>AGCTCTATAAATTCA |
| 394 | 44 | AAGTTGCCGAGTGATGGGGTGGGGGGACATTCAGACGGCAACTT |
| 395 | 97 | ACCTTTAAAGGTTTGGCAGTTGCAGAGGAAGGAGCCGAACACGTTCACGCAGGAGCCTCCGTTCTCACACTCC<br>TCTCGTTCGCACTCATTGATGTCC |
| 396 | 80 | GCGGTGTATGCCTACCGCCACCAGATTCATCGCCGGAGCCATCAGCATATGTCTCCTCTTGCTGCCTCATAATC<br>CTTTAT |
| 397 | 89 | CCTCAAGTTCCGGAATACTCACCTAGGCAAGAAAGGATCCGAGATCTAAGTGGCAATCTTTGGGAGCGTTCCA<br>GGAGGCCCAAGGAGGT |
| 398 | 150 | ATGCGGATGGCCATTGAGTCGGCAGGTGAGACTCCAGTTCCTCCAGTTAAATGCTGCGTTCGCCCCGTGGGTC<br>AGGGTCTTCACCGGTGGGTTCACCAACTCCCATTTTCCTTCCTCTTTATTTTCTTCACAGAAGCACATGCCCAG<br>AGA |
| 399 | 149 | GTGGGCGCCTGTAGTCCCAGCTACTCAGGAGGCTGAGGCAGGATAATGGCGTGAACCCGGGAGGCGGATCTT<br>ACGGTCACACATCTCGCCACACTTCACTACAGCAGAGGCGACAGAGGGGAGACTCGTCGCAAAAAAGAGGGG<br>GGTTA |
| 400 | 90 | GTCAGACCTCCAGGACTAAGATGAAGTCAAAGAGCGACCACAGCTCTCGGAGGTTATTTTGCATCGGTGAGCC<br>AGACAGAATGATCCGAT |
| 401 | 150 | ATGCGGATGGCCATTGAGTCGGCAGGTGAGACTCCAGTTCCTCCAGTTAAATGCTGCGTTCGCCCCGTGGGTC<br>AGGGTCTTCACCGGTGGGTTCACCAACTCCCATGTTCCTTCCTCTTTATTTTCTTCACAGAAGCACATGCCCAG<br>AGA |
| 402 | 86 | ATTAAAGGATGTTCTTGAAGAATTCCATGGGGCTATTTTTTGATCTACTGAGGGAATCTGATGTTCTTGGAGAA<br>CTAACTGCTTCT |
| 403 | 126 | CCTACTGATGAGTATTCTTTCTTCCTTCCCAGGATATTCTCCAGGGAAGATGCCATCCAAATGCAGGACTCCAA<br>GATTGTATGACGCATTGGGTTCCCCATTTCTTCTGCTTTTAACCAGTACTTT |
| 404 | 81 | TGGACGTGGATAAATCACTTTGTGATTTGGGGTTCTTTAGCCTTCTATGTATTTTTCTCATTCCTCCCCCAGAAGA<br>ATGAGA |

405 103 CCATTCCTAATTCTCCAGTCCACGATATTGAGTTCAACAGCAGCAAACCACTTCCACAGCCAGTGCCACCTAA
AGGGCCCGTCTGGGCTGGGTCTCCAGATGT

406 116 CGGAATCAAACACATTATGCTTTGCAGGCATCACTGAAGCTACTTGATTTTTATGAAAAGTACTTTGATATCTA
CTATCCACTCTCCAAAACTTTTCCTCCCTTAGAAAGTGGAAA

407 149 GATCTTTTATTAAAGTTATGGAAACTGAACTAAAAAGGATTTATAAATTTTACAAGATGGATTTGTTTTCTCAT
GGTGCCCCTATCTTTTCAGAGCAGATATCTTCCCTCCAAAAGGCAAAGAAACCAACTTGCAAATCTAGTCCTC
AA

408 102 AGATGGCTCCTGTGGGCTTGTGTACGAGCATGCTGCAGCGGAGGGGCCCCCTATTGTCACCCTTCTGGACTAT
GTCATTAGATTATGCTTGGCTTTCACTGA

409 76 TCCTGCAGGGTGATTCTGGGGGTCCGCTGGTATGTCAGTTCAAGAAGAATGGTCAGCAAAAAGCTGACACAG
ATGA

410 150 GAATGCACTTTTGGTTTTTGGTCATGTTCGGTTGGTCAAAGATAAAAACTAAGTTTGAGAGATGAATGCAAAG
GAAAAAAATATTTTCCAAAGTCCATGTGAAATTGTCTCCCAGATCGGAAGACACACGACTGAACTCCAGTCAC
GACA

411 83 ATTCCCCAACTGGCATTGACTTTTCTGTATTAAGCTTGAACTTATTCATGCAATCTTCTGCTAAGTTAAGATGG
ACAACTTGC

412 132 AATGCTGATCCAGCCCCTATTTATGTTTGGAAGTCCTACTGGGATCGGGGTACTTAGGAGTTTGTGAGGAAGGA
GGAAGGCTCTCACTTTCATCTGAATCTGAGGATGAGGTATCTGTTTCTATGATTTCCCT

413 118 GGAATGGACTTTGTTGCCCAACAGAAAATGAGAACTCAAACAGAGGAGCTACACTATAAATACACTGTATGT
GATAAAAGCTTCCACCCTCTCCTCTTCCACCCTCAGTGGATGATAA

414 149 CTCTGCCCAGGCCCGGGCACATGCCCCCGCCCAGCCGGCGGCGACAGCAGCAGCAGCCAGACCAGTCCCTGCGAC
CCCACCTCGGAGGCAACTCCATTTCGGACCTGGGACTTCTGACGCGGATCCTCAGCTACTTCTCACCTCGGGG
CTGT

415 78 GAAGTTTTGCTATTACACAACAATGACCAGAACGCAGAGGAAGATCAAGATCATTTCCACGTGCATGGAATC
GTGGCC

416 109 GGAGCTGAAAAGATTGCAGGATTAAGCCAGATTTACAAAATGGGAAGCTTGCCTGAAGCTGTTGATGCTGCC
AGGCCGAAGTCAAAGCGGCATTGCTCGTTCATGGAGT

417 137 TACGAGGGCCTTTGATGCTTAGTGGAATGTGTGTCTAACTTGACTCTCTTCCTCTTTGGGCCCGAAGAATTCAT
AGGTTCATTTTTAATTTCTTTCTCATAGACATAGTCAGAGGGTCTGAGTAGTTGACTATGAAA

418 148 CATTCCATAGTATTGATGGTAATTACCACTGGTTCTGAACCCAGTTTTGTCCATGGTACATGAATCCTCAATTC
ATGAATATGTCCACTTAAAAAAGTGAATGGTAATTACCATCAATACTATGGAATGCATTTTGAAACTTAAGGA
T

419 150 GCATAGAAACATGATCGAAGTGATCATTTACTTTATCATGTCATAAAACACAAGTGCTTATTGTAGAACAGTG
ATATAATTTAAATGGCAAGGAAATTTGTTATGAATCACAAATTTCCTTGCCATTTAAATTATATCACTGTTCTA
CAA

420 111 ACCAAGCTCAAACCTGTGGGTGACAAGTCTGGTTTCTCTGCAACACAACTCCGGTGACCACGTCTCCTGT
ATGGACAGGGACTGGGTCGTCCATCATGAACAGCGTCT

421 85 CCATCTGCAGGTACCTGGTGGACGTGCGGCAGGAGCAGTGGCTGCTGCCCGGAGCCGTACCGAGGCAGTGGC
ATGGTCAGCTGCT

422 140 CACCACGCAAAGTGATGTGTAAGTGTGGGTGTTGCTCTCTTGGGGTGAGTGTGGGGGAGAGAGGGTCATCCTC
ACGAGCACAGGCATCTCGTCGTCCGACATCGAGCTGGCTGGCTTGTCTCTGGCGGAGGGGGCGGCAA
ATATGATGGAATTGACAGCAACTTTGAACCTGAGGTTTGTAAAGAGCCATCACTGACCATGTGACAAATTTTG

423 150 CCAAACGCATCATCAACTAAGCGTATTTCTTCATTAGAAGAAGGAATTGGGACAATGCTAAATTAAAAGAAA
AGAAA

424 150 GAAATAATCTGGTTACCATCTACTCCTTCATGAGTCTGTTCTAAAGTCACAGGGATTCTCTGGCAGGGGCGTGT
TGGGAATCTGACCGCCCATGTGCATCCGAATATGTTGCTGCAGCATCACGGCATTAGTGAACTTCTTCTGGCA
GAT

425 134 CAGGAACAGGACTCTGCTGCCCAACACAACCTTGACCTATGACACACTTTCCATAGGCCTTCCTCACCACATC
ACAGAACTCAGAGAAGAGGCTGTCTTCTTGTTTGATACACAGCTCGTCTCTTAGGAACACC

426 150 AATTTCTTCTAGGTAATTGATCTGGGTGGTGAGCCAATTAAAAGCAGTGACTACTTTGGTAGGAGGGATGTCT
GTAGAGAAAAGGTCTATTTCAGTATCAGCCAGCAAAACAGCCTTCATTCCCCTAGAGCTGAGCACTTGTTTAC
AGAA

427 150 GCTGCCAATTCTGCAACAACTCAGGTTCTGATTGGGAACAACATTCGATTAACTGTACCTTCTGACTCATACAT
CCTCTTCTGCACATCTTTTAATGCATCTTCTAATTGCTTTATCTGTTCATCTGAATCTCCTTGACCCTTTCACAT
CCCAGGAGCGTCTGGCAACAGCTTTGCAGAAGCTGGAGGAAGCTGAGAAGGCAGATGTAAAAGGAAGCAGG

428 150 TTATGACAGGCCATCCTATATTCTTCATATTGGGAACTACAGTTCCCAAAGCTGCCATCAAGCAGTGCTTCAGA
AACCT

429 93 TGGTGGCCCATCCTGTTTTTTAAAGTTTCAGCTGTAAAACATATCCATGGAAAATACTTCACTCCAAAGATATC
CATAATATTGGCCATCGTA

430 76 AGAAGTCAGCTTCTAAGTTGAGTATGTTGGTGAGACCAATTTGGGGAAGTATGAAGGTTTAAAAAAGGTATGC
AAT

431 85 GCTCCCAGCATGTCCTCGCTCAACTCCTTGGCCTCTTCTTGCCTTTCAGTCTCTGGCGAGCAAAGTCCTCAAAT
ACAATGTCGTC

432 111 CCTGCCCAACTGGTCTCCACCTACCACCACCTGGAGTCTGTCATCAACACAGCCTGTTTCACCCTCTGGACCCA
GCCTGACCCGAGGCTCCTCCACCACCTCCCACAGCTC

433 150 GGACATATTTGACAGTGATTGGTACACTTCTCGAAATCTAATTGGGGGCGCTGAATCATTGTGATCAAATACA
ACGTTAATGACAAGTTTTCATTCCATGAAGTCATCATAGAGACAGCGTTGACGCTTGGAGAGCCTGCAGCGGA
TAAC

434 104 ATGTGCAGGCCATGACTGCTCCCTGCGTCTCTGGAGCCTGGACATCCACAAACTCCTCAAAGATCCTATGGGC
CTTAGAGACCAGTTTTGCAGTTGACCTGGTC

| 435 | 68 | AGTCACCCCCAGCCTGCAGGAACCAGGTGGAGGCTGATCAGTGGCTTGTCCATCTCTTCTGGCATTTT |
| 436 | 137 | CTTGGTCCTGAGCAGCCAACACACCAGCCCAGACAGCTGCAAGTCACCATGGATCTAGCTCAGGTCATTCCTC |
| | | TACCCATGGGCAACATGGTTCTACATCAGGACAGTCATCGAGCTGTGGCCAACATGGAGCTAGT |
| 437 | 70 | GAGGCTCAAGAAGCAGAATGTCAGGCGTTGAGCAGCTCCATTTCCTGCTGGAGCTGTAGCCTCAAGGCCT |
| 438 | 88 | ATCAAAGTATGAGGGGAAATGGCAGAAAACAGTATCAAGATTCACCTAATCAAAAGAAAAATATACAACAA |
| | | AGGGAACAAGAAAGGTA |
| 439 | 113 | ATCTCCTCGGCATAAAAAAAGTGATTCTTCAGGTCAGGAGATAGAGTAGTCTTTGTAGGAGATGGGGACGCA |
| | | GTACCAGGTGAGTCAGTTCCTCCATCGGAAGTACTTGGTGA |
| 440 | 116 | ATTTTGTTTTAGCTTCCTTCCCTCCCAAGGCTGACTCCCTGGATAGTGGAGTGGTGAGCACCATTGATTTTAGTG |
| | | TTACTGTGCAAAGGTGATAATGAGAAAGAATTAGTTCTGTGT |
| 441 | 94 | AAAAGTTTATCCAGGAAAACATGTGAGTACTTCTTTGTATCTTGTCTGTCAAGAGCAGCTTCAACAGCATCAA |
| | | AGAACTCTTCTTCATTAATCA |
| 442 | 141 | TGGGCCTGGCATGTGTCTTGTCCCTTTGGAACACAGGTGGGTCTAGCTGGAGTTCTCCTTGACCTTGTGTGACA |
| | | TCCCTACAGAACCGGGATCCATTTCTTGTCCTCGTTGTAGAGGAAGGACTGCTTTCCTGGAAGCTCT |
| | | TGAAACCTGTAGCAAGCCAGGCTGGACTTGGCTTTATTGATCAGCTGTCCTGCATGGCTTGCTACAGGTTTCAC |
| 443 | 150 | CAGTCAGTGCTTGCAAAACCTTTGTCATTTTCTAAAAATAAAAAAGCTAAAAGTTATTGTAGAAGAAGGGGCT |
| | | TTT |
| | | AAAAGAGAATGTAATCTTTTTTCAGCAAGGAATGCTCCCCGCCCATGAGTTTTGATGGGAAAATTATTTTGGAA |
| 444 | 150 | GAGAAGAACAAAGTTTCTATGGCTCCAGGCATCAGTGGAGAGCTGATGAAGGAAGTGCAGTAGCTGGAATCA |
| | | AGCTA |
| | | ACCAGGTGACTGATATTTTCCAGCATCACATCTCTGTACAAGGCCCTCTCGGCAGGCACCATGATGGCCCATT |
| 445 | 150 | CCTTCGTAGTGAAGTACACAGCCACGTCCTCAAAAGCCACTGGGATCGGAAAGAGCACACGTCTGAACTCCA |
| | | GTCAC |
| | | GACACGTGCTGCTGAACTTTCACTTGCTTGTTAATGATGCTGTGTTACAGGCTCTTCTCACAGCCTTCCTGCTG |
| 446 | 143 | AGCACACACACACAGGCCGTATCTATCCGGATAAACCGCCAGGCAGCCTGCTTGCCATCCATGGTCAGC |
| 447 | 73 | AATGGATCTCTTTCTCCACCCATCCCGAGACAGGGCTTGGAGATGCTCATCTGTTAATGGCTACTCAAAGTGT |
| | | GGTTACCACTACAGATGGCACAGGGTATTCTGCACAAAAGGTAAGAGCTAGGGCAGTTGGGCTAGCTGCTCC |
| 448 | 149 | AGTTCCTGCTCCATATGCTCCTGCAGCTCACTGGGCCTCAGCAGGACCTGGCAGATGGGGCAAATTGGGGCCT |
| | | GGCT |
| 449 | 57 | ATCTACGAATTCTGAATGAACTTCCTACCTTCACGGGCACCGTCCGCCCATCATGCA |
| | | ATACTTTGGTTGTTCCTTTGATCTGACCAAAGTGAAGGATTCCAGTTTTGAACAACACAGTGTCCAAATAATGG |
| 450 | 149 | TCAAGGATAATGCAGGAAAAATTAGATCGGAAAGAGCACACGACTGAACTCCAGTCACGACAAGAGATCTCG |
| | | TAT |
| 451 | 73 | GAAAGAGGAATGTGGTACCTTGAAGGCAGTAGATCCACTGCAATCCAAAGACGTCAGTCCTGCCTGGAAACC |
| | | A |
| 452 | 96 | AGGACCAACCATGGGCCTGGGGAAAAACGTTCAGCATTGGTAAAATTGCTGTAGCTGTATTAGAAGAAACCAC |
| | | AAGAGAGAGGACAGATGTTATTGA |
| 453 | 58 | GAGCCATAATGTCGACCTGCCGGAGTCGAGGATTCAGGATAAGAGAGGAGAGTGTCTT |
| 454 | 127 | CAGAGGCAACCGTTTCTAAGTTCCAGCCACTCTTCATAGAGCTCTCCACCTTGGCTGAATCCTGCGTGGGGAT |
| | | GTCAGAGAGGTTTAGTGTCTTAAGAAACTCATCCTTCATGCTCTGGAGCAGTGT |
| | | TACACTGTCCGTGTCACCTGCCTCTACCCTGGGGGTGGCTCCTCTACGCTGACTGGCCGGAAGTCCTCATTGAT |
| 455 | 150 | GTAGGTGAAGCTGGTGACAGGGACAGCCTTCCTAGGGGGCACAGCTGGGCACAGCGGCTCCTGTGCCTCATA |
| | | GCCA |
| | | GCCCAGCTAATTTTTGTATTTTTAGTAGAGACGGGGTTTCACCATGTTGGTCAGGCTGGTCTCGATCTCGTGAC |
| 456 | 150 | CTCGTGATCCACCCGCCTCGGCCTCACGAACCAATAATGAGAAAACAGCAATAATCCACACACCCACCCAGCTCGC |
| | | TTT |
| 457 | 120 | ATGTGTTTCCATTGATTGTTTTCTCATTTCTTTGAAGACATGTGGTCCTCTGTGGGCTGTCCAGCAGGAGGTGC |
| | | CTGAGGCTGCTGGGGATTGCCTCCTGCTGGAGGTGGGGGACCTTGA |
| 458 | 150 | ATTATTTTCTTTAGTATGATTCTGACTGAAATGGAAACTCATGAGGATGCATGGCCTTTTCTACATTCATTCTGT |
| | | CTTTAGTGAGTAATAAACTGATGTTCTCATGCTGTAATGAGCTGGCATGAGTATTTGTGCCACATGGCTCCACA |
| | | T |
| 459 | 88 | CCCCAAGGAGAGGAGCAGTCCCATCATCATCCAGCTGCTGTCAGACAGGTGTCTGCTCCTTGCTCAGGCCTGG |
| | | GGGCAGTGCCTTGTT |
| 460 | 106 | CTGCCCTTTTGTTAACAGGAAGGCCTAGGGAGTCAGGTAAAAAACGGGGATCTACATGGGTGTTTTCTGCTTC |
| | | GACTTTCAGAGTATAAAGTCTTCGGCTCTCATA |
| 461 | 122 | CAGACCCCGTGGAGGCAGGGGAGGGAAGTTAAGGTGAGAATGGACCTGCTGCCGCTCCACATGGGCCTCCACA |
| | | AGCTGTGCAAGGGCTCCTGGGCTTGGCCCAGTGCCAACCCCAGGGATTCC |
| 462 | 81 | TATTCGGATGATAGTCAGAGGGACTCTTGGCAGCTGTAGCAGCAGCAGAGAAAGGAAAGCATTCATTCAGCG |
| | | TTCATTCCA |
| 463 | 139 | CCCAATGACCACCTCAGTCTCCTCCCTAAGTGCCTCCCAGCCACCAGAACCTTCCCTTCCCCTAGAACACCCCT |
| | | CACCCTCACCCACCTGGAAGGCCAGGGCCAGGATGGTCCTTGCTGCAGGCACATCCCCTGCCAGC |
| 464 | 64 | CGATGTGTGCTGAAATCTTGCATTTGTTGCCTTTGTCCTGGACTGCCTCAAACACCAGGTCATT |
| | | GGAGAAGGCGGTGCTGGACGAGCTGGGCGACGCACGGGGACCCGGCTGCAGCCCCTGACCCGGGGCCTCTT |
| 465 | 150 | CGGAGGGAGCTGAGGGCCGCGCCCAGGGCGGCAAGCGCTGGGGGAAGCCTACTCCGCATCCTTGGCCTCCGA |
| | | GCTAGA |
| | | CAAGCATGTCCCCGGGCCAGGCAAGCCGGCTCTGTCTAAAAAGAACCCCGGCGTGGTGGCCTACCAAGGAGG |
| 466 | 118 | CTACTGGAGATCTCAAATCAAGTGTCTCCACGCACGGCATTTCA |
| | | CTTCCAAGAGCCTTTCCTCCCCATATCCACTTTGATGGCTGGAGTCATGGCTATGATTTCTGGTACTCCACTCTT |
| 467 | 149 | GCCAAGAACCCTCTGCACATGGCCTGGGTTCGGGTAATCAGCTGGGCCAGCTTCTCATCTCGCATCTCCTCTAA |
| 468 | 106 | GGACCAAAGTGGTTCCCTGCCCTGCCTGCTCCTGGCCAAGCACTCTCAACCCCTCAGTGACCCACTTCGAAGC<br>ACTTCTGTTCACCTTGTTGATGATACAGTTTTT |
| 469 | 72 | ACAATTGAAAAGGTTTCCTGTGAAATGAGATACAAGGCTACAACTTACGGGTGCAGGGCATAGAGGCAGCAT |
| 470 | 125 | ACGCTTCCAGAACGCCTGCCGCGACGGCCGCTCGGAAATCGTAAGTCGGCTGGCCCGGACAGGAAATCCCGC<br>CACAGGGCAGGAAGTCCCGCCCGCAAACGCTCCAGTCGCCGGCGTAGCGGTT |
| 471 | 89 | ATGGAGATGCCCAGAGTGCTACCCTCAACTTTCTACAGGCCGCGGCTGCCCACTTTGTGACCTGGGAGAAAAG<br>AGGGTGAGAATGTTTA |
| 472 | 150 | CCCACAGTCCAGCCACAACTTTTAGGTCCCGCCCACAGCCGCCGGCCCAGGCTCTGGCCACGCCCCCCCCCCG<br>CCCCCCCCGCCCCCCCCCCCCCCCCCCCCCCCCCCCCCCCCCCCCCCCCACACAAAAACAAACACACACAAAAAAA<br>CACA |
| 473 | 150 | TCTCTTTGTCGAAGCGGAGGTTGTGGAGTCTCAGCAGGCAATGGAAAAGGAGTTAGGATTGAATTTGATGATG<br>AAAATGATATAAATGTTGATGATCTTAAAGCAAGTTCTAAGATTTTCTGAACTTAGTTCTGCCAAAACAATAA<br>ATGT |
| 474 | 150 | TTCTCACCCTGTCTGCTTTTACTTTTTCTCTTTCTTCCTCCCTTTCTCCTTCTCCCCTCCTTCTTTTTCATTATTCTG<br>TTCTGTATTCCTTTCCCCTTCATTCTCACCCTGTCTGCTTTTACTTTTTCTCTTTCTTCCTCCCTATCTCCA |
| 475 | 137 | CATAATTCTGCCTTCAAACAAGCCATTCAAGCCTCTCACAAAAAGGTATCAAAAATGGCTTGTTTGAAGGCAG<br>AATTATGGTTGATATTTTCACTTCTCTTTGAGGAATAAAAAATTGATACACTGAAAGAGCTCCC |
| 476 | 91 | TTTCACCTCCTGGCCAGACCACGGTGTTCCCGACACCACTGACCTGCTCATCAACACCTACCCCAAGTCTGGTA<br>AGTGAGGAGGGCCACCC |
| 477 | 121 | CCGGCGGCTAGGCGCGGGAGAAGTGCGGAGGAGCCATGGGCGCCGGGAGCTCCCTGCGTCCCCCTCAGCCAT<br>GTGGCCCCTGGACCTGCCCTGCAGCCGCCGCCTGCCAGGTCCGGAGCGC |
| 478 | 150 | GTCTCAAATTCCTAGCCTCAAGTGATCCTCCCATCTTGGCCTCCTAAAGTGCTGGGATTACAGGCATGAGCCAC<br>TGTGCCCCGCCAAGAGTATAATATTTAATTATACTAGATTTTATTTATATATCTTGTGACACTGAGGCAGTGAT<br>GA |
| 479 | 120 | TGAAAAGGCATTATGTGAAGGCGAAATAATTGATAAGCGAAGAGAGAGGGAAACTCTAGAGAAGGCTGAAA<br>AAAACGCCTAGAGGCACGTGGAGTAGCCCCAGACAATTGAGTGCAGCAA |
| 480 | 149 | GCGGTTTCCTGCTTGCTGATTGCGTACGGAGCAAGTAAACCAAACGGTGAGTGTCCTCTCCCTCCATCTTCTGT<br>CAGGTTCAGCTTCCCATTGGCGAACGCCCGTCTGGTGAACTCGCCTGCCTCCGCCGGTCGAAGCCCTGGCACG<br>CT |
| 481 | 149 | AGAGATCGCGCCACTGCACACCAGCCTGGGCGACAGAGTGAGACTCTGTCTCAAAAAAAAAAAAAAAAAAAA<br>AAAAAAACAAAAAAAAGAAAGCAAAAGACCACGACTGAACACCACTAACGACACGAGAGGAAGGATTGCGA<br>CTGCTGA |
| 482 | 107 | CTCACTGTTGGCTCTCTTAACTGTGTCCACTCCTTCATGGTGTCAGAGCACTGAAGCATCTCCTGATTAATTTTT<br>AGTTCTTCTCCTGGTCCACTTTCCTTATGAAT |
| 483 | 150 | GGCTCCGGCTGGCCCTGAGTAGCCGAGAGGATGAGCTGGTCCGCACGCAGGCCTCCCTGGAGGCCATCCGAG<br>CTGAGAAGGAGACCTGGAACGAAAGGGCACTTTGGACTGGAAACAGCCACAGAGCAGGAGCCACAGGGCGC<br>GCTGGAT |
| 484 | 150 | AAGAACTCAACATTCACGTATTTCACCTCTGACCATGGAGGACATTTAGAGGCAAGAGATGGACACAGCCAG<br>TTAGGGGGATGGAACACGTAGGAAACATACAGAGGGACTCGAAGGGTCACATAGGACTGCACTAGGTTTAGG<br>ACCTAA |
| 485 | 150 | ACTTCGGGCACGGATGGCCAGCCTGCGTCAGGGCTGCGGGGACCTCCGAGGTTTGGTCAGCACCTTTACCCAG<br>AGCTGTCAGGGTTCTTACAGGTTTCTATCCACCAGCGGTGAGGTGTCTCCTCTGTGTCCTCTCCCACCTGAAAA<br>ATA |
| 486 | 102 | CAAACTCAAGTATTTTTTTGCTGTGGACACAGCCTACGTGGCCAAGAAGCTAGGGCTGCAGGGTACAAAAAGC<br>CAATTTCAGAAGTAAGTGGACAAGGCAGA |
| 487 | 150 | GAGCCTGTCGGTGATGCTGGTGCGAAACCGCGACGAGGTGCAGGCGCTGGCCTTCGACGAGCAGCGGCGGCC<br>ACTCCCGCGCCGCGCAGCGAGGGAGACGCGGACGAATGCCCGGGGCAGCGTACAAAAACTAAAGAGGACGA<br>AACAGCA |
| 488 | 113 | CATCAGATGCCGAGCATCAGTGTCAGTCCCCCCAACACGAGCCCCTTCAGCTCCCTGCACCTGTTCGGGTGCT<br>CCGTGCTGTTCCGCCATCTCCCCAGGAACACCTCTGGCTT |
| 489 | 125 | TGCCTCCAGCACCCTGAAGATGCTGTGTGATATGTCCTCCTTACCAAGTTGGTGATGGCTGCCTCAGTGGTACT<br>TAATGCTGATATGCTCACAATTTCTGTATAAAATAGAAGTTGAGAAAGGGA |
| 490 | 150 | AGTTGTGTGGCCGGGACCCTACTCTCACTGATGAGCTGCTGAATATTCTCACAGAGCTAACTCAACTCAGTAA<br>GACCACCAATGAAACAAAATTTTAGCACACACATACCTGTGGAACCATTGCAACCCTCTGGTTTCTTTTAGGT<br>GACC |
| 491 | 122 | AAACATGGCAGAAATGGAGAAAGAAGGGAGACCTCCCGAAAATAAACGGAGCAGGAAGCCGGCTCACCGCG<br>ATCAAGTTCAACACCAACAGCACTTTGACTCTTCTCAGTGAAGCCATCAAA |
| 492 | 149 | GCGTCCAGTTGGTACGCAAGTTTGCCCAGTCCATCTTGCAATCTTTGGATGCCCTCCACAAAAATAAGATTATT<br>CACTGCGATAGATCGGGAAGAGCACACGTCTGAACTCCAGTCACGACACGAGATCGCGTATGCCGTCTTCTGC<br>GT |
| 493 | 150 | CCACTGCTCTTTCCTCCTCCCCAGCTGAACCTCCACAGTCCTCTACACTCTTCCAGGAGCAGCAGAAAATGAAC<br>ATAAATGTCCCCGCAAAACAACAAAAAAACATCCAATCAGGGTTTGTCCCCAGCTGCACTGCAATACAAGTTC<br>CAA |
| 494 | 114 | TTATAAATGTAGCAAATGTGAGAAGAGCTTTTGGCATCACTTAGCGCTTTCAGGACATCAGAGCAGGCGGGAC<br>AGCGGGTGGCTGTCCAGGGCTCCAATCAGCTCCGAGCAGGC |
| 495 | 150 | CAACTGAGTTTGAATTCAATCTCTATTACTTAAAGTTCTTTGGCTTTCCACTATGTGTCTAATCGCTTTAAAACAT<br>TTCTCCTTGCGCACCTGCTGCATGCAAAATCTTCAAAAGCGGCAGTTTTGGTCGTACCTGATATAAACATAAA<br>AT |

```
          TGCATTGGATAAAACAGAAGAGCTCGAGGTGAGCAACGGCCACTTAGTGAAGCGTCTGGAAAAAATGAAAGC
496  150  AAATCAATAACCTCTCTCTTATCTACCACAGGACTTGAGTCAGCATTATCCATCAAAGGAATATAAGTAGTTG
          CATGA
497  100  GAGAAATAGTCAGCCTAGCAGTTCTACCATGATCAGCGTGCTTCGAGCGGGTGGCGCCGCTCGGCCAGGCCCC
          CGCGGCCCAGCACTTCGCCGCTGAACT
          GAGAGGAAGCCGAAGCTAAAGAGGAAAAGAAAGTGGAGGAAAAGAGTGAGGAAGTGGCTACCAAGGAGGA
498  150  GCTGGTGGCAATGGAGGGTCTGCAATAATCTGAATAGCCTCTTCCTCTGAAATTGGCTGCATTGGAGACAATG
          GAGGCAA
          GTCAATAATTTTGACCAAGACTTTACCCGGGAAGAGCCGGTACTCACCCTTGTGGACGAAGCAATTGTAAAGC
499  150  AGGGTCAGAGCAGTCACTGGGTTTACCCGGAGGTACTTGCAGTTGGGCTGACCATAGTCAGCAAGGTAGGTTG
          ACCA
          GCTATATATGATTTCTTCCAAGTGGCCTCCAGCAAGGAGCGGTATAAGCTGACAGTTGGGAAATACAGAGGCA
500  150  CGGCAGGAAAGGCGGCCCGAACCCGCCGCCGCTGCCGCTAACGCCAACAACACCAGCAACGCTGCTCGCCCC
          GGGGC
          CCAGAAGCCGGGCCTGGAGCTGGCCCCGGCCGAGCCCGCGTACCCGCCGGCGGCGCCGGAGGAGCCGTTCGA
501  122  TGATGTTCAGCAGAATGTCCCACGCCACCACCTGGAGCTCCTTCCTATAC
          GCTGAGACAGGGTCTCACTCTGTCTCAGCAAAAAAAAAAAAAAAAAAAAAGAAAAAACAAATACAAAAG
502  150  TGGGAAAAAAAAAAAATGTCCTAAATGAAAAAAAAAAAATTTTCTAAAACGGAAGAGCACACGTATTAACACCA
          GAACCCA
          GAGTATGTGAATCTGTCAAGCAGTTTTCCCACCATCCATTACATCCAGGGGGGATTCCTTCGGCATCTTAAT
503  150  TCACTAATAAAAGTCTCGAAGTAACCCCAATCCTGATTGGAATTTGCAAGCAAAGCTTCTTCTGTATTGCACTT
          GT
          GATCCAAACAGCACACCAATGAACTCAACTCACGACACGAGATCTCCTGTGCCGTCGTCAGCATCCTAGACCA
504  150  GCCGGTGGCCAGGGGGGTGCGCGGTGTACGCGGGAGCGGGGTAGGGATAGGGGTGAGGCGGCGGGTGCCGC
          TGCTCC
          CTGGACTTGAGCAAGCTGTCACCATCTTCTTCTTCTTCCTCATCCTCATCCAGCTCCAGCTCCCGGCGTGGGCT
505  150  GTCCGGGGGGGCCGCCCCCCAGGGGCGGCGGGGAGAGCACACGTCGGAACGACAGTAACGACCACAGAGAAG
          GTTAT
          AGAATTTTCTATTTTTCTACACTACTCTGTGCTATATGTCATCCCATTTAAAAAAAATCTGCACAATTATGATTC
506  104  AGTTGAAAACAAAGATAGTACCACTATAA
          GAGCACTCTCAAAAGTCTTACTTTAAAAAAAACAAAAACAAAAACAAAAAATGGCCGGGCACGGTGGCTCAC
507  150  GCCTGTAATCCCAGCACTTTGGGAGGCCAAGGAGGGGGGATCATTAGGTCAGGAGTTAGAGAGCACCATCTC
          AATAAT
          CATAGCAAATTACAGAAACGGAGGTAAGAACGTAAGAAATCGCAAAGCACGTGACGTAATTTGCTAT
508  66   GTCTCTTCAGTGGGGATGGTGGGTTTTTCCGTGGGGATGGTGGGTTTTTCCGTGGAGATGGTGGGTTTTTCTGT
          GGGAATGGTGGGCTTTTCTTTGGGGACTGTGGGCTTTTCTGTTGTGACACTGGGTTTTTCTATGGAAACTGTAG
509  150  AT
          GACCTCGTGATCCGCCCGCCTCGGCCTCCCAAAGTGCTGGGATTACAGGCGTGAGCCACCGCGCCCGGTCAAT
510  150  AAAGGACTCTTAAATCGTCTCAAAGTGTGGCGTTCTCTCTAACACCCCTGGGCACAACAAGAGGTTGTGGTGA
          GCGA
          GTCTTAAGACAGTTGCAGCTTTTTAACTGTTACTAGTTTTAAGAATAAAAATTTACGTTTGACCTTTGTAACAG
511  140  GAGTACAGTAATCAGAGGATTGTATTTTCCCAGATGGATGTAGCATTTTCCCATCTATTGTAAACA
          GGGGCAACAACTGAGTGCAACTGAGATTCTGTTGGAAGGAAGACGAGAACTTCATCCTGTGCTTTCTCCATCA
512  150  CCTGCCCTCTTGGTCTCCATCATGGCACTCCAGTCAGGAGACAATCACAAATTGGAATGCAAGTGGGTGTGT
          AGAC
          AAGAACAATCTGCCAGTTATCAAGACAGGAGGCAAATGAAGGGACTTCGACTGAAGTCTTTCCTGGTGGCTG
513  123  AGATCACAATTTCCCCCTTTTTTAGGGCTAGAGCCAAGGTGAATTCTTTCC
          CTTATAGCAGCAGCCAGTCTAGACTGATAAGATTCAATGTCAGCTTCCAGTCTTTTCTTGCTTTCTTTTTCCTTC
514  149  AACAGTTCGGCATTGAGCCTTGTATTCTCAGCCTAGATCGGAAGGAGCACACGTCTGAACTCCAGTCACGACA
          A
          CAAGTGATGGCACATTCCTCCAGTGAGTCCTCAGGGACTTTGCTTCCTCCTTCGTTGGAGACTTGGCCTTTTCC
515  123  GGGGACTTGACCTCAGCTGGAGATTTTGCTTCTTCCTTCACTGGGGACT
          CCACAAGGGAATTTGAAGAGATTCTGTGAGAGAGATATGAAAATATTTTATAGTTAAAACACCAATATATTCAA
516  139  CTTTACACTTCTTGTGTTTTTCATAAACATTCCTGTTCTTTTCCTTTCAGTAATAAAATTTAAAAA
          CATCCCCCAGAGTGCCGAAGCTGCAGCCGAGGCCACCAAGAATATGGAAGCTGGAGCCGGAAGAGCTTGGCC
517  87   ACGGCGCCACACACCC
          CCCACACTGGGTTAGTTGATCCTCTGGGGTCCTCTGCCTGGCCATCTTCCTCTCAAGATTTGCCCTAAGGGTCA
518  118  CGGCCTCTATGAAGCTTTCCCTGACCCTCCCACACTGGGTTAGT
          ATTATTTTTTTAGTAAAATTGTTTGAAGTTATTGAAACAAGAGAGTCAGACATTTGCAAAATATTACAATATAC
519  110  AGAATCCACCATCTTTTAGGAGGTTTTTTCCAGTATA
          CCACACACCAACAATTATAGCAGGTCCACAGAACATAACAACATCTCTTCATCAGACTGTAGTTTTGGAATGC
520  113  ATGGCCACAGCCTTTCCCCTCAGAAGCAGCATTCTTTGAA
          ACGCCCTCTGGCCTTTCCTCGGGTTCCTTCCTCCCCCATGTCCAGGAAGGGCCTGGGGGGCGGCTC
521  65   TCCAATGAAACAACAAGAATCATATTTCTTGGCCACCATTTTAAATCATGTTAATCCATTTCTCAGAATACTGC
          ATCTGCAAAAAAAAAAAAAAAAAAAAAAAAAAGGCGTTAAAAAAAAAAAAAATGTTTTAAAACCAAAAAAA
522  150  AACCC
          AAGTACTGGAAACCGAAACCCTGGGAGCTCTGGGGTCTGGCTGCCTCACACTTGGGGTGGAGAATGTGGCAT
523  103  CCCTCGCTGACAGGATGGGAGAGTCCCGGCCC
524  104  TGACCGGGTTTCTTTGATGCGACCTCAGGAAGGAGATGAGATTTGTGGCTCTTGATTCCTCCTTGAAGTCTCAGT
          TTTGATCAAATCTTCTTTAGAGTGTGATTT
```

```
525   150   CCAGGAACAACGGCCACCCAGACGACAGGCCCACGTCCAACCCCAGCAAGCACCACAGGCCCAACCACCCCA
            CAGCCAGGACAACCCACGAGGCCCACGGAAACGCTCTGTGGAAGACACAAAGCCTTTGGCCTGCTGGGCTCT
            CCGGGA
526   137   GATCTGGGGCTAGTGTAGAAGGTGGTTGATTTCTCACTGAGGCTTGGTGTTGTGGGACTGCCGGGGCAAGAGA
            AGACTGCCTAGAATTAAGTGTGGAGGGTCTTATAGGATACTCTTCGTTTTGTACATTTCCAGCA
527   87    GAAGGTCAGGATTTTAAGAAGAAGAAATGGAGCAGAATGCATCCTAGATCTGGGCTACCATATGAGACAGCC
            CAGATCTAGGATGCA
528   150   CCTCTCAAAGTGCTAGGATTATAGGCGTGAGCCACTGAGGCCTGCCTGCCTGCAGTTTTCATTGAGGTGTTTTT
            TTTTTTTTTTTTTTTTTTTTTATACGGAGGAGTGAAAAAGTACCCCAAGCAGGAGGACAGTGGCAGTAAAATTGC
            TC
529   150   ACTTCTTTGTGATGTGTGTATTCAACTCACAGAGTTTAACCTAACTTTACAAAGAGCAGATTTGAAACACACTG
            TTTGTAAAGTATGCAAGTGGAGATATCAACCGAAGAGAACACATCTGAACACCAGGCACTAACATAAAAAAA
            AAAC
530   150   ACTGAGATCCTACCATTGCACCCCGGCTTGGGAGACAGAGAGAGACTCTATCTCCAAAAAAAAAAAAAAAAAA
            AAAAAGGGGAAAGTAATTGTTGGAAGGGGAGGGAAGGGCAACACTGGGAAAAACCTGACACCACAGGAGA
            GGGGGTT
531   149   TCAGTGGGATGTACATGGGGGAGCTGGTGAGGCTTATCCTGGTGAAGATGGCCAAGGAGGAGCTGCTGAAGG
            AGCCGTTCGATGATGTTCAGCAGAATGTCCCACGCCACCACCTGGAGCTCCTTCCTATACTTCTTGATGAGCCT
            GGT
532   102   GTTTTATCCTGGGTGTCATTTTGGATCAAAAAAGATGCTACGCCAGCAAGAACAGCATTATAAGGCCAGACGT
            GACAATAGGAGAGATCCCTAGCTCCATCA
533   150   TTTCTCTGGAAGCTCACCCACTGCAGCCTTCAGCCGAGTCAGCTCCTGGTAGATCTCCTGCATCTTAGATTTCT
            CTGGAAGCTCTCTGGGATGTCGCAAACATCCAGCTGTGACCACTGGCTGAATTCAAAGGTGAAATTGTTATTA
            ACT
534   149   TGGGGTGTTTGTTTTTTTCCTGTAAATTAGTCTGAGTTCATTGTAGATTCCTGATATTAGCCCTTTGTCAGATG
            GATAGACTGAGAAGTTTTCTCCCATTCTGTAGGATGTCTGTTCACTCTGATTAAAGTTTCTTTTGATGTGCGAA
            AGAGACGGGGTTTCACCGTGTTAGCCAGGATGGTGTCGATCTCCTGACCTCGTGATCCACCTGCCTCAGCCTC
535   150   CCAAAGTGCTGAGATGACAGGCGTGAGGCACCGCACCAGTCCTTTTTTTTTTTTATTATTTGAGAATGAGTTTCA
            CTA
536   149   GTGTAAGGAAGGACTCACATTCTTTTTTTTTTTTTTTTTTCAACTGGGGCCTTGTTATGTTGCCCATACAGGAGT
            GCAATTGTTTTATCATGGCTTACATCAGCCATTGTCTTATGGGCCCAATCAAACATCATGCCTCACCCTCCAGA
537   73    CCACTCACTGCAACCTCTGCCTCCCAGCTTCAAGCAATTCTCCTGCCTCAGCTTCTGGAGTGGCTGGGAGCAC
            ATCCTATGCCAGAATAACAGAGATAGATAGATACATAGATAGATAGATAGATAGATAGATAGATAGATAGATAGAT
538   150   AGATAGATAGACAGATGATAGATAGATAGATAGATAGATAGATAGATATTGATTGATAGATGAAAAACTGTA
            ACCACA
            TAATATCTCTTAATAAATCCCTTTCTGTGACCTGCACGTACACATCCAGATGGCCGGTTCCTGCCTTAACTGAT
539   150   GACATTCCACCACAAAAGAAGTGAAAATGGCCTGTTCCTGCCTTAACTGATGATATTACCTTGTGAAATTCCTT
            CT
            TCTTGCTCAAATTTTCCATCTTGGACCATTGGACTCTCTTTCGGTTGGCTCCTGTGTCCTTTGTCCTTTTTTTTT
540   150   TTTTTTTTTTTTTTAAAGTTTAAAAAAATTTTGCCCAGGCGGGCGGACCGTTTTGTTATAACCAAAAAAAAGC
            GTGTAAGGAAGGACTCACATTCTTTTTTTTTTTTTTTTTTTAAAAAGGGGCCATGTTATGTTGCCCAGACAGGAG
541   149   TTAAAAGGTTTTATAAATGCATACAGCAGCCATTGTCTTCTGGGCCCAAGCAAACATAAAGCCACAGCCACCA
            GA
            CCCTCCATATATGTATATATACATGCACACATCCTCTCATCTACCAATCTATCATCCATCCATCCATCCTCCATC
542   150   CTTTTATTCATCCATCCATCCTCCCATCCATATACATATATATACATACATACATCCTCTCATCTGCCAATCTAT
            CATTCCACTCCATTCCATTCCATCCCATCCACTCCATTCCATTCCATCCCAATCGGGTTGATTCGATCGCAATAGAAAA
543   150   CATTCCATTCCATTCCATTCCTTTCCTTTCCATTCCGTTCCATTCCATTAGGGTTGATTCGATCGCAATAGAAAA
            AAACCTTTTTTCATTCAGCAGTTTGGAATCACTGATTTGTTAGGATCAGAGACGAGAAAGGTAGTAGTGAAT
            AAATGAGCCTTAAGGTGAAAAAGCAAATATTTCAGATAAGAGCTAGAACGAAGCTGTCTGAGAACCTGCTGG
544   149   GTGA
            CTCAGTGCAGATCACTGGTTTTATCAGAAGGTCAGTGGTTGGAGACATCCATTTTCTGAAATCAGGTAAAGTG
545   150   TTGAATGCTGTATATAGCACCCGAGAGGTGAGGCAGATATTTATCAATTTAGGTACACAGGTGTCAGTTTGAT
            TATT
            CTTTTTCATCTCTTATGCCAGTATATAGTTGGTTTTTGAAGTCAGGTGTGTAGGCAGGCTTATTTCCTAGGCATA
546   150   TCATGACAGGAAAATGAAGAACGCTTTTTTTTTTTTTTTTTTTTTAAGGGTTTTTTCCCATGTCGCCCCAGCAGGA
            CAGGTTATGAAACCAGTTAGTTTTTGTAATTTTTTTTTTTTTTTTTTTTTTTTTTTTTTTTTGACGTGTTT
547   149   TAACCGTTTTGCCAAGGCTTGGTCCGAGGGAACCACCGGCCCGCGGCCACAAAACGGAAGAGCACCCGACA
            GGCGGAGGTTGCAGTGAGCTGAGATCGCGGCCACTCCACTCCAGCCTGGGCGACAGAGCGAGACTCTGTCTC
548   150   AAAAAAAAAAAAAAAAAAAAAAAAAAAAAAGGGAAAACAACACGGGAACGAAAGACAAAACGAATAAAAACAA
            GGAACAAC
            CCCCATTCTTCCCTCTCTCTGTCCTCAGAACACTGCCTCATATCCTTCCCTGGTCCCTGGCTCTCTGAGTCCCTC
549   150   GTTTTTTTTTTTTTTTTTTTTTTTTTAGATAGATAGGAAAAGCAAACCTCTGAAAAACACTCACAACCACAGA
            GGCAGAGTTAGACTGTCGGAAAAAAAAAAAAAAAAAGAAATGTATTGAAGTAAAAAATAGACAAAAAAA
550   150   TTTTAAAATTGGAGAGTAATTTTGATAAATTCTGAGGGCAGTTGGAGAAGAGAACGGAAAAGCAAAATTAAG
            AACTCCA
            CAACTCAATAAGGTGAAAAGTCCAAACCTGACAGGTCGAGAATAAATAAATACAGCATCTAGATCATTATTG
551   149   TGCCAGGGGAAAAACTGGAAAAACTGACTTAGACATTGGCATTAAACTTTAATATCTAACGGGGGCTTTGGTT
            TTGA
```

552 148 ACTCGATTGTTATGGAATGGAATGGAATGGAATGGAATGGAATGGAATGGAATGGAATTAACCCGAATACAA
TGGAATTGAATGGACTGGAACGGAATGGAGTGGAATGGAATGGAATCAACCCGAGTGCAGGGGAATGGAAT
GGAAT

553 147 TCTTCCCAAGGGAGGAGGAGCTCAAGTGTCGGGAACTGTCTAACTTCAGGTTGTGTGAGTGCGTTAAAAAAAA
AAAAAAAAAGAAGCCAAAAACAACATTTTTAAAACGAAATACCAACCGAATAAAAACCAGACACAACAAGA
GAA

554 149 ACTTCAGTACCACAGGTTAAGGGAGTGGTAGTTGAGTGACAGCTATGCTTCTGCTGTGTAACCTTTAGCAAAC
CAAATAATTTTCTAATGCAAAACTTTTTTTCTGATCAAATCTACAGTACTATTATTCTAAATTCCTTTTCAATAT
A

555 150 CCAGTCCTCCATCTCTGGAGTGACATGGTCAATGCTGCATCTTCACAAAGCAGGGACACTAATCCCATTTAAG
AGGGCTCCATCTTCAGACCTAATTATCTTGCAAAGGCACCACCTTTAACATATGAATCAGGAGGGTGGGAATA
CACT

Table S4. Geographical classification of DNA sequences from the subset of ray-finned fishes

| Group | Sequence ID | Number | Average affinity | E-value | Blast Mode |
|---|---|---|---|---|---|
| Non-native | #001 #002 #003 #007 #009 #010 #012 #013 #014 #015 #016 #164 #165 #166 #168 #169 #017 #049 #176 #179 #180 #181 #182 #183 #190 #191 #192 #193 #194 #195 #196 #199 #205 #208 #223 #225 #226 #233 #236 #240 #246 #253 #254 #258 #259 #260 #261 #262 #272 | 49 | 80.34% | ≤ 4.00E-9 | the MS mode |
| Native | #004 #011 #018 #019 #020 #021 #022 #023 #024 #025 #026 #207 #028 #029 #030 #031 #032 #033 #034 #035 #036 #037 #038 #039 #040 #041 #042 #043 #044 #045 #046 #047 #048 #050 #051 #052 #053 #055 #056 #057 #058 #059 #060 #061 #062 #063 #064 #065 #066 #067 #068 #069 #070 #071 #072 #073 #074 #075 #076 #077 #078 #079 #080 #081 #082 #083 #084 #085 #086 #087 #088 #089 #090 #091 #092 #093 #094 #096 #097 #098 #099 #100 #101 #102 #103 #104 #105 #106 #107 #108 #109 #110 #111 #112 #113 #114 #115 #116 #117 #118 #119 #120 #121 #122 #123 #124 #125 #126 #127 #128 #129 #130 #131 #132 #133 #134 #135 #136 #138 #139 #140 #141 #142 #143 #144 #145 #146 #147 #148 #149 #151 #152 #153 #154 #155 #156 #157 #158 #159 #161 #170 #171 #172 #173 #177 #200 #207 #209 #210 #213 #215 #216 #217 #219 #228 #229 #241 #242 #230 #237 #238 #243 #244 #245 #247 #250 #251 #252 #264 #265 #266 #267 #268 #269 #270 #273 #275 #276 #280 #281 | 180 | 83.85% | ≤ 3.00E-14 | |
| Other ray-finned | #008 #054 #137 #150 #160 #162 #163 #167 #174 #211 #218 #227 #232 #277 | 14 | 77.51% | ≤ 5.00E-12 | |
| Non-specific | #175 #178 #184 #197 #198 #202 #203 #204 #206 #212 #220 #221 #222 #235 #248 #278 | 16 | 65.45% | ≤ 3.00E-9 | the lowest mode |

The transposase sequences are underlined in the table while the mitogenomic and ribosomal sequences are presented in bold. All sequences can be found in Table S3.

Table S5. Results from DNA Extraction and Sequencing

| ID | Fossil portions | Sample weight (g) | aDNAmix (ng/mL) | Sequences amount |
|---|---|---|---|---|
| 1 | texture | 30 | 33 | 22253 |
| 2 | texture | 28 | 54 | 30932 |
| 3 | texture | 40 | 99 | 445752 |
| 4 | texture | 50 | 141 | 735110 |
| 5 | texture | 25 | 35 | 24976 |
| 6 | non-texture | 50 | 73 | 52 |
| 7 | non-texture | 50 | 82 | 583 |